\documentclass[11pt]{article}
\usepackage[left=1in,top=1in,right=1in,bottom=1in,head=0in,nofoot]{geometry}

\usepackage{setspace,url,bm,amsmath} 
\usepackage{scalerel}

\usepackage{xcolor}
 \usepackage{multirow}
 \usepackage{arydshln}
 \usepackage{mathtools}
\usepackage{titlesec} 
\titlelabel{\thetitle.\quad} 
\titleformat*{\section}{\bf\Large\center}

\usepackage{float}
\usepackage{graphicx} 
\usepackage{bbm}
\usepackage{latexsym}
\usepackage{caption}
\usepackage[margin=20pt]{subcaption}
\usepackage{enumerate}
\graphicspath{ {./Graphs/} }
\def\AB{{\textsc{ab}}}
\def\AC{{\textsc{ac}}}
\def\ABC{{\textsc{abc}}}
\def\BC{{\textsc{bc}}}

\def\rss{\textsc{rss}}
\def\tta{\textup{A}}
\def\ttb{\textup{B}}
\def\ttc{\textup{C}}
\DeclareMathOperator{\diag}{\text{diag}} 
\def\prg{\medskip\noindent\textbf}

\def\mes{{multi-armed experiments}}
\def\mess{{multi-armed experiments }}
\def\tosrp{\tilde\Omega_{*,\textup{u},\pp}}
\def\hybr{\hy\la \hbr\ra}

\def\hybrs{ \hy\la\hbrs\ra}
\def\hybs{\hy\la b_*\ra}
\def\hybi{\hy\la \bi \ra}
\def\hyba{\hy\la b \ra}
\def\la{\langle}
\def\ra{\rangle}
\def\hsxy{\hat S_{xY}}

\def\rpp{R_\perp}
\def\rpt{R^\T_\perp}

\def\xpp{(X_1 + \cdots + X_{Q})}

\def\hssr{\hat\Psi_{*, \rr}}
\def\ccinvs{(\chi_*^\T\chi_*)^{-1}}
\def\Ms{M_*}
\def\murq{\mu_{\rr,q}}

\def\hsig{\hat\Sigma}
\def\hsigr{\hsig_\rr}
\def\hsigrq{(\hsigr)_{[Q]}}
\def\hsis{\hsi_*}
\def\hesi{\hep_{*,i}}
\def\hsxq{\hat S^2_x(q)}
\def\hsxyq{\hat S_{xY(q)}}

\def\qit{{q\in\mt}}

\def\ninv{N^{-1}}
\def\cspt{\csp^\T}
\def\csmt{\csm^\T}

\def\mlr{M_{\lin,\rr}}

\def\chil{\chi_\lin}

\def\hepri{\hep_{\rr,i}}

\def\rb{A}
\def\mub{B}
\def\apo{average potential outcomes}
\def\cpoc{correlations between potential outcomes and covariates} 

\def\torp{\tilde \Omega_{\rr,\pp}}
\def\zikz{\zik^0}
\def\zimkz{\zimk^0}

\def\ust{\us^\T}
\def\co{\go}

\def\go{correlation-only}
\def\gor{correlation-only restriction}
\def\so{separable}
\def\sor{separable restriction}
\def\sr{separable restriction}

 \def\pmo{\{-1, +1\}}
\def\hylr{\hy_{\lin,\rr}}
\def\hyfr{\hy_{\fisher,\rr}}
\def\hynr{\hy_{\nm,\rr}}

\def\ttrp{\tilde\tau_{\rr,\pp}}

\def\ttrmk{\ttau_{\rr,\mk}}
\def\ttsp{\ttau_{*,\pp}}
\def\ttnp{\ttau_{\nm,\pp}}
\def\ttfp{\ttau_{\fisher,\pp}}
\def\ttlp{\ttau_{\lin,\pp}}

\def\tts{\ttau_{*}}
\def\ttn{\ttau_{\neyman}}
\def\ttf{\ttau_{\fisher}}
\def\ttl{\ttau_{\lin}}

\def\ttfrp{\ttau _{\fisher, \textup{u},\pp}}
\def\ttnrp{\ttau _{\neyman, \textup{u},\pp}}
\def\ttlrp{\ttau _{\lin,\textup{u} ,\pp}}
\def\ttsrp{\ttau _{*,\textup{u},\pp}}

\def\ttsmk{\ttau_{*, \mk}}
\def\ttnmk{\ttau_{\nm, \mk}}
\def\ttfmk{\ttau_{\fisher, \mk}}
\def\ttlmk{\ttau_{\lin, \mk}}

\def\ttnrmk{\ttau_{\nm, \textup{u},\mk}}
\def\ttfrmk{\ttau_{\fisher, \textup{u},\mk}}
\def\ttlrmk{\ttau_{\lin, \textup{u}, \mk}}

\def\ttsrmk{\ttau_{*, \textup{u}, \mk}}
\def\ttau{\tilde\tau}
\def\murs{\mu_{\rr,\sss}}
\def\us{U_\sss}
\def\vrrs{V_{\rr,\sss}}

\def\tos{\tilde \Omega_*}
\def\ca{c_\A}
\def\cb{c_\B}
\def\cab{c_\AB}
\def\csk{c_{\sss,\mk}}
\def\cmk{c_\mk}
\def\ndt{nondegenerate linear transformation of the  regressors}
\def\hyrz{\hy_{\rr,0}}

\def\hgrz{\hg_{\rr,0}}
\def\hyrs{\hy_{\rr,\sss}}
\def\hbrs{\hg_{\rr,\sss}}
\def\hgrs{\hg_{\rr,\sss}}

\def\kik{{k\in\mk}}

\def\zik{Z_{ik}}
\def\zimks{\zimk}

\def\tz{\tau_0}
\def\tzp{\tau_{0,\pp}}
\def\tzm{\tau_{0,\mmm}}
\def\czm{C_{0,\mmm}}
\def\czmk{c_{0, \mk}}

\def\tsmk{\tau_{\sss,\mk}}
\def\tsm{\tau_{\sss,\mmm}}

\def\csmt{C^\T_{\sss,\mmm}}
\def\csm{C_{\sss,\mmm}}
\def\csmk{c_{\sss, \mk}}

\def\tzmk{\tau_{0, \mk}}

\def\epni{\ep_{\neyman,i}}
\def\vrr{V_\rr}

\def\cst{C_\sss^\T}
\def\cs{C_\sss}
\def\csp{C_{\sss,\pp}}
\def\ts{\tau_\sss}
\def\tsp{\tau_{\sss,\pp}}
\def\ta{\tau_\A}
\def\tb{\tau_\B}

\def\sss{{\textsc{s}}}

\def\tab{\tau_\AB}

\def\black{\color{black}}

\def\mrm{\mathbb R^m}
\def\mc{\mathcal C}

\def\preci{\preceq_\infty}
\def\opo{\Op(1)}
\def\oo{O(1)}
\def\ei{E_\infty}
\def\covi{\cov_\infty}

\def\brq{\gamma_{\rr,q}}
\def\xir{\xi_\rr}

\def\ary{A\rhy}
\def\cxi{x_i}

\def\lr{{\rr}}

\def\czmk{c_{0, \mk}}

\def\hsigl{\hat\Sigma_\lin}

\def\hsigs{\hat\Sigma_*}

\def\heli{\hat\ep_{\lin,i}}
\def\heri{\hat\ep_{\rr,i}}
\def\helri{\hat\ep_{\lr ,i}}
\def\hefi{\hat\ep_{\fisher,i}}
\def\heni{\hat\ep_{\nm,i}}
\def\chin{\chi_\nm}

\def\hsr{\hsi_\rr}

\def\czp{C_{0, \pp}}
\def\cz{C_0}

\def\pk{\mathcal P_K}
\def\mfm{\mathcal F_\mmm}
\def\mfp{\mathcal F_\pp}
\def\mfpp{\mathcal F_\pp'}

\def\tmk{\tau_\mk}

\def\tauAB{\tau_\AB}

\def\reslp{(R\tl - r)}

\def\mur{\mu_{\lr }}

\def\zimk{Z_{i, \mk}}

\def\rols{\rls}
\def\rolss{\rlss}
\def\rls{\textsc{rls}}
\def\rlss{{\textsc{rls} }}

\def\mb{\mathcal{B}}
\def\vlr{V_\rr}
\def\vrr{V_\rr}

\def\uy{U_Y}

\def\dz{\Delta_0}
\def\es{\emat\otimes\sxx}
\def\esinv{(\es)^{-1}}
\def\ry{R_Y}
\def\rg{R_\gamma}
\def\rhy{\rho_Y}

\def\rhg{\rho_\gamma}

\def\rhyt{\rho_Y^\T}
\def\rhgt{\rho_\gamma^\T}
\def\beginp{\begin{pmatrix}}
\def\endp{\end{pmatrix}}

\def\meani{N^{-1}\sumi}
\def\ryt{R_Y^\T}
\def\rgt{R_\gamma^\T}

\def\coremat{\Phi}
\def\emat{\Pi}
\def\einv{\emat^{-1}}

\def\tl{\theta_\lin}
\def\nfl{* =\neyman,  \fisher, \lin}

\def\nf{* = \neyman, \fisher}

\def\wiq{{ 1(Z_i = q)}}

\def\ccinvs{(\chi_*^\T\chi_*)^{-1}}
\def\Ms{M_*}

\def\hefi{\hat\ep_{\fisher,i}}
\def\heni{\hat\ep_{\nm,i}}

\def\chin{\chi_\nm}

\def\htlr{\htau_{\lr }}

\def\prodk{\prod_{k=1}^K }
\def\hsi{\hat\Psi}
\def\hyg{\hy\la\gamma\ra} 

\def\succi{\succeq_\infty}

\def\hglq{\hg_{\lin,q}}

\def\sri{\Sigma_\rr}

\def\vpbi{V^\para_{b,\infty}}

\def\vrbi{\vr_{b,\infty}}
\def\hvrbi{\hat V^\perp_{b,\infty}}
\def\hvpbi{\hat V^\para_{b,\infty}}

\def\precre{Assume complete randomization and Condition \ref{asym}.}
\def\cre{complete randomization}
\def\rem{the ReM in Definition \ref{def::rem-factorial}}
\def\creasym{complete randomization and Condition \ref{asym}}
\def\prerem{Assume the ReM in Definition \ref{def::rem-factorial} and  Condition \ref{asym}.}

\def\meaniq{N_q^{-1}\sum_{i:Z_i = q}}
\def\sumiq{\sum_{i:Z_i = q}}

\def\ccl{\chil^\T\chil}
\def\chil{\chi_\lin}
\def\thl{\theta_\lin}

\def\hthlr{\hth_{\lr }}

\def\hbsr{\hg_{*, \rr}}

\def\hbnr{\hg_{\neyman, \rr}}

\def\vsr{V_{*, \rr}}

\def\vl{V_\lin}

\def\hthlr{\hth_{\lr }}

\def\sz{s_0}

\def\A{\textsc{a}}
\def\B{\textsc{b}}
\def\AB{\textsc{ab}}

\def\gp{\bar{\bar \gamma}}

\def\htl{\htau_\lin}
\def\ab{\textsc{ab}}

\def\A{\textsc{a}}
\def\B{\textsc{b}}

\def\para{{\mkern3mu\vphantom{\perp}\vrule depth 0pt\mkern2mu\vrule depth 0pt\mkern3mu}}

\def\pp{{\scaleto{+}{4pt}}}
\def\mmm{{\scaleto{-}{4pt}}}

\def\plim{\textup{plim\,}}

\def\tsx{\hat S_x^2}

\def\pq{e_q}
\def\mri{M_{\rr,\infty}}

\def\bbr{\mathbb R}

\usepackage{scalerel}
\def\hd{\hat\delta}

\def\htr{\htau_\rr}

\def\lmd{\Lambda}

\def\my{\mathcal{Y}}

\def\vxd{V_{x\delta}}
\def\bi{b_\infty}

\def\dbi{D_{b, \infty}}
\def\vbi{V_{b,\infty}}
\def\db{D_b}

\def\ccinvl{(\cll^\T\cll)^{-1}}

\def\cff{\chi_\fisher}
\def\cll{\chi_\lin}
\def\cnn{\chi_\nm}
\def\mss{\chi_*}

\def\heps{\hep_*}

\def\hep{\hat\ep}
\def\hthn{\hth_\nm}
\def\hthf{\hth_\fisher}
\def\hthl{\hth_\lin}
\def\hths{\hth_*}

\def\mk{\mathcal{K}}

\def\core{(\emat^{-1}-\po)}
\def\remd{ \|\hbd\|_\mm \leq a}

\def\hys{\hy_*}
\def\hysr{\hy_{*,\rr}}

\def\po{1_{Q\times Q}}
\def\hsi{\hat\Psi}

\def\Vs{V_*}
\def\Vl{V_\lin}

\def\mhl{{Mahalanobis }}
\def\mhld{{Mahalanobis distance }}
\def\dsim{\stackrel{\cdot}{\sim}}

\def\rs{\rightsquigarrow}
\def\hts{\htau_*}
\def\neyman{\nm}

\def\sqrtn{\sqrt N}

\def\htn{\htau_\nm}

\def\mn{\mathcal{N}}
\def\hyn{\hy_\nm}
\def\hyf{\hy_\fisher}
\def\hyr{\hy_\rr}

\def\hgrq{\hg_{\rr,q}}
\def\hbr{\hg_{\rr}}
\def\hgr{\hg_{\rr}}

\def\hxt{\hx^\T}

\def\hv{\hat V}
\def\sqrtn{\sqrt N}

\def\hy{\hat Y}\def\hgl{\hg_\lin}
\def\yb{\bar Y}
\def\by{\bar Y}
\def\ep{\epsilon}
\def\wiq{1(Z_i=q)}
\def\eliq{\ep_{\lin,i}(q)}
\def\eli{\ep_{\lin,i}}

\def\eniq{\ep_{\neyman,i}(q)}
\def\eni{\ep_{\neyman,i}}
\def\yiq{Y_i(q)}

\def\olss{{\textsc{ols} }}

\def\begini{\begin{itemize}}
\def\endi{\end{itemize}}
\def\sxyq{S_{xY(q)}}
\def\sxxinv{(S_x^2)^{-1}}
\def\sumi{\sum_{i=1}^N}

\def\rr{\textup{r}}
\def\hthr{\hth_\rr}
\def\mr{M_\rr}

\def\ab{\textsc{ab}}

\def\yb{\bar Y}
\def\by{\bar Y}
\def\ep{\epsilon}

\def\yiq{Y_i(q)}

\def\olss{{\textsc{ols} }}

\def\begini{\begin{itemize}}
\def\endi{\end{itemize}}
\def\sxyq{S_{xY(q)}}
\def\sxxinv{(S_x^2)^{-1}}
\def\sumi{\sum_{i=1}^N}

\def\rr{\textup{r}}
\def\hthr{\hth_\rr}
\def\mr{M_\rr}

\def\hep{\hat\epsilon}
\def\hth{\hat\theta}

\def\rank{\textup{rank}}

\def\gq{\gamma}

\def\ml{\mathcal{L}}

\def\sm{Supplementary Material}

\def\hg{\hat\gamma}

\def\ma{\mathcal{A}}
\def\tr{\text{tr}}
\def\mm{{\scaleto{\mathcal{M}}{4.2pt}}}
\def\sxx{S^2_{x}}
\def\htx{\htau_x}

\newcommand{\sxy}{ S_{ x Y}}

\def\ms{* = \nm, \fisher,\lin}

\def\hyf{\hY_\fisher}
\def\hyl{\hY_\lin}
\def\olss{{\textsc{ols} }}

\def\hb{\hat\beta }
\def\hbf{\hat\beta_\fisher}

\newcommand{\htau}{ \hat\tau}
\def\htf{ \hat\tau_\fisher}

\def\hx{\hat x}
\def\hX{\hat X}
\def\bx{\bar x}

\newcommand{\vdd}{ V_{\delta\delta}}

\def\mt{\mathcal{T}}

\newcommand{\mN}{\mathcal{N}}

\newcommand{\Op}{O_\textsc{p}}
\newcommand{\op}{o_\textsc{p}(1)}

\newcommand{\C}{\textsc{c}}

\newcommand{\lin}{\textsc{l}}
\newcommand{\nm}{\textsc{n}}

\newcommand{\bg}{  \gamma }

\newcommand{\hbd}{\hat{\delta}}

\newcommand{\sumn}{\sum_{i=1}^N}
\newcommand{\sumq}{\sum_{q \in\mt} }

\newcommand{\hY}{\hat Y}

\newcommand{\pr}{{\mathbb P}}
 \newcommand{\ot}[1]{1, \ldots,#1}
 \newcommand{\zt}[1]{0, \ldots,#1}

\newcommand{\ols}{\textsc{ols}} 
 
\newcommand{\ehw}{\textsc{ehw}} 
\newcommand{\ehws}{\textsc{ehw} }

\newcommand{\argmin}{\text{argmin}}

\def\T{{ \mathrm{\scriptscriptstyle T} }}

\newcommand{\hbb}{ \hat{ \beta}}

\newcommand{\cov}{\text{cov} }

\newcommand{\mI}{ I}

\newcommand{\hbt}{\hat{\tau}}

\newcommand{\bt}{ \tau}

\newcommand{\fisher}{\textsc{f}}

 \newcommand{\bY}{ \bar Y}

\newcommand{\bbY}{\bar{ Y}}

\newcommand{\proj}{\textrm{proj}}
\newcommand{\res}{\textrm{res}}

\newcommand{\asim}{\overset{\cdot}{\sim}}
\newcommand{\approxi}{\overset{\cdot}{\approx}}

\newcommand{\vr}{ V^{\perp}}

\newcommand{\mL}{{\mathcal{L}}}

\def\begina{\begin{eqnarray*}}
\def\enda{\end{eqnarray*}}
\def\beginy{\begin{eqnarray}}
\def\endy{\end{eqnarray}}
\def\begine{\begin{enumerate}}
\def\ende{\end{enumerate}}

\usepackage{amsthm}
\usepackage{amssymb}
\usepackage{amsmath}
\usepackage{color}

\usepackage{comment}
\theoremstyle{definition}

\newtheorem*{theorem*}{Theorem}
\newtheorem{theorem}{Theorem}
\newtheorem*{rmk*}{remark}
\newtheorem{proposition}{Proposition}
\newtheorem{lemma}{Lemma}
\newtheorem{example}{Example}

\newtheorem{condition}{Condition}

\newtheorem{definition}{Definition}
\newtheorem{remark}{Remark}
\newtheorem{corollary}{Corollary}
\newtheorem*{corollary*}{Corollary}

\usepackage{natbib} 
\bibpunct{(}{)}{;}{a}{}{,} 

\usepackage{etoolbox} 
\apptocmd{\sloppy}{\hbadness 10000\relax}{}{} 

\usepackage{multibib}
\newcites{sec}{References}

\usepackage{color}
\usepackage{listings}
\usepackage{hyperref}
\usepackage{booktabs}

\usepackage{lscape}

\RequirePackage[normalem]{ulem}

\allowdisplaybreaks

\begin{document}

\singlespacing

\title{\bf  
 Covariate adjustment in multi-armed, possibly factorial experiments
} 
\author{Anqi Zhao and Peng Ding
\footnote{Anqi Zhao, Department of Statistics and Data Science, National University of Singapore, 117546, Singapore (E-mail: staza@nus.edu.sg). Peng Ding, Department of Statistics, University of California, Berkeley, CA 94720 (E-mail: pengdingpku@berkeley.edu).
Zhao was funded by the Start-Up grant R-155-000-216-133 from the National University of Singapore. 
Ding was partially funded by the U.S. National Science Foundation (grant \# 1945136). 
}
}
\date{}

\maketitle

\begin{abstract}
Randomized experiments are the gold standard for causal inference and 
enable unbiased estimation of treatment effects. 
Regression adjustment provides a convenient way to incorporate covariate information for additional efficiency. 
This article provides a unified account of its utility for improving estimation efficiency in \mes. 
We start with the commonly used additive and fully interacted models for regression adjustment  in estimating average treatment effects ({\textsc{ate}}), and clarify the trade-offs between the resulting ordinary least squares (\ols) estimators in terms of  finite sample performance and asymptotic efficiency. 
We then move on to regression adjustment based on restricted least squares (\rls), and establish for the first time its properties for inferring {\textsc{ate}} from the design-based perspective. 
The resulting inference has multiple guarantees. First, it is asymptotically efficient when the restriction is correctly specified. Second, it remains consistent as long as the restriction on the coefficients of the treatment indicators, if any, is correctly specified and separate from that on the coefficients of the treatment-covariate interactions. Third, it can have better finite sample performance than the unrestricted counterpart even when the restriction is moderately misspecified.  
It is thus our recommendation when the \olss fit of the fully interacted regression risks large finite sample variability in case of many covariates, many treatments, yet a  moderate sample size. 
In addition, the newly established theory of \rlss also provides a unified way of studying \ols-based inference from general regression specifications. 
As an illustration, we demonstrate its value for studying \ols-based regression adjustment in factorial experiments. 
Importantly, although we analyze inferential procedures that are motivated by \ols, we do not invoke any assumptions required by the underlying linear models.
\end{abstract}


\newpage

\onehalfspacing

\section{Introduction}

\subsection{Multi-armed experiment and covariate adjustment}
\label{sec::experiment+covariates}

Multi-armed experiments enable comparisons of more than two treatment levels simultaneously, and are intrinsic to applications with multiple factors of interest {\citep[see, e.g.,][]{cb, cdl, rahul}.}
They have been extensively used in agricultural and industrial settings \citep[see, e.g.,][]{box, wh}, and are becoming increasingly popular in social and biomedical applications \citep[see, e.g.,][]{duflo, jens, imai, blackwell}. 
Ordinary least squares (\ols) regression remains the dominant strategy for subsequent inference of treatment effects, 
delivering not only point estimates but also their standard errors. The flexibility of model specifications further provides a convenient way to incorporate covariates 
for additional efficiency. 

The additive and fully interacted regressions are two commonly used strategies for covariate adjustment by \ols. 
In particular, the additive regression regresses the outcome on the treatment and covariates to adjust for covariate imbalance across treatment groups \citep{Fisher35}. The fully interacted regression regresses the outcome on not only the treatment and covariates but also their interactions to further accommodate treatment effect heterogeneity \citep{Lin13}.
The theoretical superiority of the fully interacted regression is well established  under treatment-control experiments \citep[see, e.g.,][]{Lin13, LD20}.
Similar discussion, however, is largely missing for   experiments with more than two treatment arms, except for some preliminary results in  \cite{Freedman2008b}, \cite{Lin13},  \cite{lu2016covariate}, and \cite{schochet2018design}.
To fill this gap, we clarify the validity and relative efficiency of the additive and fully interacted regressions for estimating average treatment effects ({\textsc{ate}}) under \mess from the {\it design-based} perspective, which conditions on the potential outcomes and evaluates the sampling properties of estimators over the distribution of the treatment assignments \citep{Neyman23, Freedman08a, Lin13, DasFact15, CausalImbens}.
The results are similar to those under the treatment-control experiment.  
The \olss estimator from the fully interacted regression is consistent and ensures efficiency gains over the unadjusted regression asymptotically. 
The additive regression, on the other hand, is always consistent yet only ensures efficiency gains when the correlations between potential outcomes and covariates are constant across treatment levels.  
In addition, we establish the asymptotic conservativeness of the associated Eicker--Huber--White (\ehw) robust covariance estimators for estimating the true sampling covariances under both the additive and fully interacted regressions. 
This constitutes our first contribution on the design-based justification of regression adjustment. 

Despite being theoretically superior, the fully interacted specification includes all interactions between treatment indicators and covariates, and can incur substantial finite sample variability when there are many treatment levels, many covariates, yet only a moderate sample size. The additive regression, in contrast, can have better finite sample performance than both the unadjusted and fully interacted regressions under such circumstances as long as the treatment effects are not too heterogeneous across covariate levels. The choice between the additive and fully interacted specifications is thus a trade-off between finite sample performance and  asymptotic efficiency.
Of interest is whether there can be some better middle ground. 

To this end,  we propose restricted least squares (\rols) as an alternative way to infer {\textsc{ate}} from the fully interacted regression, and establish its sampling properties under the design-based framework.
The resulting inference has multiple guarantees. First, it is asymptotically efficient when the restriction is correctly specified. Second, it remains consistent as long as the restriction on the coefficients of the treatment indicators, if any, is correctly specified and separate from that on the coefficients of the treatment-covariate interactions. Third, it can have better finite sample performance than its unrestricted counterpart even when the restriction is moderately misspecified. 
It is thus our recommendation for regression adjustment in \mess when the \olss estimator from the fully interacted regression risks large finite sample variability. We also prove a novel design-based Gauss--Markov theorem for \rls, clarifying the asymptotic bias-variance trade-off between \olss and \rlss under constant treatment effects. 
These design-based results on \rlss are not only of theoretical interest in themselves  but also provide a unified framework for studying regression-based inference from general specifications. 
This constitutes our second contribution on the design-based theory of \rlss for estimating {\textsc{ate}}.  The classical theory of \rlss assumes a correct linear model with homoskedastic errors under correct restriction. Our theory, in contrast, is design-based and allows for not only heteroskedastic errors but also misspecification of both the linear model and the restriction.

We then move on to factorial experiments \citep{box, wh, DasFact15, ZDa, pashley2021causal}, and illustrate the value of our theory for studying covariate adjustment under this special type of multi-armed experiments.  Specifically, factorial experiments concern multiple factors of interest, and assign experimental units to all possible levels of their combinations. The special structure of the treatment levels enables convenient factor-based regression analysis \citep{wh, lu16bFact, lu2016covariate, ZDa}, which fits the observed outcome on indicators of the factor levels by \olss and interprets the resulting coefficients as the factorial effects of interest.  The design-based theory on covariate adjustment under factorial experiments has so far focused on {\it factor-saturated} regressions that include all possible interactions between the factor indicators \citep{lu2016covariate, ZDb}. The resulting specifications have model complexity that increases exponentially with the number of factors, risking substantial finite sample variability even with a moderate number of factors and covariates.

To address this issue, we consider {\it factor-unsaturated} specifications that include only a subset of the interactions between the factor indicators when conducting regression adjustment, and clarify the properties of the resulting \olss estimators from the design-based perspective. The choice between the factor-saturated and factor-unsaturated specifications, as it turns out, boils down to a trade-off between asymptotic bias and variance, extending the result in covariate-free settings \citep{ZDa}.  
The resulting theory includes most commonly used specifications in practice as special cases, and offers guidelines on the causal interpretations of their results.  This constitutes our third contribution on causal inference with factorial experiments.

Importantly, although the regression-based covariate adjustment was originally motivated by linear models, we do not invoke any of the underlying assumptions but view  regression as a purely numeric procedure based on \olss or \rls. 
We evaluate the sampling properties of the resulting point estimators and standard errors over the distribution of the treatment assignments. 
All our theories are as such design-based and hold regardless of how well the regression equations represent the true data-generating process \citep[see, e.g.,][]{Freedman08a, Freedman2008b, Schochet10,  Lin13, PostStratYu, CausalImbens, LassoTE16, schochet2018design, fogarty2018regression,  liu2020regression, abadie2020sampling, guo2021generalized}.

\subsection{Notation and definitions}
\label{sec::notation+definition}
Let $0_{m}$ and $0_{m\times n}$ denote the $m\times 1$ vector and $m \times n$ matrix of  zeros, respectively. 
Let  $1_{m}$ and $1_{m\times n}$ denote the $m\times 1$ vector and $m\times n$ matrix of ones, respectively.
Let $I_m $ denote the $m\times m$ identity matrix. 
We suppress the dimensions when they are clear from the context. 
Let $1(\cdot)$ denote the indicator function. 
Let $\otimes$ denote the Kronecker product of matrices. 
For a set of real numbers $\{u_q: q \in \mt\}$, let 
$\diag(u_q)_{q\in\mt}$ denote the diagonal matrix  with $u_q$'s on the diagonal. 
For two $m \times m$ square matrices $M_1$ and $M_2$, write $M_1 \leq M_2$ or $M_2 \geq M_1$ if $M_2 - M_1$ is positive semi-definite. 

For a finite population $(Y_i, u_i)_{i=1}^N$ with $Y_i \in \mathbb R$ and $ u_i \in \mathbb R^m $, let $Y_i \sim u_i$ denote the linear regression of $Y_i$ on $u_i$ free of any modeling assumptions. 
Let $\hat\ep_i$ denote the fitted residual of unit $i$  from \ols. The \ehws covariance estimator of the coefficient vector of $u_i$ equals $(U^\T U)^{-1} U^\T \diag(\hat\ep_i^2)_{i=1}^N U  (U^\T U)^{-1}$, where $U = (u_1, \ldots, u_N)^\T$.

Let $\plim$ denote the probability limit of a sequence of random elements, and let $\rs$ denote convergence in distribution.  
For $\hth_1$ and $\hth_2$ that are both consistent and asymptotically normal for estimating parameter $\theta \in \mathbb R^m$ as the sample size $N$ tends to infinity: $\sqrtn (\hth_k - \theta) \rs \mn(0_m, V_k)$ for $k = 1,2$,  we say 
\begine[(i)]
\item $\hth_1$ is {\it asymptotically more efficient} than $\hth_2$ if  $V_1 \leq V_2$ and the inequality is strict for some data-generating process;
\item $\hth_1$ and $\hth_2$ are {\it asymptotically equally efficient} if $V_1 = V_2$. 
\ende

\section{Causal inference with \mes}\label{sec:setup}
\subsection{Potential outcomes and treatment effects}
Consider an experiment with $Q \geq 2$ treatment levels, $q \in \mt = \{\ot{Q}\}$, and a study population of $N$ units,  $i = \ot{N}$.
Let $Y_i(q)$ be the potential outcome of unit $i$ if assigned to treatment level $q$ \citep{Neyman23}. 
Let $\bar Y(q) = N^{-1} \sumn Y_i(q)$ be the finite population average, vectorized as 
$\bbY = (\bar Y(1),   \dots, \bar Y(Q))^\T$. 
The goal is to estimate the finite population average treatment effect (\textsc{ate}) 
\beginy\label{eq:tau}
\tau = C \bY
\endy 
for some prespecified contrast matrix $C$ with rows orthogonal to $1_Q$. 

The above setting is general and includes many commonly used designs as special cases. 
We use the treatment-control experiment, the three-armed experiment, the $2^2$ factorial experiment, the $2^K$  factorial experiment with general $K \geq 2$, and the general $Q_1\times \cdots \times Q_K$ factorial experiment as running examples to illustrate the main ideas. 
We give their definitions in Examples \ref{ex:exp_trt}--\ref{ex:exp_general} below for concreteness. 
The theory in Sections \ref{sec:setup}--\ref{sec:rls} applies to general experiments with arbitrary structures of treatment levels.
The special structures of factorial experiments further allow for the more flexible factor-based inference, which we detail in Section \ref{sec:factor}.

With a slight abuse of notation, we use customized indexes for treatment levels in Examples \ref{ex:exp_trt}--\ref{ex:exp_general} below to match the conventions in the literature. They have a one-to-one mapping with the $q \in \{1, \ldots, Q\}$ indexing in lexicographical order.

\begin{example}\label{ex:exp_trt} The treatment-control experiment has $Q=2$ treatment levels, indexed by $q \in \mt = \{ 0, 1\}$. 
The individual treatment effect is $\tau_i  = Y_i(1) - Y_i(0)$, and the {\textsc{ate}} is $\tau = N^{-1} \sumi  \tau_i = \bar Y(1) - \by(0) = C \by$, with $C = (-1, 1)$ and $\by = (\by(0), \by(1))^\T$. 
\end{example}

\begin{example}\label{ex:3-arm} Consider a three-armed experiment with one control level and two active levels, indexed by $q \in \mt = \{ 0 \ \text{(control)}, 1 \ \text{(active)}, 2\ \text{(active)}\}$. 
The {\textsc{ate}} of the two active levels relative to control are $\tau_{q} = \by(q) - \by(0)$ for $q=1,2$. 
We vectorize them as
\begina
\tau = \beginp \tau_1\\\tau_2\endp = \beginp \by(1) - \by(0) \\\by(2) - \by(0) \endp = 
\beginp
-1 & 1 & 0\\
-1 & 0 & 1 
\endp 
\beginp
\by(0)\\\by(1)\\\by(2)
\endp = C \by,
\enda 
where $C = (-1_2, I_2)$ and $\by = (\by(0), \by(1), \by(2))^\T$. 
\end{example}

The setting of Example \ref{ex:3-arm} extends immediately to general multi-armed experiments with a single factor of $Q > 2$ levels,
as in the setting of the one-way analysis of variance. This is among the most common types of multi-armed experiments in practice, featuring no special structure of the treatment levels. Factorial designs, on the other hand, accommodate multiple factors of interest in one single experiment and define treatment levels as their combinations.

\begin{example}\label{ex:exp_22} 
The $2^2$ factorial experiment is a special type of the four-armed experiment with two binary factors of interest, $\text{A}, \text{B} \in \pmo $,  and in total $Q=2^2 = 4$ treatment levels as their combinations: $ q  \in \mt = \{ (-1,-1)$, $\, (-1,+1)$, $\,  (+1,-1)$, $\, (+1,+1)\}$.
We abbreviate $(-1,-1)$ as $(--)$, etc.~when no confusion would arise. 
Let $\by = (\by(--), \by(-+), \by(+-), \by(++))^\T$ be the vector of average potential outcomes in lexicographical order of the treatment combinations.
 The standard main effects and interaction effect equal 
 \begina
 \tau_{\A} &=& 2^{-1}\left\{\bY(+ -) + \bY(+ +)\right\} - 2^{-1}\left\{\bY(- -) + \bY(-+)\right\} = \ca^\T\by, \\
 \tau_{\B} &=& 2^{-1}\left\{\bY(- +) + \bY(+ + )\right\} - 2^{-1}\left\{\bY(- -) + \bY(+ -)\right\} = \cb^\T\by, \\
  \tau_{\AB} &=& 2^{-1}\left\{ \bY(- - )+\bY(++)\right\} - 2^{-1}\left\{\bY(-+ ) + \bY(+ -)\right\} = \cab^\T\by,
  \enda
   respectively, with $c_\A = 2^{-1}(-1,-1,1,1)^\T$, $c_\B = 2^{-1}(-1, 1,-1,1)^\T$, and $ c_\AB = 2^{-1}(1, -1, -1, 1)^\T$ \citep{DasFact15}. 
The main effect of a factor compares the average potential outcomes when the factor is
at level $+1$ and level $-1$, respectively.
The interaction effect compares the average
potential outcomes when the two factors take the same and different levels, respectively.
\end{example}

\begin{example}\label{ex:exp_2k}  The $2^K$ factorial experiment is a special type of the $2^K$-armed experiment with $K$ binary factors of interest, $k = \ot{K}$, and in total $Q = 2^K$ treatment levels as their combinations: $q \in \mt = \{(z_1,\dots, z_K): z_k = -1, +1\}$, where $z_k$ indicates the level of factor $k$.
The treatment-control and $2^2$ factorial experiments in Examples \ref{ex:exp_trt} and \ref{ex:exp_22} are both special cases with $K=1$ and $K =2$, respectively. 
There are $2^K-1$ standard factorial effects corresponding to the main effects and two- to $K$-way interaction effects, respectively \citep{wh}. 
\end{example}

\begin{example}\label{ex:exp_general}
The general $Q_1\times \cdots \times Q_K$ factorial experiment has $Q = \prod_{k=1}^K Q_k$ treatment levels, as the combinations of $K \geq 2$ factors of interest, $k = \ot{K}$, each of $Q_k \geq 2$ levels. 
It is a special type of the $Q$-armed experiment, and includes the $2^K$ factorial experiment in Example \ref{ex:exp_2k} as a special case with $Q_k = 2$ for all $k$. 
\end{example}

\subsection{Assignment mechanism, estimators, and regression formulation}
We focus on complete randomization defined below.

\begin{definition}
[complete randomization]\label{def::complete-randomization}
For prespecified, fixed $N_q > 0 \ (q\in\mt)$  with $\sumq N_q = N$,  the experimenter randomly draws an allocation from all possible allocations that have $N_q$ units  receiving treatment level $q \ (q\in\mt)$. 
\end{definition}

Let $Z_i \in \mt$ denote the treatment level received by unit $i$. 
The observed outcome equals 
\beginy\label{eq:obs}
Y_i = Y_i(Z_i) = \sumq \wiq  \, Y_i(q)
\endy for unit $i$. 
Let $\hat Y(q) = N_q^{-1}\sumiq  Y_i$  be the average observed outcome under treatment level $q$.
The sample-mean estimator of $\by$ equals $\hyn = (\hY(1), \dots, \hY(Q) )^\T$ and suggests
\beginy\label{eq:htau_n}
\htn=C\hyn
\endy
as an intuitive choice for estimating $\tau$. 

Let $\pq  = N_q/N$ denote the proportion of units under treatment level $q$.
Let $ S = (S_{qq'})_{q, q'\in\mt}$ be the finite population covariance matrix of $\{Y_i(q): q \in \mt \}_{i=1}^N$ with $S_{qq'} = (N-1)^{-1}\sumn\{Y_i(q) - \bar Y(q)\}\{Y_i(q') - \bar Y(q')\}$.
Under complete randomization, 
$\htn$ is unbiased for $\tau$ with sampling covariance $\cov(\htn) = N^{-1} CV_\nm C^\T$, where $V_\nm =   N\cov(\hyn ) = \diag( S_{qq} /\pq )_{q\in\mt}  -  S$. 
Define
$$
t_i = (1(Z_i = 1),\dots,1(Z_i = Q))^\T  \in \mathbb R^Q 
$$
as the treatment indicator vector for unit $i$.
Then $\hyn$ equals the coefficient vector of $t_i$ from the \olss fit of 
\beginy\label{eq:lm_n}
Y_i  \ \sim \  1(Z_i = 1) + \cdots + 1(Z_i = Q)  \ \ \Longleftrightarrow  \ \ Y_i \  \sim \ t_i
\endy
over $i = \ot{N}$ without an intercept. We call \eqref{eq:lm_n} the {\it unadjusted treatment-based regression}, taking the treatment indicators as regressors.

The presence of covariates allows for opportunities to further improve estimation efficiency. 
Let $ x_i = \left(x_{i1}, \dots, x_{iJ}\right)^\T$ denote the $J\times 1$ covariate vector for unit $i$.
To simplify the presentation, assume centered $(x_i)_{i=1}^N$ throughout the paper with mean $\bx  = N^{-1} \sumn  x_i = 0_J$ and finite population covariance $\sxx  = (N-1)^{-1}\sumn  x_i x_i^\T$.  
Specifications 
\begin{eqnarray}
&& Y_i 
\ \sim \  t_i  + {\cxi },\label{eq:lm_f}\\
&&  Y_i 
\ \sim \   t_i +  \sumq  \wiq \, \cxi  \
\ \Longleftrightarrow \  \ Y_i \ \sim  \ t_i + t_i \otimes \cxi \label{eq:lm_l}
\end{eqnarray}
give two intuitive ways to adjust for covariates on the basis of \eqref{eq:lm_n}. Refer to them as the {\it additive} and {\it fully interacted treatment-based regressions}, respectively, depending on whether the regression equations include the interactions between $t_i$ and $\cxi$ or not. 
We denote by $\hyf \in \mathbb R^Q$ and $\hyl \in \mathbb R^Q$ the coefficient vectors of $t_i $ from the \olss fits of \eqref{eq:lm_f} and \eqref{eq:lm_l}, respectively, as two covariate-adjusted variants of $\hyn$ from \eqref{eq:lm_n} for estimating $\by$.
As a convention, we use the subscripts ``\neyman", ``\fisher", and ``\lin" to signify quantities associated with the unadjusted, additive, and fully interacted regressions, respectively. Example \ref{ex:treatment_control} in Section \ref{sec:reg} will clarify their respective connections with \cite{Neyman23}, \cite{Fisher35}, and \cite{Lin13}. 
Replacing $\hyn$ with $\hys$ in \eqref{eq:htau_n} yields 
\begina
\hts = C \hys \qquad (* = \fisher, \lin) 
\enda
as two covariate-adjusted estimators of $\tau$.  
Of interest is their validity and efficiency relative to $\htn$ from the design-based perspective. 
We address this question in Section \ref{sec:reg} after clarifying the intuition behind the regression formulation in Section \ref{sec:derived linear model} below. 

\subsection{Derived linear models and target parameters}\label{sec:derived linear model}

{\it Derived linear models} \citep{kempthorne1952design, hinkelmann} provide the intuition for using the coefficient vectors of $t_i$ from \eqref{eq:lm_n}--\eqref{eq:lm_l} to estimate $\by$.

In particular, consider the {\ols} fit of $Y_i(q) \sim 1$ over $i = \ot{N}$ for $q\in\mt$. This is a theoretical \olss fit with the $Y_i(q)$'s only partially observable,  and yields the fitted model $Y_i(q) = \bY(q) + \eniq $ with  $\eniq = \yiq - \yb(q)$. 
Plugging this fitted model in \eqref{eq:obs} implies the {\it unadjusted derived linear model} of the observed outcome: 
\beginy\label{eq:dlm_n}
Y_i  \ = \  \sumq \wiq \, \yb(q) + \epni  \ = \  t_i^\T \by + \epni  
\endy
with $\eni = \sumq \wiq  \eniq$. 
This motivates the unadjusted regression \eqref{eq:lm_n}. 

The presence of covariates motivates extending \eqref{eq:dlm_n} to include regression adjustment. 
Consider the \olss fit of $Y_i(q) \sim 1+ \cxi $ over $i = \ot{N}$ for $q\in\mt$. This is also a theoretical \olss fit, and yields the fitted model 
$
 Y_i(q)  = \by(q) +  \cxi ^\T \gamma_q + \eliq$. 
We have 
$\gamma_q = (\sxx)^{-1}S_{xY(q)}$, where
 $\sxyq =  (N-1)^{-1} \sumn  \cxi \{Y_i(q)-\by(q)\}$ denotes the finite population covariance of $\{x_i, Y_i(q)\}_{i=1}^N$.
For simplicity, we call $\gamma_q$ the {\it population correlation} of $Y_i(q)$ and $\cxi$. Plugging this fitted model in \eqref{eq:obs} implies the {\it covariate-adjusted derived linear model} of the observed outcome: 
\beginy\label{eq:dlm_l}
Y_i 
&=&\sumq \wiq\, \by(q) + \sumq \{\wiq \, \cxi\}^\T \gamma_q + \eli  \nonumber\\
&=& t_i^\T\by + (t_i\otimes x_i)^\T \gamma + \eli
\endy
with $\eli = \sumq \wiq  \eliq$ and $\gamma = (\gamma_1^\T, \ldots, \gamma_Q^\T)^\T$. 
This motivates the fully interacted regression \eqref{eq:lm_l}. 
We call $\by$ and $\gamma$ the {\it target parameters} of $t_i$ and $t_i\otimes x_i$ in \eqref{eq:lm_l}, respectively, and refer to $\thl = (\by^\T, \gamma^\T)^\T$ as the target parameter of \eqref{eq:lm_l} as a whole. 

The unadjusted regression \eqref{eq:lm_n} can be viewed as a {\it restricted variant} of \eqref{eq:lm_l}, assuming that the coefficient vectors of the $\wiq \, \cxi $'s are all zero.
Likewise can we view the additive regression \eqref{eq:lm_f} as a restricted variant of \eqref{eq:lm_l}, assuming that the coefficient vectors of the $\wiq \, \cxi $'s are all equal.
This motivates the definitions of the zero and equal correlation conditions below.
They provide not only the heuristics for the functional forms of \eqref{eq:lm_n} and \eqref{eq:lm_f} relative to \eqref{eq:lm_l}  but also the sufficient conditions for the asymptotic efficiency of $\htn$ and $\htf$, respectively (see Section \ref{sec:reg}, Lemma \ref{lem:reg}).

\begin{condition}[zero correlation] \label{cond:zero}
$
\gamma_1 = \dots = \gamma_Q = 0_J$.  
\end{condition}

\begin{condition}[equal correlation]\label{cond:equal}
$
\gamma_1 = \dots = \gamma_Q
$. 
\end{condition}

The variation in $\{\gamma_q:q\in\mt\}$ measures the heterogeneity in treatment effects that is explained by covariates \citep{Ding2019}.
Condition \ref{cond:equal} stipulates that the population correlations between potential outcomes and covariates are constant across treatment levels, implying homogeneous treatment effects. 
Condition \ref{cond:zero} is stronger than Condition \ref{cond:equal} and stipulates uncorrelatedness for all levels. 
Condition \ref{cond:sa} below gives a sufficient condition for  Condition \ref{cond:equal}.

\begin{condition}[constant treatment effects]  
\label{cond:sa}
For all $q, q'\in\mt$, 
the individual treatment effects $Y_i(q) - Y_i(q')$
are constant across $i = \ot{N}$. 
\end{condition}

Importantly, the derived linear models \eqref{eq:dlm_n} and \eqref{eq:dlm_l} are purely numeric decompositions of the observed outcome without any assumptions on the data-generating process. 
We use them as props to motivate the zero and equal correlation conditions  and introduce the target parameters. 
The design-based framework conditions on the potential outcomes and covariates,  and attributes the
randomness in $Y_i$ solely to the randomness in the treatment assignments $Z_i$'s. 
The covariances of $(\ep_{*, i})_{i=1}^N \ (* = \nm, \lin)$ are accordingly fully determined by the joint distribution of the $Z_i$'s, and are in general neither jointly independent nor homoskedastic under complete randomization.

The classical Gauss--Markov model, on the other hand, conditions on the $Z_i$'s, and attributes the randomness in $Y_i$'s to the sampling errors due to the study population being a random sample from some hypothetical superpopulation.  
The covariance of the error terms is specified by model assumptions, with joint independence and homoskedasticity being two commonly invoked options.  
\cite{Freedman08a} pointed out that randomization does not justify these assumptions.  
The theory we are about to present, nevertheless, suggests that $\hys \ (\nfl)$, as purely numeric outputs from \ols, can still deliver valid design-based inferences when coupled with the \ehws covariance estimators.  We elaborate on the details in Section \ref{sec:reg} below. 

\section{Regression adjustment by ordinary least squares}
\label{sec:reg}  

\subsection{Asymptotic efficiency of $\htau_\textup{\scriptsize L}$ over  $\htau_\textup{\scriptsize N}$ and  $\htau_\textup{\scriptsize F}$}

We establish in this subsection the validity and asymptotic relative efficiency of $\hts \ (\nfl)$ for inferring $\tau$. 
The result extends \cite{Fisher35}, \cite{Freedman08a}, and \cite{Lin13} to \mes, and complements \cite{Freedman2008b} on the asymptotics of the fully interacted regression. 
See \cite{tsia}, \cite{bugni2018inference, bugni2019},  \cite{negi2021revisiting}, and \cite{ye} for analogous results under alternative superpopulation frameworks.

Condition \ref{asym} below is standard  for  finite population asymptotics under complete randomization  \citep{DingCLT}. 

\begin{condition}\label{asym}
As $N\to \infty$, for $q\in\mt$, 
(i) $\pq   = N_q/N$ has a limit in $(0,1)$;
(ii) the first two finite population moments of $\{Y_i(q), x_i, x_iY_i(q): q \in \mt \}$ have finite limits;
both $\sxx$ and its limit are nonsingular;
 (iii) 
$\meani   Y_i^4(q)  = O(1)$, $\meani  \|x_i\|_4^4 = O(1)$, and $\meani  \|x_iY_i(q)\|_4^4 = O(1)$.
\end{condition}


Recall that $\gamma_q$ denotes the population correlation of $Y_i(q)$ and  $x_i$. 
Let $\gp = \sumq e_q \gamma_q$ be a weighted average. 
%
Let $S_\nm = (S_{\neyman,qq'})_{q,q'\in\mt}$, $S_\fisher = (S_{\fisher,qq'})_{q,q'\in\mt}$, and  $S_\lin = (S_{\lin,qq'})_{q,q'\in\mt}$ be the finite population covariance matrices of $\{Y_i(q): q\in \mt\}_{i=1}^N$, $\{ \yiq - \gp^\T x_i: q\in \mt\}_{i=1}^N$, and $\{\yiq - \gamma_q^\T x_i: q\in \mt\}_{i=1}^N$, respectively, with $S_\nm= S$.
Let 
\beginy\label{eq:v_star}
V_* =  \diag( S _{*,qq} /\pq )_{q\in\mt} -  S_* \qquad ( \nfl),
\endy 
which have finite limits under Condition \ref{asym}. 
To simplify the presentation, we will also use the same symbols  to denote their respective limiting values when no confusion would arise.

Let $\hsi_*$ be the \ehws covariance estimator of $\hys$  from the same \olss fit for $\nfl$. We first make a moderate contribution by providing a unified theory for the design-based properties of $\hys$ and $\hsi_* \ (\nfl)$  in Lemma \ref{lem:reg} below.  
The result clarifies the consistency and asymptotic normality of $\hys$ for estimating $\by$, and ensures the asymptotic conservativeness of $\hsi_*$ for estimating the true sampling covariance. 

\begin{lemma}\label{lem:reg}
{\precre} Then
\begine[(i)]
\item\label{item:wald} $\sqrt N (\hys -\bbY) \rs \mN(  0, \Vs)$ for $\nfl$ with $N \hsi_* - \Vs =  S_* + \op$, where $S_* \geq 0$;
\item\label{item:eff}   $V_\lin \leq V_\nm$ and $V_\lin \leq V_\fisher$;
\item\label{item:equiv} Under Condition \ref{cond:equal}, $V_\fisher = V_\lin \leq V_\neyman$; under Condition \ref{cond:zero}, $V_\nm = V_\fisher = V_\lin$. 
\ende
\end{lemma}

Lemma \ref{lem:reg}\eqref{item:wald} justifies the large-sample Wald-type inference of $\tau$ based on $\htau_* = C\hys$ and  $ C \hat \Psi_* C^\T$ as the point estimator and estimated covariance, respectively, for $\ms$.
Lemma \ref{lem:reg}\eqref{item:eff} ensures the asymptotic efficiency of $\htl$ over $\htn$ and $\htf$. 
The additively adjusted $\htf$, on the other hand, may be less efficient than the unadjusted $\htn$, especially when the experiments have unequal group sizes and heterogeneous treatment effects with respect to the covariates \citep{Freedman08a}. 
Lemma \ref{lem:reg}\eqref{item:equiv} gives two exceptions.
First, the equal correlation condition ensures that $\htf$ is asymptotically equally efficient as $\htl$, rendering the inclusion of interaction terms unnecessary.
Second,  the zero correlation condition ensures that $\htn$  is asymptotically equally efficient as $\htf$ and $\htl$, rendering regression adjustment unnecessary.

Lemma \ref{lem:reg} unifies the existing theory on regression adjustment under the treatment-control experiment as a special case. We review in Example \ref{ex:treatment_control} below the results from  \cite{Neyman23}, \cite{Fisher35}, and \cite{Lin13}, and clarify their connections with \eqref{eq:lm_n}--\eqref{eq:lm_l}.

\begin{example}\label{ex:treatment_control}
Consider the treatment-control experiment from Example \ref{ex:exp_trt}.
We have 
$\hts = (-1, 1)\hys$ for $\nfl$, where the $\hys$'s are the coefficient vectors of $t_i = (1(Z_i = 0), 1(Z_i = 1))^\T = (1-Z_i, Z_i)^\T$ from the \olss fits of \eqref{eq:lm_n}--\eqref{eq:lm_l}, respectively.
Then $\htn = \hy(1) - \hy(0)$ equals the difference-in-means estimator, and can also be computed as the coefficient of $Z_i$ from the \olss fit of $Y_i \sim 1 + Z_i$. \cite{Neyman23} showed that $\htn$ is unbiased for $\tau$. 

In the presence of covariate information, \citet{Fisher35} suggested to estimate $\tau$ by the coefficient of $Z_i$ from the \olss fit of 
$Y_i \sim 1 + Z_i + x_i$. 
\citet{Lin13} proposed to include also the interactions between $x_i$ and $Z_i$, and estimate $\tau$ by the coefficient of $Z_i$ from the \olss fit of  $Y_i \sim  1 + Z_i +  \cxi  + Z_i  \cxi $. 
The one-to-one correspondence between $(1, Z_i)^\T$ and $t_i$ ensures that the resulting estimators equal $\htf$ and $\htl$, respectively.
This clarifies the connections of \eqref{eq:lm_n}--\eqref{eq:lm_l} to \cite{Neyman23}, \cite{Fisher35}, and \cite{Lin13}, respectively. 
The properties of $\hts \ (\nfl)$ follow readily from Lemma \ref{lem:reg}.

\end{example}

As it turns out, the asymptotic efficiency of $\htl$ extends beyond Lemma \ref{lem:reg}  to a class of general linear estimators of $\tau$ that includes $\hts \ (\nfl)$ as special cases. We relegate the details to the \sm.

\subsection{Trade-off between $\htau_\textup{\scriptsize L}$ and $\htau_\textup{\scriptsize F}$}\label{sec:tradeoff}

Despite the guaranteed gains in asymptotic efficiency, the fully interacted regression \eqref{eq:lm_l} involves  
$ Q+JQ$ estimated coefficients, subjecting $\htl$ to large finite sample variability when the sample size $N$ is moderate relative to $Q+JQ$.
Simulation evidence, on the other hand, suggests that $\htf$ can be more efficient than $\htl$ in finite samples as long as the equal correlation condition is not severely violated.
The choice between the additive and fully interacted regressions is thus a trade-off between finite sample performance and asymptotic efficiency.
This, together with the asymptotic efficiency of $\htf$ under the equal correlation condition, grants $\htf$ a triple guarantee: it is consistent and asymptotically normal for general potential outcomes, ensures asymptotic efficiency if the $\gamma_q$'s are all equal, and can have better finite sample performance than $\htl$ as long as the $\gamma_q$'s are not too different.  \citet{Fisher35}'s analysis of covariance, as a result, can be viewed as a compromise between unadjusted inference and the fully interacted adjustment in moderate samples. See \cite{Schochet10} for empirical evidence based on eight large social policy experiments.

Recall from Section \ref{sec:derived linear model} that the additive regression \eqref{eq:lm_f} can be viewed as a restricted variant of \eqref{eq:lm_l}, assuming that the coefficient vectors of $\wiq  \, x_i$'s are all equal. 
This motivates a more general approach to regression adjustment by restricted least squares,
which estimates the coefficients of \eqref{eq:lm_l} subject to some prespecified linear restrictions.
We establish the design-based properties of the resulting inference in Section \ref{sec:rls} below. 
 
%
%

%
\section{Regression adjustment by restricted least squares}\label{sec:rls}
%
\subsection{Restricted least squares}\label{sec:rols_trt_overview}
Restricted least squares (\rols) is a standard tool for fitting linear models, enabling convenient encoding of prior knowledge on model parameters.
Its theoretical properties are well studied under the classical Gauss--Markov model \citep{theil, rao, greene}.  
The corresponding theory, however, is so far missing under the design-based framework,  where the errors are intrinsically dependent and heteroskedastic (c.f$.$~Section \ref{sec:derived linear model}). 
This section fills this gap and clarifies the design-based properties of \rlss for regression adjustment. 
The resulting theory is not only of theoretical interest in itself but also provides  a unified way to study \ols-based inference from general regressions. 
We focus on \rols-based inference for general \mess in this section, and demonstrate its value for studying \ols-based regression adjustment in factorial experiments in Section \ref{sec:factor}. 
The classical theory of \rlss assumes a correct linear model with homoskedastic errors under correct restriction.
Our theory, in contrast, is design-based and allows for not only heteroskedastic errors but also misspecification of both the linear model and the restriction.

Recall $\theta_\lin = (\by^\T,\gamma^\T)^\T$ as the target parameter of  the fully interacted regression \eqref{eq:lm_l}, motivated by the derived linear model \eqref{eq:dlm_l}. 
Let $\chi_{\lin,i} = ( t_i^\T,t_i^\T \otimes \cxi^\T  )^\T$ denote the regressor vector of \eqref{eq:lm_l}. 
The \olss fit of \eqref{eq:lm_l} can be viewed as estimating $\theta_\lin$ by
\begina
\hthl = (\hyl^\T, \hgl^\T)^\T= \argmin_{\theta} \sumi (Y_i -\chi_{\lin,i}^\T \theta)^2,
\enda 
where $\hyl$ and $\hgl = (\hg_{\lin,1}^\T, \dots, \hg_{\lin,Q}^\T)^\T$ denote the coefficient vectors of $t_i$ and $t_i \otimes \cxi$, respectively, with $\hg_{\lin,q}$ corresponding to  $\wiq \cxi $. 

The \rlss fit, on the other hand, estimates $\thl$
subject to some prespecified linear restrictions.

%
%
\begin{definition}[restricted least squares]\label{def:rls}
 For some prespecified restriction matrix $R \in \mathbb R^{m\times (Q+JQ)}$ that has full row rank $m \leq Q+JQ$ and vector $r \in \mathbb R^m$, 
the \rlss fit of \eqref{eq:lm_l} yields
\beginy\label{eq:rls}
\hthlr = (\hyr ^\T, \hgr^\T)^\T= \argmin_\theta \sumi (Y_i -\chi_{\lin,i}^\T \theta)^2 \qquad \text{subject to} \ \ R\theta = r,
\endy
where $\hyr $ and $\hgr = (\hg_{\rr,1}^\T, \dots, \hg_{\rr,Q}^\T)^\T$ denote the coefficient vectors of $t_i$ and $t_i \otimes \cxi$, respectively, with $\hgrq $ corresponding to  $\wiq \cxi $. 
\end{definition}

By \eqref{eq:rls}, we can view $\hthr$ as an alternative estimator of $\thl$ that satisfies $R\hthr = r$.
This is as if we estimate $\theta_\lin$   subject to the prior belief of 
\beginy\label{eq:restriction}
R \theta_\lin 
= r,
\endy
and suggests $\htr = C\hyr$ as an alternative option for inferring $\tau$.
The goal is to quantify its sampling properties relative to $\htl$.

To simplify the presentation, we 
will use \eqref{eq:restriction} to represent the restriction in \eqref{eq:rls} in terms of the target parameter $\theta_\lin$, 
and refer to \eqref{eq:rls} as the \rolss fit subject to {\it working restriction} \eqref{eq:restriction}.
The working restriction \eqref{eq:restriction} is accordingly a purely numeric input for \rolss that may or may not match the truth.
We say \eqref{eq:restriction}  is {\it correctly specified} if it indeed matches the truth.

Examples \ref{ex:rols_neyman} and \ref{ex:rols_fisher} below establish 
both 
$\hyn$ and $\hyf$ 
as special cases of $\hyr $ with specific choices of $(R,r)$.

\begin{example}\label{ex:rols_neyman}
 $\hyn = \hyr $, where $\hyr $ is the coefficient vector of $t_i$ from the \rolss fit of \eqref{eq:lm_l} subject to the {\it zero correlation restriction} that $\gamma_q = 0_J$ for all $q$. A matrix representation of this restriction is $R\tl =  \gamma = 0_{JQ}$ with $R = (0_{JQ\times Q}, I_{JQ})$.
\end{example}

\begin{example}\label{ex:rols_fisher}
$\hyf = \hyr $, where $\hyr $ is the coefficient vector of $t_i$ from the \rolss fit of \eqref{eq:lm_l} subject to the {\it equal correlation restriction} that $\gamma_1 = \cdots = \gamma_Q$.
A matrix representation of this restriction is $R\tl   =(\gamma_2^\T - \gamma_1^\T, \ldots, \gamma_Q^\T - \gamma_1^\T)^\T=  0_{J(Q-1)}$ with $R = (0_{J(Q-1)\times Q}, (-1_{Q-1}, I_{Q-1}) \otimes I_J)$.
\end{example}

Recall from Section \ref{sec:derived linear model} that regressions \eqref{eq:lm_n} and \eqref{eq:lm_f} can be viewed as restricted variants of \eqref{eq:lm_l}, assuming the zero and equal correlation restrictions, respectively. 
Examples \ref{ex:rols_neyman} and \ref{ex:rols_fisher} illustrate the numeric correspondence between the \rlss fit and the \olss fit of the corresponding restricted specification, establishing $\htn$ and $\htf$ as special cases of $\htr$. See Lemma S3 in the {\sm} for a more general result.

Importantly, the equal correlation {\it restriction} differs from the equal correlation {\it condition} in Condition \ref{cond:equal}.
Echoing the comments after \eqref{eq:restriction}, we view the equal correlation restriction as a purely numeric input for \rlss that may or may not match the truth. Condition \ref{cond:equal}, in contrast, represents our assumption about the true data-generating process. 
The equal correlation restriction is  hence correctly specified if and only if Condition \ref{cond:equal} holds.
Likewise for the correspondence between the zero correlation restriction and Condition \ref{cond:zero}.

The restrictions in Examples \ref{ex:rols_neyman} and \ref{ex:rols_fisher} involve only $\gamma$.  
Example \ref{ex:rols_no inter} below imposes restrictions on both $\bY$ and $\gamma$.

\begin{example}\label{ex:rols_no inter}
Consider the $2^2$ factorial experiment from Example \ref{ex:exp_22}. Define $\tau_{\ab,i}=  2^{-1} \{Y_i(--)+ Y_i(++)\} - 2^{-1}\{ Y_i(-+) + Y_i(+-) \}$ as the individual interaction effect for unit $i$.  Assume that prior knowledge suggests that $\tau_{\ab,i}= 0$ for all $i$. This implies
 \beginy
 \tau_\AB = 0, \qquad 
 (\gamma_{\mmm\mmm}+ \gamma_{\pp\pp}) - (\gamma_{\mmm\pp} + \gamma_{\pp\mmm})  = 0_J, 
\label{eq:rest_l_22}  
 \endy
which can in turn be used as a working restriction for fitting \eqref{eq:lm_l} by \rls. A matrix representation of the restriction \eqref{eq:rest_l_22} is $R\theta_\lin   = 0_{J+1}$ with $
R = 
\diag(
c^\T_\ab,  c_\ab^\T\otimes  I_J)$.
\end{example}

\subsection{Design-based theory of restricted least squares} \label{sec:rols_trt}

Write \eqref{eq:dlm_l} in matrix form as 
$
Y = \chil\thl + \ep_\lin$, 
where $Y = (Y_1, \dots, Y_N)^\T$, $\chi_\lin = (\chi_{\lin,1}, \dots, \chi_{\lin,N})^\T$, and $\ep_\lin = (\ep_{\lin,1}, \dots, \ep_{\lin,N})^\T$. 
The Gauss--Markov model assumes that $Y$ has expectation $\chil\thl$ and covariance $\sigma^2 I_N$, 
and ensures the efficiency of $\hyr$ among all linear unbiased estimators when the restriction is correctly specified \citep{theil, rao}.
The design-based framework violates these assumptions and leaves the sampling properties of $\hyr$ unclear. 
This is our focus for this subsection. 

To this end, we first review in Lemma \ref{lem:hthr} below the numeric expression of $\hthlr$ free of any modeling assumptions. For simplicity, we assume that $\chi_\lin^\T \chi_\lin$ is nonsingular throughout; see \citet{greene} for more general formulas. 

\begin{lemma}
\label{lem:hthr}
$\hthlr 
= (I-\mr R) \hthl + \mr r$ 
with $\mr=  \ccinvl R^\T\{R \ccinvl R^\T\}^{-1}$.
\end{lemma}

The rest of this section is organized as follows. Section \ref{sec:restrictions} introduces two types of restrictions as our focus for the main paper.
Sections \ref{sec:noy} and \ref{sec:sep} give the sampling properties of $\hyr$ under the two restriction types, respectively. 
Section \ref{sec:ehw_rols} then proposes a novel estimator of the sampling covariance of $\hyr $. 
Throughout Section \ref{sec:rls}, we focus on not only the consistency of $\hyr$ but also its asymptotic normality and  robust covariance estimation  for large-sample Wald-type inference. 

Recall that $\hg_{\rr,q}$ denotes the coefficient vector of $1(Z_i = q) x_i$ from the \rlss fit of \eqref{eq:lm_l}.   Theorem S3 in the {\sm} 
ensures that $(\hg_{\rr,q})_{q\in\mt}$ all have finite probability limits, denoted by $ \gamma_{\rr,q} = \plim \hg_{\rr,q}$, under complete randomization and Condition \ref{asym} regardless of whether the working restriction  $R\thl = r$  is correctly specified or not. 
Let $S_{\lr } = (S_{\lr ,qq'})_{q,q'\in\mt}$ be the finite population covariance matrix of  $\{\yiq - \brq^\T x_i : q\in \mt\}_{i=1}^N$,  analogous to $S_* \ (\nfl)$.
Let  $
\vlr =  \diag( S _{\rr,qq} /\pq )_{q\in\mt} -  S_{\rr}$, analogous to $V_* \ (\nfl)$. Lemma \ref{lem:v_r} below gives a numeric result on $\vrr$ that underlies the asymptotic efficiency of $\hyr$ relative to $\hyl$. 

\begin{lemma}\label{lem:v_r}
$\vrr \geq V_\lin$, where the equality holds if $\gamma_{\rr,q} = \gamma_q$ for all $q\in\mt$.
\end{lemma}

\subsubsection{Two types of restrictions}\label{sec:restrictions}

To simplify the presentation, we focus on the following two types of restrictions in the main paper due to their prevalence in practice and their ability to convey all main points of the general theory.
We relegate the general theory to the {\sm}.

\begin{definition}\label{def::restrictions} 
\begine[(i)]
\item 
A {\it \so} restriction restricts $\bY$ separately from $\gamma$, with \eqref{eq:restriction} reduced to 
\beginy\label{eq:rest_sep}
\rhy\by = r_Y ,\qquad  \rhg \gamma  = r_\gamma
\endy  
for some prespecified $\rhy$ and $\rhg$.
Without loss of generality, assume that $(\rhy, r_Y) = ( 0_Q^\T, 0)$ if there is no restriction on $\by$, and that $\rhy$ has full row rank if otherwise; likewise for $(\rhg, r_\gamma)$.  

\item A {\it \go} restriction restricts only $\gamma$, with  \eqref{eq:rest_sep}  reduced to 
\beginy\label{eq:rest_noy}
 \rhg \gamma = r_\gamma.
\endy
\ende
\end{definition}

\medskip

The {\gor} is a special type of the separable restriction with $(\rhy, r_Y) = ( 0_Q^\T, 0)$. 
In Definition \ref{def::restrictions}, $R = \diag(\rhy, \rhg)$ and $r = (r_Y^\T , r_\gamma^\T)^\T$ for a {\sor} with restrictions on both $\by$ and $\gamma$; 
$R = (\rhy,0)$ and $r = r_Y$ for a {\sor}  with only restriction on $\by$;
$R  = (0, \rho_\gamma)$ and $r = r_\gamma$ for a {\gor} with non-empty restriction on $\gamma$. 
An empty restriction is always correctly specified by definition.

Examples \ref{ex:rols_neyman} and \ref{ex:rols_fisher} are special cases of the {\gor}, whereas Example \ref{ex:rols_no inter} exemplifies the more general {\sr}. A common choice of $\rho_Y$ is a contrast matrix with rows orthogonal to $1_Q$, imposing restrictions on a set of finite population {\textsc{ate}}, namely $\rhy\by$.  The $\tau_\AB= 0$ in Example \ref{ex:rols_no inter} is an example with $\rhy = \cab^\T$. More generally, the assumptions of no higher-order interactions for analyzing factorial experiments all fall into this category; see \cite{ZDa}. 
We will give more examples in Section \ref{sec:factor}.

\subsubsection{$\hyr $ subject to the {\gor}}
\label{sec:noy}

\black
\begin{theorem}\label{thm:noy}
Assume complete randomization, Condition \ref{asym}, and \rlss subject to \eqref{eq:rest_noy}. Then 
\begine[(i)]
\item\label{item:noy_consis} $\sqrtn (\hyr  - \by )  \rs \mn(0_Q, \vlr)$, where $\vlr \geq V_\lin$  by Lemma \ref{lem:v_r}; 
\item\label{item:noy_eff} $\vrr = \vl$ if \eqref{eq:rest_noy} is correctly specified.
\ende
\end{theorem}

Theorem \ref{thm:noy} establishes two theoretical guarantees of $\hyr$ from the {\gor}.
Theorem \ref{thm:noy}\eqref{item:noy_consis} ensures that it is always consistent and asymptotically normal regardless of whether \eqref{eq:rest_noy} is correctly specified or not.
Theorem \ref{thm:noy}\eqref{item:noy_eff} ensures that it attains the same asymptotic efficiency as $\hyl$ when \eqref{eq:rest_noy} is  indeed correct.
Simulation in Section \ref{sec:simulation} further suggests that $\hyr $ can have better finite sample performance than $\hyl$ as long as the restriction is not severely misspecified. 
Intuitively, the restriction on $\hgr$, namely $\rhg \hgr = r_\gamma$, reduces its variability relative to $\hgl$  from \olss in finite samples.
Such reduction in variability, despite having no effect on the asymptotic efficiency of $\hyr$ when \eqref{eq:rest_noy} is correctly specified, 
improves its precision in finite samples.
This gives the third guarantee of $\hyr$  from the {\gor}. 
It is our recommendation for mitigating the conundrum of many covariates and many treatments when the sample size is moderate.

Additional restriction on $\by$, on the other hand, promises the possibility of additional efficiency over $\hyl$.
We elaborate on the details below. 

\subsubsection{$\hyr $ subject to the {\sr} with $\rhy \neq 0$}\label{sec:sep}
Let $\emat = \diag(e_q)_{q\in\mt}$, and let 
\beginy\label{eq:umu}
&&U =I_Q - \einv \rhy^\T   (\rhy \einv \rhyt)^{-1} \rhy, \quad
\mur =- \einv \rhy^\T   (\rhy \einv \rhyt)^{-1} (\rhy\by - r_Y)
\endy
for $\rhy \neq 0$.

\begin{theorem}\label{thm:sep}
Assume complete randomization, Condition \ref{asym}, and  \rlss subject to \eqref{eq:rest_sep} with $\rhy \neq 0$. 
Then 
\beginy\label{eq:sep_clt}
\sqrt N(\hyr  - \by - \mur) \rs \mathcal N(0_Q, U \vlr  U^\T),
\endy
where $\vlr \geq \vl$ by Lemma \ref{lem:v_r}.  In particular, (i) $\mur = 0$ if $\rhy\by=r_Y$ is correctly specified; (ii) $\vlr = \vl$ if $\rhg\gamma = r_\gamma$ is correctly specified.
\end{theorem}

The asymptotic normality in \eqref{eq:sep_clt} implies $\hyr  - \by = \mur + \op$ such that $\hyr$ is consistent if and only if $\mu_\rr =o(1)$.  
This is in general not true unless the restriction on $\by$ is correctly specified.
Given that we can never verify the correctness of $\rhy\by = r_Y$ exactly when $\rhy \neq 0$, this illustrates one advantage of imposing restriction on only $\gamma$.  

Theorem \ref{thm:sep}(i) ensures the consistency and asymptotic normality of $\hyr$ when the restriction on $\by$ is indeed correctly specified.
Theorem \ref{thm:sep}(ii) ensures that $U\vl U^\T$ gives a lower bound of the asymptotic covariance of $\hyr$ when the restriction on $\gamma$ is correctly specified. 
Whereas there is in general no definite order between $U\vl U^\T$ and $\vl$, one exception is under the constant treatment effects condition which ensures $U\vl U^\T\leq\vl$.
Theorem \ref{thm:gm} below builds on this intuition,  and establishes the asymptotic bias-variance trade-off between $\hyr$ and $\hyl$ when the restriction on $\by$ is non-empty. 
The result extends \citet[][Theorem A5]{ZDa} on unadjusted estimators to the covariate-adjusted variants, and ensures the asymptotic efficiency of $\hyr$ over $\hyl$ under constant treatment effects when the restriction is correctly specified.

 
\begin{theorem}\label{thm:gm}
Assume complete randomization, Conditions \ref{cond:sa}--\ref{asym}, and \rolss subject to \eqref{eq:rest_sep} with $\rho_Y\neq 0$ being a contrast matrix with rows orthogonal to $1_Q$. 
\begine[(i)]
\item \label{item:bv} If $\rhg\gamma = r_\gamma$ is correctly specified, then $\hyr - \by  = \mur + \op$ and $\hyr$ has smaller asymptotic covariance than $\hyl$. 
\item \label{item:gm}
If both $\rhy \by = r_Y$ and $\rhg\gamma = r_\gamma$ are correctly specified, then  $\hyr$ is consistent, asymptotically normal, and asymptotically more efficient than $\hyl$. That is, 
$
\sqrt N(\hyr  - \by) \rs \mathcal N(0_Q, U \vl  U^\T)$
with $U \vl  U^\T \leq \vl$. 
\ende
\end{theorem}

Assume constant treatment effects and correctly specified restriction on $\gamma$. 
Theorem \ref{thm:gm}\eqref{item:bv} states the reduction in asymptotic covariance by arbitrary restriction on contrasts of $\by$ at the cost of possibly non-diminishing bias. 
Further assume that the restriction on $\by$ is also correctly specified. 
Theorem \ref{thm:gm}\eqref{item:gm} ensures the asymptotic efficiency of $\hyr$ over $\hyl$. 
As it turns out, the efficiency of $\hyr$ extends to a class of general linear consistent estimators of $\by$. This gives  the design-based counterpart of the classical Gauss--Markov theorem for \rls.  We relegate the formal statement to Theorem S2 in the {\sm}.

Juxtapose Theorems \ref{thm:noy}--\ref{thm:gm}. The restrictions on $\by$ and $\gamma$ have distinct consequences on the sampling properties of $\hyr$. 
The restriction on $\by$, on the one hand, incurs non-diminishing bias when misspecified, yet 
promises lower asymptotic covariance than $\hyl$ under constant treatment effects.
The choice of whether to restrict $\by$ is thus a trade-off between asymptotic bias and variance. 
The restriction on $\gamma$, on the other hand,  retains consistency regardless of whether correctly specified or not, but undermines asymptotic efficiency when misspecified. 
Simulation studies further suggest that it can improve the finite sample performance of $\hyr$. 
The choice of whether to restrict $\gamma$ is thus a trade-off between finite sample performance and asymptotic efficiency. 


\subsubsection{Robust covariance for restricted least squares}\label{sec:ehw_rols}
Recall that $\hthlr   = (\hyr^\T, \hgr^\T)^\T
= (I-\mr R) \hthl + \mr r$ from Lemma \ref{lem:hthr}. 
Let
\begina
\hsigr  =  \ccinvl \big\{ \chi_\lin ^\T  \diag(\hat\epsilon_{\lr,1}^2, \dots, \hat\epsilon_{\lr,N}^2) \chi_\lin \big \}\ccinvl
\enda 
be a variant of the \ehw, also known as  the  {\it sandwich}, robust covariance estimator of $\hthl$, 
where we use the \rolss residuals $\helri = Y_i - t_i^\T \hyr - (t_i \otimes x_i)^\T \hgr$ in the middle. 
We define 
\begina
(I-\mr R ) \hsigr  (I-\mr R )^\T = (I-\mr R ) \ccinvl \big\{ \chi_\lin ^\T  \diag(\hat\epsilon_{\lr,1}^2, \dots, \hat\epsilon_{\lr,N}^2) \chi_\lin \big \}\ccinvl  (I-\mr R )^\T
\enda as the {\it double-decker-taco robust} covariance estimator of   $\hthlr$  and 
use its upper-left $Q \times Q$ submatrix, denoted by $\hsr$, to estimate the sampling covariance of $\hyr$.

Recall the definition of $S_\rr$ from the beginning of Section \ref{sec:rols_trt}. Theorem \ref{thm:ehw_rr} below gives the probability limit of $\hsr$. 

\begin{theorem}\label{thm:ehw_rr}
{\precre}
\begine[(i)]
\item\label{item:ehw_noy} Under \rlss subject to \eqref{eq:rest_noy}, we have $N \hsr = \vlr + S_{\lr } + \op$  with $S_\lr \geq 0$.
\item Under \rlss subject to \eqref{eq:rest_sep} with $\rhy \neq 0$, 
we have $$N\hsr = U\vlr U^\T +  U \big\{S_{\lr } + \diag(\murq^2 /e_q)_{\qit} \big\} U^\T + \op,$$
where $\murq$ denotes the $q$th element of $\mur$ with $U  \{S_{\lr } + \diag(\murq^2 /e_q)_{\qit}\} U^\T \geq 0$. \\
Further assume that $\rhy \by = r_Y$ is correctly specified. Then  $N\hsr= U\vlr U^\T +  US_{\lr }U^\T  + \op$ with $US_{\lr }U^\T \geq 0$.

\ende
\end{theorem}

Recall that $\sqrtn (\hyr  - \by )  \rs \mn(0_Q, \vlr)$ under \rlss subject to \eqref{eq:rest_noy}  from Theorem \ref{thm:noy}, and $\sqrt N(\hyr  - \by) \rs \mathcal N(0_Q, U \vlr  U^\T)$ under \rlss subject to \eqref{eq:rest_sep} when $\rhy \neq 0$ and $\rhy\by = r_Y$ is correctly specified from Theorem \ref{thm:sep}. Theorem \ref{thm:ehw_rr} thus ensures that $\hsr$ is asymptotically conservative for estimating the true sampling covariance of $\hyr $ under both scenarios. 
This justifies the large-sample Wald-type inference of $\tau$ based on $\htlr = C\hyr$ and $C \hsr C^\T$ as the point estimator and estimated covariance, respectively, when $\rhy\by = r_Y$ is correctly specified.

In addition, recall that the unadjusted and additive regressions can be viewed as restricted variants of \eqref{eq:lm_l}, assuming the zero and equal correlation restrictions, respectively.  
The $\hsr$ introduced above provides an alternative way to estimate the covariance of $\hys \ (\nf)$ in addition to the $\hsi_*$'s from the \olss fits in Lemma \ref{lem:reg}. 
Let $ \hsi_{\nm, \rr}$ and $ \hsi_{\fisher, \rr}$ be the values of $\hsr$ from fitting \eqref{eq:lm_l} subject to the zero and equal correlation restrictions, respectively. 
Theorem \ref{thm:ehw_rr}\eqref{item:ehw_noy} and Lemma \ref{lem:reg}\eqref{item:wald} together ensure that $\hsi_*$ and $\hssr$ are asymptotically equivalent for $\nf$. Proposition S4 in the Supplementary Material 
gives a stronger result on their numeric equivalence, i.e., $\hsis=\hssr \ (\nf)$,  free of any assumptions.
This illustrates the equivalence between the double-decker-taco covariance estimator from the \rlss fit and  the \ehws covariance estimator from the \olss fit of the corresponding restricted specification. See Lemma S3 in the Supplementary Material for a general result.

\subsection{Concluding remarks and an extension to rerandomization}\label{sec:rem_rmk}

Theorems \ref{thm:noy}--\ref{thm:ehw_rr} summarize our main results on the design-based properties of \rlss for regression adjustment in \mes.
The \rolss estimator $\htlr = C \hyr $ includes $\hts \ (\nf)$ as special cases and ensures a triple guarantee for inferring $\tau$: it is (i) asymptotically efficient when the restriction is correctly specified, (ii) consistent when the restriction on $\by$ is correctly specified and separate from that on $\gamma$, and (iii) can have better finite sample performance than  the unrestricted $\htl$ when the restriction is not severely misspecified.
These, together with the asymptotic conservativeness of $C\hsr C^\T$ for estimating the true sampling covariance,  suggest the advantage of \rlss in finite samples.

The correctness of the restriction is central to the theoretical guarantees of $\htr$. 
This can be assessed by testing the null hypothesis of $H_0: R\thl = r$. 
Let 
$
\hsig_\lin$ be the \ehws covariance estimator of $\hthl$ from the \olss fit of \eqref{eq:lm_l}. 
We propose to test $H_0$ with  
$
W = (R \hthl - r)^\T (R   \hat\Sigma_\lin R^\T)^{-1}(R \hthl - r)$
as the test statistic and compute a one-sided $p$-value by comparing $W$ to $\chi^2_m$ with $m=\rank(R)$. 
The resulting test preserves the nominal type one error rates asymptotically.
We give the details in the {\sm}.

Rerandomization gives another way 
 to incorporate covariate information in the design stage of experiments  \citep{morgan2012rerandomization}. It accepts a treatment allocation if and only if it satisfies some prespecified covariate balance criterion.
Let $\hx(q) = N_q^{-1}\sumiq x_i$ denote the sample mean of covariates under treatment level $q \in \mt$.
Contrasts of $\hx(q)$'s provide intuitive measures of covariate balance across the $Q$ treatment groups, and allow us to form covariate balance criterion based on the \mhl distance of their concatenation to the origin \citep{branson, AOS}.
The resulting inference based on $\htr$ inherits all guarantees from inference under complete randomization and, in addition, ensures less loss in asymptotic efficiency when the restriction is misspecified. 
Due to space limitations, we relegate the theory to the {\sm}.  
The result extends the existing literature on rerandomization and highlights its value for additional protection against model misspecification.

\section{Regression adjustment in factorial experiments}\label{sec:factor}
\subsection{Overview of factorial experiments}\label{sec:factor_overview}
Factorial experiments are a special type of \mes, featuring treatments as combinations of two or more factors, each of two or more levels. 
The treatment-based regressions provide a principled way of studying factorial experiments, enabling inference of arbitrary {\textsc{ate}} by least squares. 
Despite their generality  and nice theoretical guarantees, however, they are not the dominant choice for analyzing factorial data in practice.
Factor-based regressions, as a more popular approach, regress the observed outcome directly on the factors themselves and interpret the coefficients as the corresponding
factorial effects of interest.
This enables not only direct inference of the treatment effects based on regression outputs but also flexible unsaturated
specifications to reduce model complexity.

The existing literature on covariate adjustment for factor-based regressions focuses on {\it factor-saturated} specifications that include all possible interactions between the factors \citep{lu2016covariate, ZDb}. 
Recall from Example \ref{ex:exp_general} that a general factorial experiment has in total $Q = \prod_{k=1}^K Q_k $ treatment levels, with $Q  \geq 2^K$. The resulting regressions contain $Q + J$ and $Q(1+J)$ estimated coefficients under the additive and fully interacted specifications, respectively, subjecting subsequent inference to large finite sample variability even when $K$ is moderate.  
We extend the discussion to {\it factor-unsaturated} regressions and establish their properties for covariate adjustment from the design-based perspective. 
The resulting theory is not only of practical relevance in itself given the rising popularity of factorial experiments but also illustrates the value of \rolss for studying estimators from general \olss regressions.
We will focus on $2^K$ factorial experiments 
for notational simplicity. 
The result  extends to the 
general factorial experiments in Example \ref{ex:exp_general}
with minimal modification; see also \citet[][Section A]{ZDa}.

\subsection{Standard factorial effects for $2^K$ factorial experiments}
Consider a $2^K$ factorial experiment with $K$ factors of interest, $k = \ot{K}$, each of two levels. 
Inherit the notation from Example \ref{ex:exp_2k}.
The $Q = 2^K$ treatment levels are $q = (z_1,\dots, z_K) \in \mt = \pmo ^K$, where $z_k \in \pmo $ indicates the level of factor $k$. 

Let $\pk = \{\mk: \text{$\mk \subseteq [K]$ and $\mk \neq \emptyset$} \}$ denote the set of the $2^K-1$ non-empty subsets of $[K]= \{\ot{K}\}$. 
There are $2^K - 1$ standard factorial effects under the $2^K$ factorial experiment, one for each $\mk \in \pk$, 
characterizing the main effect or $|\mk|$-way interaction of the factor(s) in $\mk$ for $|\mk| =1$ and  $|\mk| \geq 2$, respectively. 
The $(\ta, \tb, \tab)$ in Example \ref{ex:exp_22} are special cases with $\mathcal P_2 = \{ \{\textup{A}\}, \{\textup{B}\}, \{\textup{A}, \textup{B}\}\}$. 
Let $ \tmk = \cmk^\T\by$ denote the standard factorial effect corresponding to $\mk\in\pk$.
Lemma \ref{lem:cs} below gives the numeric properties of $\cmk$'s  \citep{wh}.

\begin{lemma}\label{lem:cs} For all $\mk \neq \mk' \in \pk$, we have (i) $\cmk^\T 1_Q =0$; (ii) $
2^{K-1} \cmk \in \pmo^Q$;   (iii) $\cmk^\T c_{\mk'} =0$. 
\end{lemma}
 
 By Lemma \ref{lem:cs}, the standard factorial effects are {\it orthogonal} in terms of the contrast vectors that define them. 
Example \ref{ex:2k} below gives the formulas for the main effects and two-way interactions;
see \cite{DasFact15} and \citet{AOS} for formulas for the higher-order interactions.

\begin{example}\label{ex:2k}
The standard main effect of factor $k$ equals
\begina
\tau_{\{k\}} = \frac{1}{2^{K-1}} \sum_{q: z_k = +1} \by(q)   -   \frac{1}{2^{K-1}} \sum_{q:z_k = -1} \by(q),
\enda
comparing the average potential outcomes when factor $k$ is at level $+1$ and level $-1$, respectively. 
The standard interaction effect between factors $k$ and $k'$ equals
\begina
\tau_{\{k,k'\}} = \frac{1}{2^{K-1}} \sum_{q: z_k z_{k'} = +1} \by(q)   -   \frac{1}{2^{K-1}} \sum_{q:z_k z_{k'}= -1} \by(q),
\enda
comparing the average potential outcomes when the two factors take the same and different levels, respectively.
The  $(\ta, \tb, \tab)$ in Example \ref{ex:exp_22}  are special cases at $K = 2$. 
\end{example}

\subsection{Factor-saturated regressions for $2^K$ factorial experiments}\label{sec:fs}
Let $Z_{ik} \in\pmo$ indicate the level of factor $k$ received by unit $i$. 
For $\mk \in \pk$, let $\zimk = \prod_{k \in \mk} Z_{ik}$ represent the interaction between the factors in $\mk$.
The classical experimental design literature takes
\beginy
Y_i 
\ \sim \ 1+ \sum_{k=1}^K Z_{ik} + \sum_{k\neq k'} Z_{ik}Z_{ik'} + \cdots + \prodk  Z_{ik}  &\Longleftrightarrow& Y_i \ \sim  \  1 + \sum_{\mk \in \pk} \zimk \label{eq:lm_2k_n}
\endy
as the standard specification for factor-based regression analysis, 
and estimates $\tmk$ by 2 times the \olss coefficient of $\zimk$ \citep{wh, lu16bFact}. 
The $Y_i \sim 1+Z_i$ under the treatment-control experiment and the $Y_i \sim 1+Z_{i1} +Z_{i2} +Z_{i1}Z_{i2}$ under the $2^2$ factorial experiment are both special cases with $K =1$ and $K=2$, respectively. 
We call \eqref{eq:lm_2k_n} the  {\it factor-saturated} unadjusted regression, which 
includes all possible interactions between elements of $(Z_{ik})_{k=1}^K$.
The presence of covariates further motivates
\beginy
&&Y_i  \ \sim \  1 + \sum_{\mk \in \pk} \zimks + \cxi , \label{eq:lm_2k_f}\\
&&Y_i  \ \sim \ 1 + \sum_{\mk \in \pk} \zimks + \cxi  + \sum_{\mk \in \pk} \zimks \cdot \cxi \label{eq:lm_2k_l}
\endy
as the additive and fully interacted variants, respectively.

Regressions \eqref{eq:lm_2k_n}--\eqref{eq:lm_2k_l} define three factor-saturated specifications for factor-based  analysis of the $2^K$ factorial experiment, paralleling \eqref{eq:lm_n}--\eqref{eq:lm_l} under the treatment-based formulation. 
The upper panel of Table \ref{tb:specs_2k} summarizes them.
Let $c_{\emptyset} = 2^{-(K-1)} 1_Q$ be an orthogonal complement of the linear span of $\{\cmk: \mk \in \pk\}$ 
by Lemma \ref{lem:cs}.
The one-to-one correspondence between $t_i$ and $\{1, \zimks: \mk\in\pk\}$ ensures that $2^{-1}\tmk$, $2^{-1}(c^\T_{\emptyset} \otimes I_J) \gamma$, and $2^{-1}(\cmk^\T\otimes I_J) \gamma$ give the target parameters of $\zimk$, $x_i$, and $\zimk \cdot x_i$ for $\mk \in \pk$ in \eqref{eq:lm_2k_l}, respectively; see the proof of Proposition S5 in the {\sm}. 
This gives the intuition behind using 2 times the \olss coefficient of $\zimk$ for estimating $\tmk$.

Let $\ttsmk \ (\nfl)$ be 2 times the coefficients of $\zimk$ from the \olss fits of \eqref{eq:lm_2k_n}--\eqref{eq:lm_2k_l}, respectively, vectorized as 
\begina
\tts = \{\ttsmk: \mk\in\pk\}.
\enda
Let $\tos$ be the \ehws covariance estimator of $\tts$ from the corresponding \olss fit. 
As a convention, we use $\nfl$ to signify the unadjusted, additive, and fully interacted specifications, respectively, and use the tilde $(\tilde{\color{white}\tau})$ to signify outputs from factor-based regressions. 
Let
\begina 
\ts = \{\tmk: \mk \in\pk\} = \cs\by 
\enda
be the vectorization of $\tmk$'s in the same order of $\mk$ as in $\tts$.
Then $\cs$ is a $(Q-1)\times Q$ contrast matrix with $\{\cmk: \mk \in\pk\}$ as its row vectors.
Proposition \ref{prop:2k_saturated} below follows from the invariance of \olss to {\ndt}, and justifies the large-sample Wald-type inference of $\ts$ based on  $(\tts, \tos)$.  
The results on $\ttn$ and $\ttl$ are not new \citep{lu2016covariate, ZDa}, whereas that on $\ttf$ is. 

Recall $\hsi_*$ as the \ehws covariance estimator of $\hys \ (\nfl)$.

\begin{proposition}\label{prop:2k_saturated}
$\tts  = \cs \hys$ and $\tos = \cs \hsi_* \cst$  for $\nfl$. 
\end{proposition}

Proposition \ref{prop:2k_saturated} is numeric,  and ensures the equivalence between the factor-saturated specifications and their treatment-based counterparts for estimating $\ts$. 
The relative efficiency between $\tts\ (\nfl)$ follows immediately from Lemma \ref{lem:reg}, with $\ttl$ being asymptotically the most efficient.

\subsection{Factor-unsaturated regressions}
Despite the conceptual straightforwardness and nice theoretical guarantees, 
the factor-saturated regressions \eqref{eq:lm_2k_n}--\eqref{eq:lm_2k_l} involve $2^K$, $2^K+J$, and $2^K(1+J)$ estimated coefficients, respectively, subjecting subsequent inference to large finite sample variability even when $K$ is moderate \citep{ZDa}.  
Often, only the main effects and two-way interactions are of interest, with the higher-order effects believed to be small or nonexistent. 
A common practice is to include only the relevant terms, and estimate the effects of interest from  first- or second-order  specifications like
$Y_i \sim 1+ \sum_{k=1}^K Z_{ik}$ or $Y_i \sim 1+ \sum_{k=1}^K Z_{ik} + \sum_{k\neq k'} Z_{ik} Z_{ik'}$. 

More generally, assume that we are interested in only a subset of the $2^K-1$ standard factorial effects, 
summarized by 
\begina
\tsp   = \{\tmk   : \mk \in \mfp \} = \csp\by 
\enda
for some $\mfp\subsetneq\pk$ with $\mfm = \pk \backslash \mfp \neq \emptyset$.
The $\csp$ is a $|\mfp|\times Q$  submatrix of $\cs$ 
with $\{\cmk: \mk \in\mfp\}$ as its row vectors.
As a convention, we use $+$ and $-$ in the subscripts to indicate {\it effects of interest} and {\it nuisance effects}, respectively.
The discussion from Section \ref{sec:fs} ensures that the vectors
\begina
\ttsp = \{\ttsmk: \mk\in\mfp\} \qquad (\nfl)
\enda 
give three estimators of $\tsp$ from the factor-saturated regressions \eqref{eq:lm_2k_n}--\eqref{eq:lm_2k_l}. They are all consistent by Proposition \ref{prop:2k_saturated} but may have substantial finite sample variability when $K$ is large.

Alternatively,  recall that we estimate $\tmk$ by 2 times 
the \olss coefficient  of $\zimk$. Now that we are interested in only the factorial combinations in $\mfp$, we can define
\beginy\label{eq:lm_n_2k_u}
&&Y_i  \ \sim \  1+\sum_{\mk \in \mfp} \zimks,\\
&& Y_i \ \sim \ 1 + \sum_{\mk \in \mfp} \zimks + \cxi , \label{eq:lm_f_2k_u}\\
&&Y_i \ \sim \ 1 + \sum_{\mk \in \mfp} \zimks + \cxi  + \sum_{\mk \in \mfp} \zimks \cdot \cxi \label{eq:lm_l_2k_u}
\endy
as the {\it factor-unsaturated} variants of \eqref{eq:lm_2k_n}--\eqref{eq:lm_2k_l}, respectively, to include only the terms relevant to $\mk \in \mfp$, and use the resulting coefficients to estimate $\tsp$. 
This gives three additional regression specifications for estimating $\tsp$, with the first- and second-order specifications above being special cases with $\mfp = \{\{k\}: k \in [K]\}$ and $\mfp = \{\{k\}, \{k,k'\}:   k \neq  k' \in [K] \}$, respectively.
They are arguably the more commonly used approach in practice,  allowing for considerable reduction in model complexity.
Of interest is the implications of such reduction on subsequent inference. We address this question in Section \ref{sec:result} below.

Let
\begina
 \tsm = \{\tmk : \mk\in\mfm \} = \csm \by 
\enda
denote the vector of nuisance effects that complements $\tsp$.
Then $\csm$ is a $(Q-1-|\mfp|)\times Q$  submatrix of $\cs$  
with $\{\cmk: \mk \in\mfm\}$ as its row vectors.
The intuition on target parameters after \eqref{eq:lm_2k_l} allows us to view \eqref{eq:lm_n_2k_u}--\eqref{eq:lm_l_2k_u} as restricted variants of \eqref{eq:lm_2k_l} subject to working restrictions
\beginy
  \nm: && \tsm = 0,\qquad \gamma = 0;\label{eq:rest_2k_main_n}\\
 \fisher: && \tsm  = 0, \qquad \gamma_1 = \cdots =\gamma_Q; \label{eq:rest_2k_main_f}\\
 \lin: && \tsm  = 0, \qquad (\csm\otimes I_J) \gamma = 0, 
 \label{eq:rest_2k_main_l}
\endy
respectively.  In particular, 
Lemma \ref{lem:cs} ensures that the $\gamma = 0$ in \eqref{eq:rest_2k_main_n} gives the concise form of  ``$(c^\T_{\mk} \otimes I_J) \gamma = 0 \text{ for all }\mk \in \{\emptyset\} \cup \pk$", corresponding to \eqref{eq:lm_n_2k_u}.
Likewise for the $\gamma_1 = \cdots =\gamma_Q$ in \eqref{eq:rest_2k_main_f} and $(\csm\otimes I_J) \gamma = 0$ in \eqref{eq:rest_2k_main_l} to give the concise forms of  ``$(c^\T_{\mk} \otimes I_J) \gamma = 0 \text{ for all }\mk \in \pk$" and ``$(c^\T_{\mk} \otimes I_J) \gamma = 0$ for all $\mk \in  \mfm$", respectively, corresponding to \eqref{eq:lm_f_2k_u} and \eqref{eq:lm_l_2k_u}.

\begin{table}[t]\small
\renewcommand{\arraystretch}{1.25}
\caption{\label{tb:specs_2k}  Six factor-based regressions under the $2^K$ factorial experiment.}
\begin{center}
 
\begin{tabular}{|c|l|c|c|}\hline
  & & {{\ols} estimator}& {{\ols} estimator}\\
Base model & \multicolumn{1}{c|}{Regression equation} & {of $\tmk$}& of $\tsp$ \\\hline
&\eqref{eq:lm_2k_n}: $Y_i \sim 1 + \sum_{\mk \in \pk} \zimk$ & $\ttnmk$  & $\ttnp$     \\\cline{2-4}
$Y_i \sim 1 + \sum_{\mk \in \pk} \zimk$ &\eqref{eq:lm_2k_f}: $Y_i \sim 1 + \sum_{\mk \in \pk} \zimk+ \cxi $ & $\ttfmk$& $\ttfp$    \\\cline{2-4}
(factor-saturated) &\eqref{eq:lm_2k_l}: $Y_i \sim 1 + \sum_{\mk \in \pk} \zimk + \cxi  + \sum_{\mk \in \pk} \zimk \cdot x_i  $ &{$\ttlmk$} & $\ttlp$   \\\hline
& \eqref{eq:lm_n_2k_u}: $Y_i \sim 1 + \sum_{\mk \in \mfp} \zimk$ & $\ttnrmk$ & $\ttnrp$  \\\cline{2-4}
{$Y_i \sim 1 + \sum_{\mk \in \mfp} \zimk$}&\eqref{eq:lm_f_2k_u}: $Y_i \sim 1 + \sum_{\mk \in \mfp} \zimk+\cxi $ & $\ttfrmk$ & $\ttfrp$   \\\cline{2-4}
(factor-unsaturated) & \eqref{eq:lm_l_2k_u}: $Y_i \sim 1 + \sum_{\mk \in \mfp} \zimk + \cxi  + \sum_{\mk \in \mfp} \zimk \cdot x_i   $  & $\ttlrmk$  & $\ttlrp$ \\\hline
\end{tabular}

\end{center}
\end{table}

\subsection{Design-based properties of the factor-unsaturated regressions}\label{sec:result}
Let $\ttsrmk \ (\nfl)$ denote 2 times the coefficients of $\zimk$ from the \olss fits of  \eqref{eq:lm_n_2k_u}--\eqref{eq:lm_l_2k_u}, respectively.
This gives three additional estimators of $\tmk$ for $\mk\in\pk$, summarized in the lower panel of Table \ref{tb:specs_2k}.
We use the subscript ``u" to signify their origins from the factor-unsaturated regressions.
Let  
\begina
\ttsrp = \{\ttsrmk: \mk\in\mfp\}
 \qquad (\nfl) 
 \enda
 be the corresponding estimators of $\tsp$. 
This, together with the $\ttsp$'s from the factor-saturated regressions, gives in total six estimators of $\tsp$,  summarized in the last column of Table \ref{tb:specs_2k}.

Let $\tosrp$ be the  \ehws covariance estimator of $\ttsrp$ from the corresponding \olss fits of \eqref{eq:lm_n_2k_u}--\eqref{eq:lm_l_2k_u}. 
Let $\hynr $, $\hyfr $, and $\hylr  $ be the coefficient vectors of $t_i$ from the \rolss fits of \eqref{eq:lm_l} subject to working restrictions \eqref{eq:rest_2k_main_n}--\eqref{eq:rest_2k_main_l}, respectively.
They are all special cases of $\hyr$ with separate restrictions on $\by$ and $\gamma$.
Let $\hat \Psi_{*,\rr}$ be the corresponding double-decker-taco covariance estimator of $\hy_{*, \rr} \ (\nfl)$ from Section \ref{sec:ehw_rols}.
Proposition \ref{prop:2k} below states the numeric correspondence between $(\ttsrp, \tosrp)$ and $(\hysr, \hat \Psi_{*,\rr})$.

\begin{proposition}\label{prop:2k}
$\ttsrp   =  \csp\hysr$ and $\tosrp   =  \csp\hat \Psi_{*,\rr}\csp^\T$  for $\nfl$. 
\end{proposition}

The design-based properties of $(\ttsrp, \tosrp)$  then follow from those of $(\hysr, \hat \Psi_{*,\rr})$ in  Theorems \ref{thm:sep}--\ref{thm:ehw_rr}. 
In particular, Theorem \ref{thm:sep} implies that $\ttsrp$ is in general not consistent  for estimating $\tsp$ unless $\tsm = 0$ is correctly specified. When $\tsm $ is indeed $0$, Corollary \ref{cor:2k_wald} below follows from Theorems \ref{thm:sep} and \ref{thm:ehw_rr}, and justifies the large-sample Wald-type inference  of $\tsp$ based on $(\ttsrp, \tosrp)$.

Recall the definition of $\vrr$ from the beginning of Section \ref{sec:rols_trt}.
Let $\vsr$ be the value of $\vrr$ associated with $\hysr$ for $\nfl$. 
Let $(\us, \murs)$ 
be the values of $(U, \mur)$ at $\rhy = \csm$ from \eqref{eq:umu}.

\begin{corollary}\label{cor:2k_wald}
Assume \cre, Condition \ref{asym}, and $\tsm = 0$. 
For $\nfl$, we have $\sqrt N(\ttsrp  - \tsp) \rs \mathcal N(0, \csp \us  \vsr  \us^\T \csp^\T)$, with $\tosrp$  being asymptotically conservative for estimating the true sampling covariance of $\ttsrp$.
\end{corollary}

Corollary \ref{cor:2k} below follows from Theorem \ref{thm:gm}\eqref{item:gm} and gives the asymptotic relative efficiency of $\{\ttsp, \ttsrp: \nfl\}$ under constant treatment effects.

\begin{corollary}\label{cor:2k}
Assume \cre, Conditions \ref{cond:sa}--\ref{asym}, and $\tsm = 0$.
\begine[(i)]
\item\label{item:1} Among $\ttsp \ (\nfl)$ from the factor-saturated regressions \eqref{eq:lm_2k_n}--\eqref{eq:lm_2k_l}, $\ttfp$ and $\ttlp$ are asymptotically equally efficient, and are both more efficient than $\ttnp$. 
\item\label{item:2} Among $\ttsrp \ (\nfl)$  from the factor-unsaturated regressions \eqref{eq:lm_n_2k_u}--\eqref{eq:lm_l_2k_u}, $\ttfrp$ and $\ttlrp$ are asymptotically equally efficient, and are both more efficient than $\ttnrp$.
\item\label{item:3} The $\ttsrp$ from the factor-unsaturated regression is asymptotically more efficient than $\ttsp$ from the factor-saturated regression for $\nfl$. 
\ende

\end{corollary}

Assume Condition \ref{cond:sa} of constant treatment effects. 
Corollary \ref{cor:2k} establishes the asymptotic efficiency of $\ttfrp $ and $\ttlrp$ among $\{\ttsp, \ttsrp: \nfl\}$ for estimating $\tsp$. 
 Intuitively, Condition \ref{cond:sa} implies the equal correlation condition, which ensures that the additive adjustment is asymptotically as efficient as the fully interacted adjustment to begin with.  This underlies the asymptotic equivalence between $\ttfp$ and $\ttlp$  in Corollary \ref{cor:2k}\eqref{item:1}, and that between $\ttfrp $ and $\ttlrp$ in Corollary \ref{cor:2k}\eqref{item:2}. The additional, correct knowledge on the nuisance effects further secures extra precision of $\ttfrp $ and $\ttlrp$ over $\ttfp$ and $\ttlp$, as formalized by  Corollary \ref{cor:2k}\eqref{item:3}.
 This illustrates the value of factor-unsaturated regressions in combination with covariate adjustment for improving efficiency.

A key limitation of the factor-unsaturated regressions is that the consistency of  $\ttsrp  $ depends critically on the actual absence of the nuisance effects. 
This can never be verified exactly in reality and subjects subsequent inference to possibly non-diminishing biases. 
This suggests the merit of the factor-saturated additive regression \eqref{eq:lm_2k_f} as a  compromise between asymptotic bias, asymptotic efficiency, and finite sample performance. The resulting inference is always consistent, and ensures asymptotic efficiency under equal correlations. 
Simulation in Section \ref{sec:simulation} further demonstrates its finite sample advantage over the asymptotically more efficient fully interacted counterpart \eqref{eq:lm_2k_l}.

One exception is when the  treatment groups are of equal size.
The resulting $\ttsrp$ is consistent even when $\tsm \neq 0$ thanks to the orthogonality of $\cmk$'s from Lemma \ref{lem:cs}. We formalize the intuition in Proposition \ref{prop:bd} below.

\begin{condition}
[equal-sized design]\label{condition::equal-group-size}
$e_q = Q^{-1}$ for all $q \in \mt$. 
\end{condition}

\begin{proposition}\label{prop:bd}
Assume complete randomization and Conditions \ref{asym}--\ref{condition::equal-group-size}. Then
\begine[(i)] 
\item\label{item:bd_1} $\sqrt N(\ttsrp - \tsp ) \rs \mn(0, \csp \vsr \csp^\T)$ for $\nfl$, with  $\tosrp$ being asymptotically conservative for estimating the true sampling covariance of $\ttsrp$;
\item\label{item:bd_15}  $\ttlp$ is asymptotically more efficient than $\ttsrp$ for $\nfl$; 
\item\label{item:bd_2}
$\ttlrp$ is asymptotically as efficient as $\ttlp$ if $(\csm  \otimes I_J) \gamma = 0$;\\
 $\ttfrp$ and $\ttlrp$ are asymptotically as efficient as  $\ttlp$ under Condition \ref{cond:equal}; \\
 $\ttsrp \ (\nfl)$ are asymptotically as efficient as $\ttlp$ under Condition \ref{cond:zero}. 
\ende
\end{proposition}

\medskip

Proposition \ref{prop:bd}\eqref{item:bd_1} justifies the large-sample Wald-type inference of $\tsp$ based on $(\ttsrp, \tosrp)$ in equal-sized designs regardless of whether $\tsm = 0$ or not. The result for $\ttnrp$ is a direct consequence of the numeric equivalence between $\ttnrp$ and $\ttnp$ under equal-sized designs: $\ttnrp =  \csp \hyn = \ttnp$; see \cite{ZDa}. We give similar results on $\ttfrp$ and $\ttlrp$ in the {\sm}. This additional protection against model misspecification echos the emphasis of the classical experimental design literature on equal treatment group sizes and orthogonality of the factorial effects. 

The asymptotic efficiency of $\ttsrp$ nevertheless still requires the associated restriction on $\gamma$ in \eqref{eq:rest_2k_main_n}--\eqref{eq:rest_2k_main_l} to be correctly specified, and can no longer exceed that of $\ttlp$, as demonstrated by Proposition \ref{prop:bd}\eqref{item:bd_15}--\eqref{item:bd_2}. 
In particular, 
the condition $(\csm \otimes I_J)\gamma = 0$ implies that the  restriction on $\gamma$  in \eqref{eq:rest_2k_main_l} is correctly specified, and ensures that $\ttlrp$ attains the same asymptotic efficiency as $\ttlp$.  
Condition \ref{cond:equal} implies that the restrictions on $\gamma$  in \eqref{eq:rest_2k_main_f} and \eqref{eq:rest_2k_main_l} are both correctly specified, and ensures that $\ttfrp$ and $\ttlrp$ attain the same asymptotic efficiency as $\ttlp$.
Condition \ref{cond:zero} implies that the restrictions on $\gamma$ in \eqref{eq:rest_2k_main_n}--\eqref{eq:rest_2k_main_l} are all correctly specified, and ensures that all three $\ttsrp \ (\nfl)$ attain the same asymptotic efficiency as $\ttlp$.
This illustrates the asymptotic bias-variance trade-off regarding the use of equal-sized designs, which ensure the consistency of $\ttsrp \ (\nfl)$ but preclude the possibility of additional asymptotic efficiency over $\ttlp$ (c.f.~Theorem \ref{thm:gm}).

\subsection{Extensions}

We focused on specifications \eqref{eq:lm_2k_n}--\eqref{eq:lm_l_2k_u} because of their intuitiveness. 
The same results extend to general specifications like 
$
Y_i \sim 1 + \sum_{\mk \in \mfp} \zimks + x_i + \sum_{\mk \in \mfp'} \zimks \cdot \cxi$ for possibly different $\mfp, \mfp' \subseteq \pk$ 
with minimal modification. 
We relegate the details to the \sm. 

In addition, we focused on specifications under the $\{-1,+1\}$ coding system for inference of the standard factorial effects.
Practitioners instead often encode the factor levels by $\{0, 1\}$.
The resulting specifications, despite prevalent in practice, cannot recover the standard factorial effects directly as regression coefficients \citep{ZDa}. 
The invariance of \olss to {\ndt}, on the other hand, ensures that all results so far extend to the $\{0,1\}$-coded regressions for estimating a different set of factorial effects with minimal modification. We relegate the details to the \sm.

\section{Simulation}\label{sec:simulation}

We now illustrate the finite sample properties of the proposed method by simulation.
Consider a $2^2$ factorial experiment with $Q = 4$ treatment levels, $q  \in  \mt = \{(--), (-+), (+-), (++)\}$, and a study population of $N=100$ units, $i = \ot{N}$. 
We choose the moderate sample size on purpose to illustrate the limitation of the factor-saturated fully interacted regression in finite samples. 

Inherit the notation from Example \ref{ex:exp_22}.
For each $i$, we draw a $J=20$ dimensional covariate vector as $x_i \sim \mn(0_J, I_J)$, and generate the potential outcomes as $Y_i(q) \sim \mathcal{N}( x_i^\T \beta_q , 1)$ for $q = (-+), (+-), (++)$ and $Y_i(--) = Y_i(+-) + Y_i(-+) - Y_i(++)$. 
The resulting $Y_i(q)$'s satisfy the condition of zero individual interaction effects from Example \ref{ex:rols_no inter}, and ensure that working restriction \eqref{eq:rest_l_22} is correctly specified with  $\tau_{\AB} = 0$ and $ (\gamma_{\mmm\mmm}+ \gamma_{\pp\pp}) - (\gamma_{\mmm\pp} + \gamma_{\pp\mmm}) = 0_J$.
We fix $\{Y_i(q), x_i: q\in\mt\}_{i=1}^N$  in the simulation and generate the treatment assignments from complete randomization with $(N_{\mmm\mmm}, N_{\mmm\pp}, N_{\pp\mmm}, N_{\pp\pp}) = (22, 23, 24, 31)$.


Consider six models for estimating $\ts =   (\ta,\tb , \tab)^\T$ by \ols, summarized in Table \ref{tb:sim}.
They are special cases of regressions \eqref{eq:lm_2k_n}--\eqref{eq:lm_l_2k_u}, respectively, with $K = 2$ and $\mfp = \{\{\textup A\}, \{\textup B \}\}$. 
We use $A_i$ and $B_i$ to indicate the levels of factors A and B for unit $i$ for intuitiveness.  The equivalence between \rlss and \olss on the corresponding restricted specification ensures that the \olss fit of \eqref{eq:lm_l_2k_u} is equivalent to the \rlss fit of \eqref{eq:lm_2k_l} subject to \eqref{eq:rest_l_22},  which is correctly specified. 
The \olss fits of  \eqref{eq:lm_2k_n}, \eqref{eq:lm_2k_f}, \eqref{eq:lm_n_2k_u}, and \eqref{eq:lm_f_2k_u}, on the other hand, are equivalent to fitting \eqref{eq:lm_2k_l} subject to restrictions that are misspecified, summarized in the last column of Table \ref{tb:sim}.
Observe that $\beta_q$ gives a good approximation to the finite population correlation $\gamma_q$ by construction for $q = (-+), (+-), (++)$. The variability in $(\beta_{\mmm\pp}, \beta_{\pp\mmm}, \beta_{\pp\pp})$  hence reflects the closeness of the ground truth to the equal correlation condition that justifies the additive regression \eqref{eq:lm_2k_f} and its factor-unsaturated variant  \eqref{eq:lm_f_2k_u}.

\begin{table}[t]
\caption{\label{tb:sim} Six factor-based regressions and their respective restrictions relative to \eqref{eq:lm_2k_l}. We use ``N'', ``F'', and ``L'' to indicate the unadjusted, additive, and fully interacted adjustment schemes, respectively, and use the suffix ``\_us'' to indicate the factor-unsaturated variants. }
\begin{center}
\begin{tabular}{|c|l|l|c|}\hline
&&& If  correctly\\
& {\color{white}(18):$ \ $} Regression equation & \multicolumn{1}{|c|}{Restriction relative to  \eqref{eq:lm_2k_l}} & specified \\\hline
N &  \eqref{eq:lm_2k_n}: $\ Y_i \sim 1+A_i +B_i + A_iB_i$ & $\gamma_q = 0_J$ for all $q\in\mt$ & no \\\hline
F & \eqref{eq:lm_2k_f}: $ \ Y_i \sim 1+A_i +B_i + A_iB_i + \cxi $ &  $\gamma_q$'s are all equal & no \\\hline
L &  \eqref{eq:lm_2k_l}: $ \ Y_i \sim 1+A_i +B_i + A_iB_i  $ && yes\\
  &$\qquad\qquad  + \, \cxi +A_i\cxi  +B_i \cxi  + A_iB_i \cxi $&&\\\hline
N\_us &  \eqref{eq:lm_n_2k_u}: $\ Y_i \sim 1+A_i +B_i$ & $\tauAB = 0$ and $\gamma_q = 0_J$ for all $q\in\mt$ & no  \\\hline
F\_us &  \eqref{eq:lm_f_2k_u}: $ \ Y_i \sim 1+A_i +B_i +  \cxi$ & $\tauAB = 0$ and $\gamma_q$'s are all equal & no \\\hline
L\_us & \eqref{eq:lm_l_2k_u}: $ \ Y_i \sim 1+A_i +B_i +  \cxi +A_i\cxi  +B_i \cxi $ & $\tauAB = 0$ and   & yes\\
&&$(\gamma_{\mmm\mmm}+ \gamma_{\pp\pp}) - (\gamma_{\mmm\pp} + \gamma_{\pp\mmm}) = 0_J$&\\
    \hline
  \end{tabular}
\end{center}
\end{table}

\begin{figure} 
\begin{center}
\includegraphics[width= \textwidth]{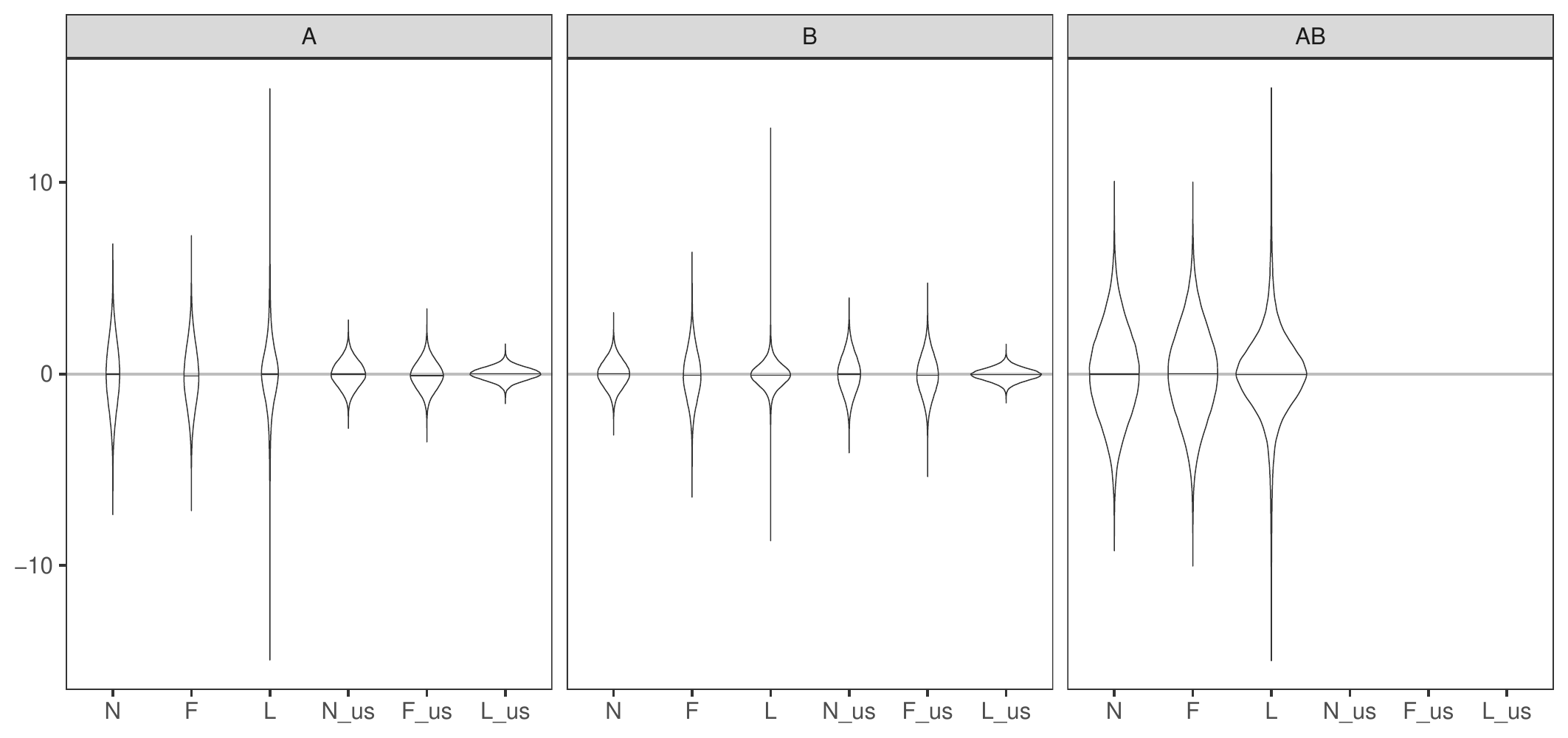}
(a) $(\beta_{\mmm\pp}, \beta_{\pp\mmm}, \beta_{\pp\pp})  = (1_J, 0_J, -1_J)$ such that the $\gamma_q$'s differ considerably.
\end{center}

 \begin{center}
\includegraphics[width= \textwidth]{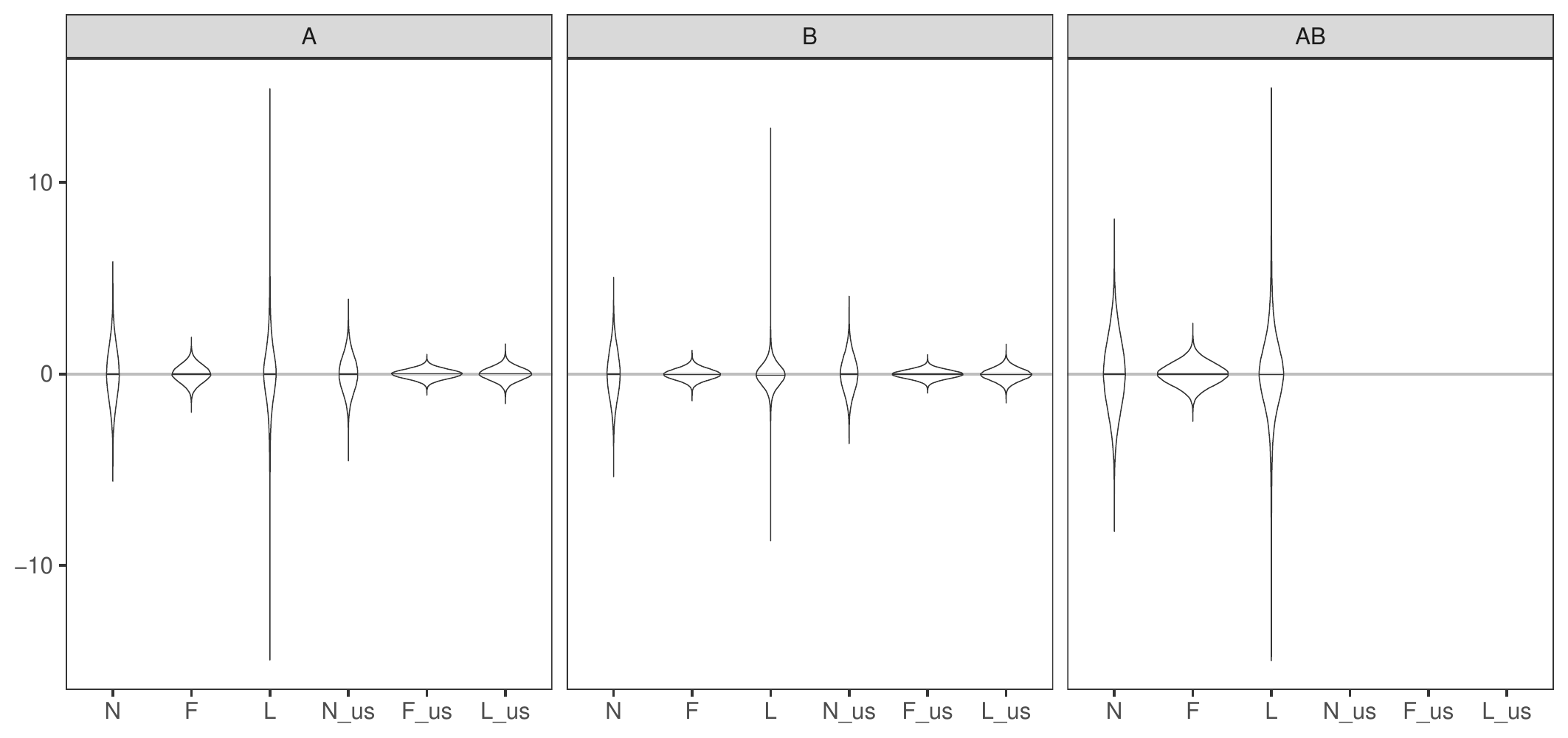}
(b) $\beta_{\mmm\pp}=\beta_{\pp\mmm}= \beta_{\pp\pp} = 1_J$ such that the $\gamma_q$'s are fairly close.
\end{center}

\caption{\label{fig} Violin plot of the differences between 2 times the \olss coefficients of $(A_i, B_i, A_iB_i)$ and the true values of $(\ta, \tb 	, \tab)$  over $100,000$ independent complete randomizations. We use ``N'', ``F'', and ``L'' to represent the factor-saturated unadjusted, additive, and fully interacted regressions, respectively, and use ``N\_us'', ``F\_us'', and ``L\_us'' to  represent their respective factor-unsaturated variants. See Table \ref{tb:sim}.}
\end{figure}

Figure \ref{fig} shows the violin plot of the differences between  2 times the \olss coefficients of $(A_i,B_i,A_iB_i)$ and the true values of $ (\ta,\tb , \tab) $ over $100,000$ independent complete randomizations. Figure \ref{fig}(a) corresponds to  potential outcomes generated from $(\beta_{\mmm\pp}, \beta_{\pp\mmm}, \beta_{\pp\pp})  = (1_J, 0_J, -1_J)$.  The heterogeneity among $\beta_q$'s suggests considerable deviation from the equal correlation condition such that regressions \eqref{eq:lm_2k_f} (``F") and   \eqref{eq:lm_f_2k_u} (``F\_us") are both substantially misspecified.  Figure \ref{fig}(b)  corresponds to potential outcomes generated from $\beta_{\mmm\pp}=\beta_{\pp\mmm}= \beta_{\pp\pp} = 1_J$. The equality of $\beta_q$'s suggests reasonable closeness to the equal correlation condition such that both \eqref{eq:lm_2k_f} and \eqref{eq:lm_f_2k_u} are approximately correctly specified.

The message is coherent across different values of $\beta_q$'s and in line with the theory. The correctly specified  factor-unsaturated regression \eqref{eq:lm_l_2k_u} (``L\_us'') shows the smallest variability in Figure \ref{fig}(a). 
The factor-unsaturated additive regression \eqref{eq:lm_f_2k_u} (``F\_us'') is only approximately correctly specified in Figure \ref{fig}(b), but already delivers even better finite sample performance than the correctly specified \eqref{eq:lm_l_2k_u} thanks to the more parsimonious model. 
The fully interacted regression (``L''), despite being asymptotically the most efficient among the three factor-saturated specifications, shows substantial variability in all cases. 
The four misspecified regressions, namely ``N'', ``F'', ``N\_us'', and ``F\_us'', on the other hand, show much stabler performance even in Figure \ref{fig}(a), where the equal and zero correlation conditions are considerably violated.
This illustrates the robustness of the proposed method to misspecification of the restriction.

\section{Discussion}\label{sec:discussion}
We recommend using restricted least squares on the fully interacted regression  for covariate adjustment in \mess when the sample size is  moderate relative to the number of covariates or treatment levels. 
Assume that the restriction on the {\apo} is correctly specified and separate from that on the {\cpoc}. 
The resulting inference is consistent for estimating the {\textsc{ate}} regardless of whether the restriction on the correlations is correctly specified or not, and ensures additional efficiency over the \olss counterpart 
if the restriction on the correlations is indeed correct under constant treatment effects. 
Simulation studies further show that it can have better finite sample performance than the \olss counterpart even when the restriction is moderately misspecified.  

When prior knowledge on the {\apo} is less reliable, we recommend imposing restriction on only the correlations. 
The resulting estimator ensures consistency yet can be at most as efficient as the \olss counterpart asymptotically. 
Importantly, all results are design-based and hold without assuming any stochastic models for the potential outcomes.

In case of experiments with multiple factors of interest, we assumed a {\it full factorial design} that includes all possible combinations of the different levels of the factors, and used factor-unsaturated specifications to incorporate prior knowledge of the absence of certain factorial effects.
Alternatively, we can accommodate prior knowledge of negligible effects by a {\it fractional factorial design} in the design stage. 
A natural question is then how these two strategies compare to each other in terms of the validity and efficiency of subsequent inference.
Due to space limitations, we illustrate intuitions about their relative strengths and weaknesses by simulation in the {\sm}, and leave the complete theory to future research.

Our theory is asymptotic. While we have also evaluated the finite sample properties of the proposed method 
 by simulation, it is of great interest to derive nonasymptotic results under the design-based framework. This is a nontrivial task because of the lack of probability tools in this framework. We leave it to future research.


\bibliographystyle{plainnat}
\bibliography{refs_factorial-X}

 \newpage

\setcounter{equation}{0}
\setcounter{section}{0}
\setcounter{figure}{0}
\setcounter{example}{0}
\setcounter{proposition}{0}
\setcounter{corollary}{0}
\setcounter{theorem}{0}
\setcounter{table}{0}
\setcounter{condition}{0}
\setcounter{lemma}{0}
\setcounter{remark}{0}

\renewcommand {\theproposition} {S\arabic{proposition}}
\renewcommand {\theexample} {S\arabic{example}}
\renewcommand {\thefigure} {S\arabic{figure}}
\renewcommand {\thetable} {S\arabic{table}}
\renewcommand {\theequation} {S\arabic{equation}}
\renewcommand {\thelemma} {S\arabic{lemma}}
\renewcommand {\thesection} {S\arabic{section}}
\renewcommand {\thetheorem} {S\arabic{theorem}}
\renewcommand {\thecorollary} {S\arabic{corollary}}
\renewcommand {\thecondition} {S\arabic{condition}}
\renewcommand {\thepage} {S\arabic{page}}
\renewcommand {\theremark} {S\arabic{remark}}

\setcounter{page}{1}

  \setcounter{equation}{0}
\renewcommand {\theequation} {S\arabic{equation}}
  \setcounter{lemma}{0}
\renewcommand {\thelemma} {S\arabic{lemma}}
   \setcounter{definition}{0}
\renewcommand {\thedefinition} {S\arabic{definition}}
   \setcounter{example}{0}
\renewcommand {\theexample} {S\arabic{example}}
   \setcounter{proposition}{0}
\renewcommand {\theproposition} {S\arabic{proposition}}
   \setcounter{corollary}{0}
\renewcommand {\thecorollary} {S\arabic{corollary}}

 \begin{center}
\bf \Large 
Supplementary Material
\end{center}


Section \ref{sec:ff_app} uses simulation to compare full and fractional factorial designs.

Section \ref{sec:yb} discusses the properties of a class of general linear estimators of $\by$. It forms the basis of all theoretical results.

Section \ref{sec:rls_app_res} presents additional results on restricted least squares. 

Section \ref{sec:2k_app} presents additional results on $2^K$ factorial experiments. 

Section  \ref{sec:rem} presents the results on rerandomization using the Mahalanobis distance.

Section \ref{sec:yb_app} gives the proofs of the results in Section \ref{sec:yb}.

Section \ref{sec:ols_app} gives the proofs of the results on treatment-based regressions. In particular, Section \ref{sec:ols_ehw_proof} gives the proof of the results from ordinary least squares, and Section \ref{sec:rls_proof} gives the proof of the results from restricted least squares.

Section \ref{sec:factor_proof} gives the proofs of the results on factor-based regressions.


\section*{Notation and definition}
Assume centered covariates with $\bar x=0_J$ throughout to simplify the presentation. 

For a sequence of random vectors $(A_N)_{N=1}^\infty$ with $A_N\rs A$,
let $\ei(A_N) = E(A)$ and $\covi(A_N) = \cov(A)$ denote the expectation  and covariance  with respect to the asymptotic distribution.
For two sequences of random vectors $\{A_N\}_{N=1}^\infty$ and $\{B_N\}_{N=1}^\infty$ where $N$ denotes the sample size, write $A_N \approxi B_N$ if $\sqrtn (A_N -  B_N) = \op$.

Recall that we quantify the relative efficiency of consistent and asymptotically normal estimators by their asymptotic covariances in the main paper. Following \cite{AOS}, we use the concept of peakedness \citep{sherman} below to generalize the comparison to estimators with nonnormal asymptotic distributions.

\begin{definition}[peakedness] 
\label{def:peakedness}
For two symmetric random vectors $A$ and $B$ in $\mrm $, we say $A$ is {\it more peaked} than $B$ if $\pr(A\in \mc) \geq \pr(B \in \mc)$ for every symmetric convex set $\mc \in \mrm $, denoted by $A \succeq B$. 
\end{definition}

For $m = 1$, a more peaked random variable has narrower central quantile ranges.
For $A$ and $B$ with finite second moments, $A \succeq B$ implies $\cov(A) \leq \cov(B)$ \citep[Proposition 4]{AOS}.
For $A$ and $B$ that are both normal with zero means, $A \succeq B$ is equivalent to $\cov(A) \leq \cov(B)$.
This relation between peakedness and covariance suggests that peakedness gives a more refined measure of asymptotic efficiency than covariance for estimators with nonnormal asymptotic distributions.
An asymptotically more peaked estimator has not only a smaller asymptotic covariance but also narrower central quantile ranges.
We formalize the idea in Definition \ref{def:eff} below.

\begin{definition}\label{def:eff}
Assume that $\hth_1$ and $\hth_2$ are two consistent estimators for parameter $\theta \in \mrm$ as the sample size $N$ tends to infinity, 
with $\sqrtn(\hth_1 - \theta) \rs A_1$ and $\sqrtn(\hth_2 - \theta) \rs A_2$ for some symmetric random vectors $A_1$ and $A_2$.  
We say 
\begine[(i)]
\item $\hth_1$ and $\hth_2$ are {\it asymptotically equally efficient} if $A_1$ and $A_2$ have the same distribution, denoted by $\hth_1 \asim \hth_2$; 
\item $\hth_1$ is  {\it asymptotically more efficient} $\hth_2$ if $A_1 \succeq A_2$, denoted by $\hth_1 \succi \hth_2$ or $\hth_2 \preceq_\infty \hth_1$. 
\ende 
\end{definition}

For two estimators $\hth_1$ and $\hth_2$ that are both consistent for parameter $\theta$, 
we have $\hth_1 \approxi \hth_2$ implies that $   \hth_1  \asim \hth_2 $ by Slutsky's theorem.

\section{Simulation comparison of full and fractional factorial designs}\label{sec:ff_app}
The discussion on factorial experiments in the main paper assumed a {\it full factorial design} that includes all possible combinations of the different levels of the factors, and used factor-unsaturated specifications to incorporate prior knowledge of negligible factorial effects.
Alternatively, we can accommodate such knowledge  by a {\it fractional factorial design} in the design stage. 

Consider an experiment with three binary factors A, B, and C. 
Assume that prior knowledge suggests that
the interaction between A and B is negligible. 
This motivates adopting a $2^{3-1}$ fractional factorial design with {\it defining relation} $\ttc= \tta\ttb$ \citep{wh}.  
Index by $\{-1, +1\}$ the two levels of each factor, abbreviated as $\{-, +\}$. Instead of considering all eight possible treatment combinations as under a $2^3$ full factorial design, the $2^{3-1}$ fractional factorial design includes only $2^{3-1} = 4$ combinations of A, B, and C, featuring a $2^2$ full factorial design of factors A and B, and the level of factor C determined by the defining relation as the product of the levels of A and B: 
\begin{center}
\begin{tabular}{ccc}\hline
A & B & C \\\hline
$-$ & $-$  & $+$\\
$-$ & $+$  & $-$\\
$+$ & $-$  & $-$ \\
$+$ & $+$  & $+$  \\
\hline
\end{tabular}
\end{center} 
We can then randomize the units to these four treatment combinations by complete randomization, and estimate the factorial effects of interest by \ols. 
A natural question is how these two strategies compare to each other in terms of the validity and efficiency of subsequent inference.

On the one hand, the fractional factorial design includes only half of the treatment combinations in the full factorial design, effectively doubling the sizes of the treatment groups.
This promises the opportunity to improve estimation efficiency.

On the other hand, the defining relation $\ttc = \tta\ttb$ implies $\tta = \ttb\ttc$ and $\ttb = \tta\ttc$.
This suggests that the main effects of factors A and B are aliased with the interaction between factors B and C and that between factors A and C, respectively, such that they are not identifiable under the fractional design unless the corresponding interactions  are zero.
Likewise for the three-way interaction between A, B, and C to be indistinguishable from the intercept given $\tta\ttb\ttc = \tta\ttb(\tta\ttb) = \textup{I}$.

The same limitation does not exist under the full factorial design, where we can use the restricted specification
$
Y_i \sim 1 + A_i + B_i + C_i + B_iC_i + A_iC_i +A_iB_iC_i 
$
to encode the prior knowledge of no interaction between A and B without affecting the estimation of the main effects. 
The full factorial design is in this light more flexible than the fractional factorial design for accommodating a specific subset of negligible effects. 

The choice between the full and fractional factorial designs is therefore a trade-off between estimation and exploration. 
The fractional factorial design promises improved efficiency for estimating the factorial effects of interest if and only if the assumption of negligible higher-order interactions is correctly specified.  In contrast, the full factorial design allows us to explore the absence or presence of higher-order interactions. 
We illustrate these intuitions by simulation in the following two subsections, and leave the complete theory to future research. See \cite{pashley2021causal} for related discussion.

\subsection{Scenario I: all interaction effects are zero}
Consider an experiment with $K=3$ binary factors, A, B, and C, and a study population of $N = 80$ units, indexed by $i = \ot{N}$. 
Index the $2^3 = 8$ possible treatment combinations by $\mt = \{(abc): a, b, c = -1, +1\}$. 
Building on Example \ref{ex:exp_22} in the main paper,
let $\tau_\A$, $\tau_\AB$, and $\tau_\ABC$ denote the main effect of factor A, the interaction effect between factors A and B, and the interaction effect between factors A, B, and C, respectively; likewise define $\tau_\B, \tau_\C, \tau_\AC$, and $\tau_\BC$.

Let $(x_i)_{i=1}^N$ be independent $\mn(0, 4)$ and $(\ep_i)_{i=1}^N$ be independent $\mn(0,1)$. 
We generate the potential outcomes as $Y_i(abc) = 4+2a + 2b + 2c + a x_i + \ep_i$, and then center $\{Y_i(abc)\}_{i=1}^N$ for each $(abc) \in \mt$ to ensure $(\tau_\A, \tau_\B, \tau_\C, \tau_\AB, \tau_\BC, \tau_\AC, \tau_\ABC) = (4, 4, 4, 0, 0, 0, 0)$.
This ensures that all interaction effects are absent in the ground truth.  

Fix the potential outcomes in the simulation. 
The $2^3$ full factorial design assigns $N_{abc} = 10$ units to each of the eight treatment combinations in $\mt$.
We consider the following three specifications for estimating the standard factorial effects by \ols: 
\begina
\texttt{m1}: &&Y_i \sim 1 + A_i + B_i + C_i +x_i, \\
\texttt{m2}: &&Y_i \sim 1 + A_i + B_i + C_i + A_iC_i + B_iC_i + A_iB_iC_i + x_i,\\
\texttt{m3}: &&Y_i \sim 1 + A_i + B_i + C_i + A_iB_i + A_iC_i + B_iC_i + A_iB_iC_i +x_i. 
\enda
Specification \texttt{m1} assumes the absence of all two-way and three-way interactions, matching exactly the ground truth. Specification \texttt{m2} assumes the absence of the interaction effect between factors A and B. Specification \texttt{m3} is factor-saturated and makes no assumptions about the absence or presence of the factorial effects. They are all correctly specified given the ground truth.  
We use the additive specification for covariate adjustment given the moderate sample size. 

The $2^{3-1}$ fractional factorial design, on the other hand, assigns $N'_{abc} = 20$ units to each of the four treatment combinations in $\mt' = \{(abc): a, b = -1, +1; \ c = ab\} = \{(--+), (-+-), (+--), (+++)\}$ subject to the defining relation $\ttc = \tta\ttb$,  and runs \olss using the same specification as \texttt{m1} under the full design.
We index the resulting regression model by \texttt{m4} to distinguish between the full and fractional designs:
\begina
\texttt{m4}: &&Y_i \sim 1 + A_i + B_i + C_i +x_i \quad \text{under the fractional factorial design}.
\enda
It assumes the absence of all two-way and three-way interactions as specification \texttt{m1},  matching exactly the ground truth. 

\begin{figure}[t]
\begin{center}
\includegraphics[scale=.5]{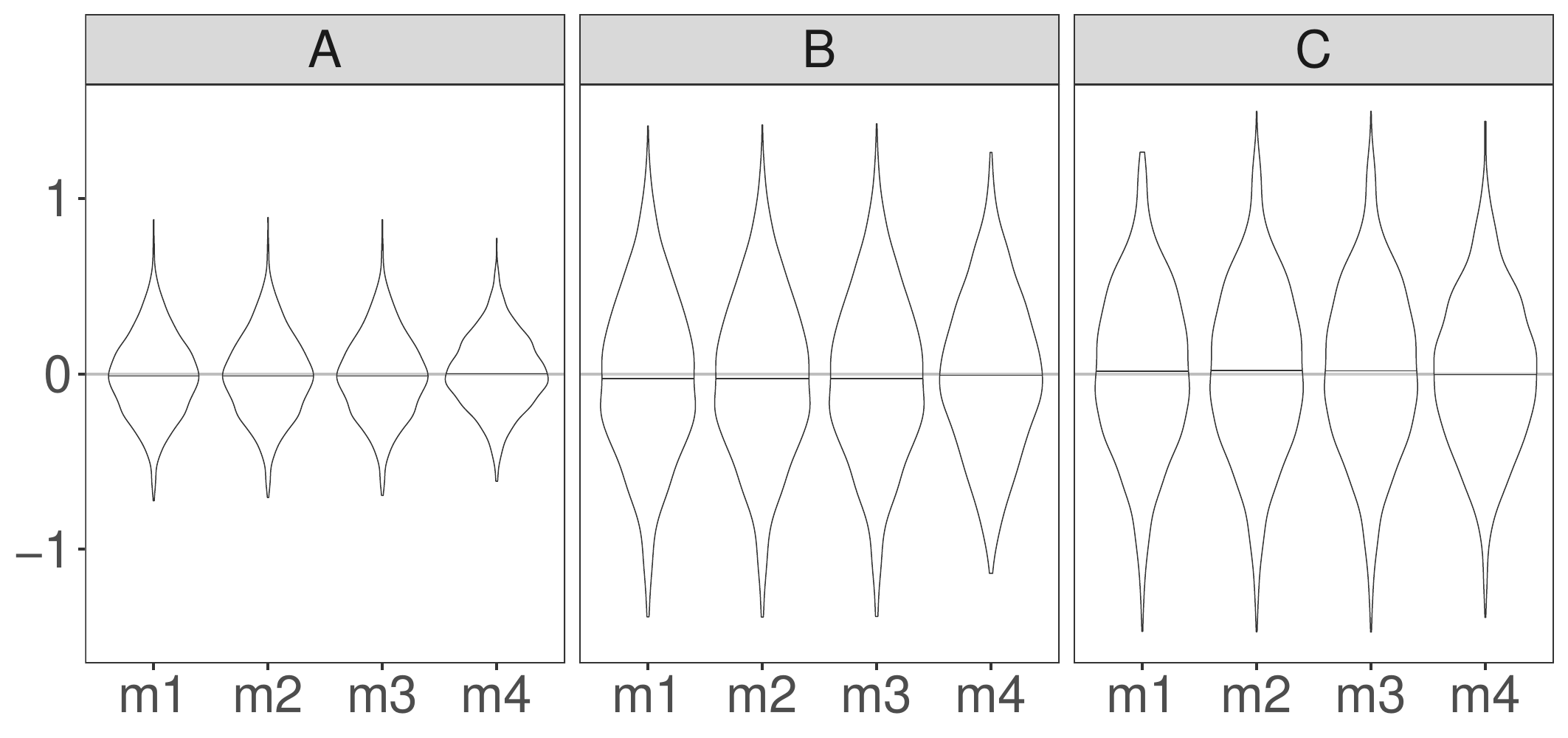}
\end{center}\caption{\label{fig:ff_1}{Violin plot of the differences between the estimated and true values} for the three main effects over 1,000 independent complete randomizations from regressions \texttt{m1}--\texttt{m4} under Scenario I. Regressions \texttt{m1}--\texttt{m3} correspond to the $2^3$ full factorial design, and regression \texttt{m4} corresponds to the $2^{3-1}$ fractional factorial design. All four regressions are correctly specified.}
\end{figure}

Figure \ref{fig:ff_1} shows the {violin plot of the differences between the estimated and true values} for the three main effects over 1,000 independent complete randomizations under the full and fractional factorial designs, respectively. 
Specification \texttt{m4} under the fractional factorial design is on average the most efficient, yielding visible improvement over specifications \texttt{m1}--\texttt{m3} for estimating the main effects of factors A and B. 
This illustrates the gain in efficiency by the fractional factorial design when its assumption on negligible higher-order interactions is correctly specified. 

\subsection{Scenario II: no two-way interaction between A and B}
Inherit most of the setting from Scenario I except that we now generate the potential outcomes as $Y_i(abc) = 4 + 2a + 2b + 2c  + ac + bc + 2^{-1}abc + a x_i + \ep_i$, and then center $\{Y_i(abc)\}_{i=1}^N$ for each $(abc) \in \mt$ to ensure  $(\tau_\A, \tau_\B, \tau_\C, \tau_\AB, \tau_\BC, \tau_\AC, \tau_\ABC) = (4, 4, 4, 0, 2,2,1)$. 
This ensures that only the interaction effect between factors A and B is absent in the ground truth, such that regressions \texttt{m2} and \texttt{m3} are correctly specified whereas regressions \texttt{m1} and \texttt{m4} are not. 

\begin{figure}[t]
\begin{center}
\includegraphics[scale=.5]{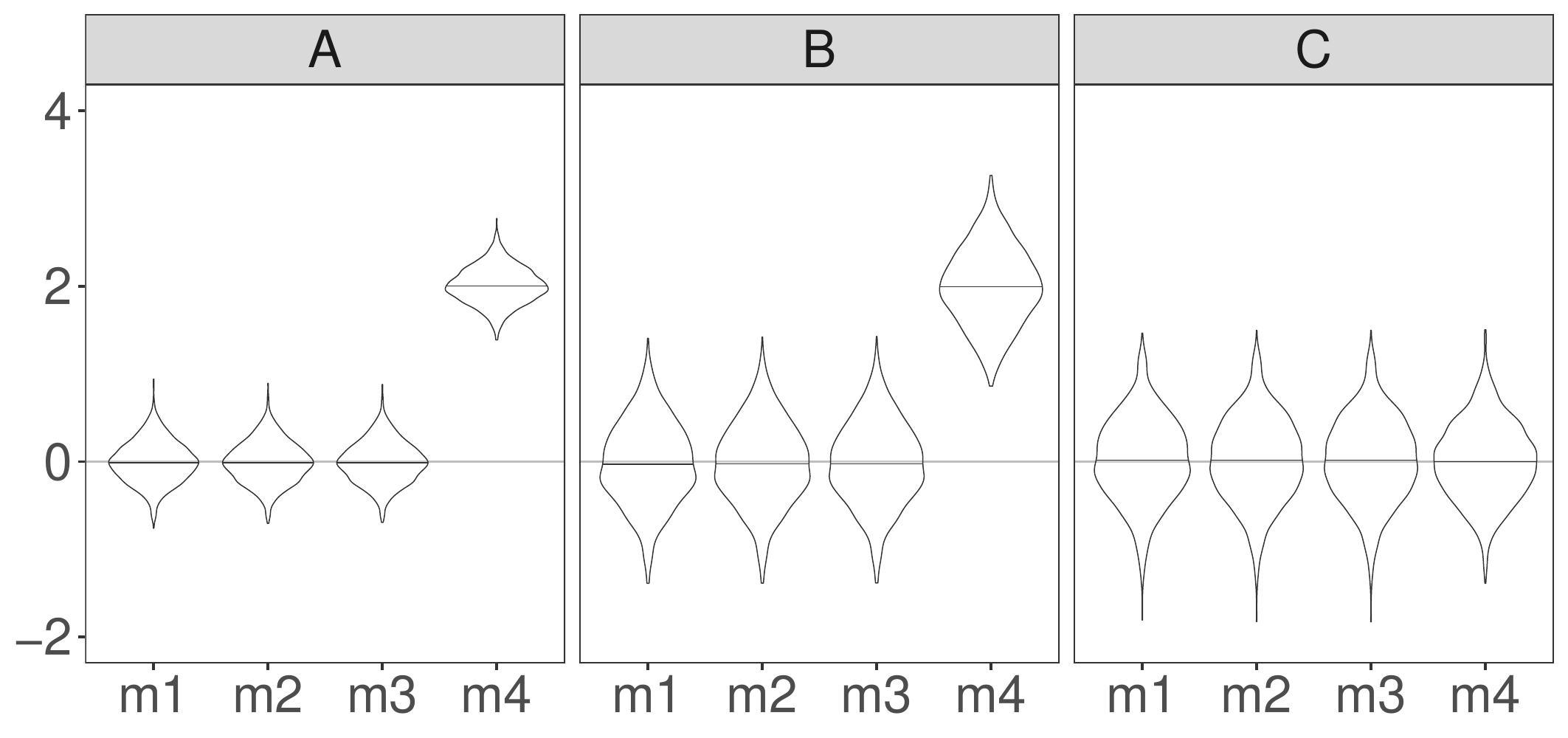}

 (a) Equal-sized full factorial design with $N_q = 10$ for all $q \in \mt$. 
\end{center}

\begin{center}
\includegraphics[scale=.5]{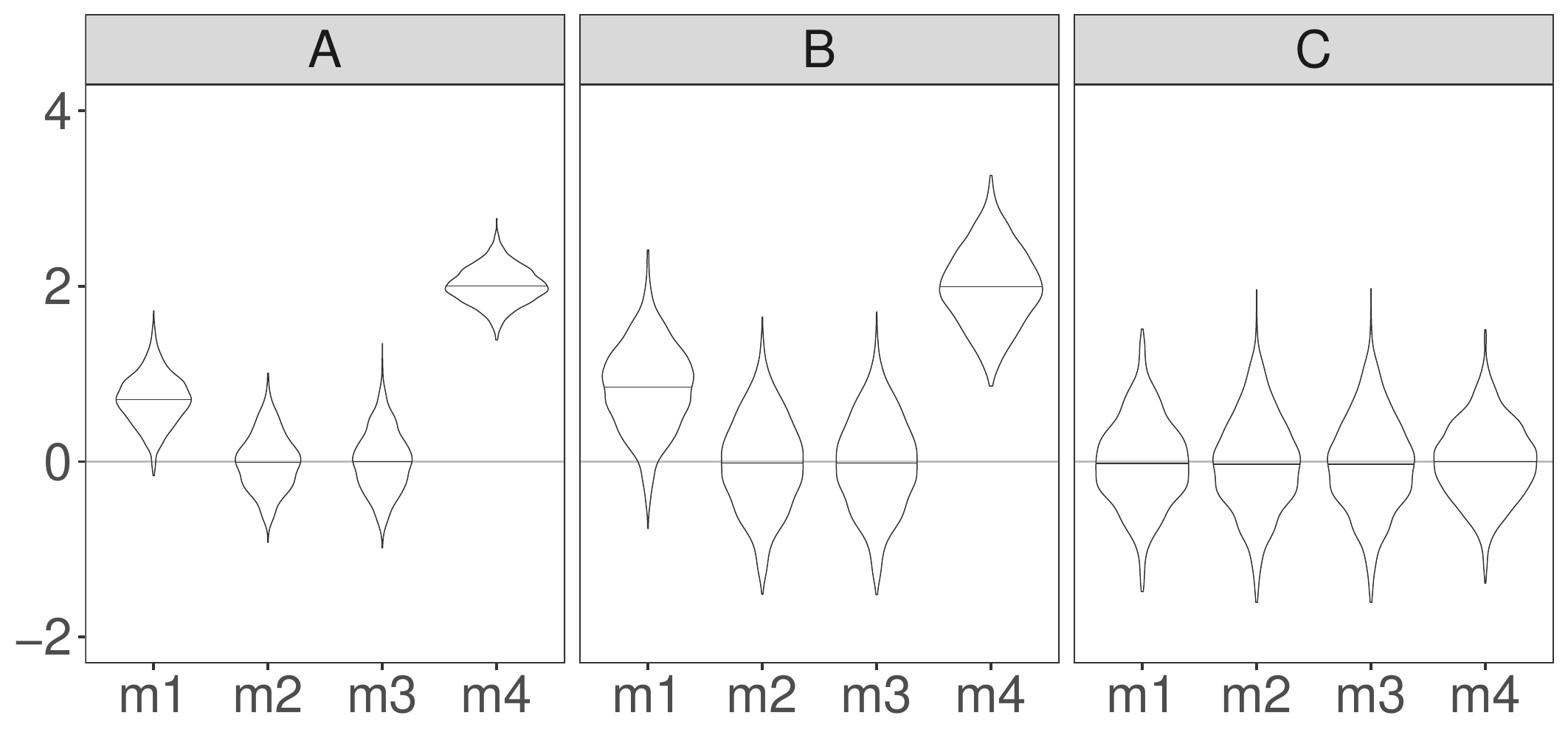}

 (b) Unequal-sized full factorial design with $(N_q)_{q\in\mt} =(5, 5, 15, 15, 5, 15, 5, 15)$.  
\end{center}\caption{\label{fig:ff_2}{Violin plot of the differences between the estimated and true values} for the three main effects over 1,000 independent complete randomizations from regressions \texttt{m1}--\texttt{m4} under Scenario II. Regressions \texttt{m1}--\texttt{m3} correspond to the $2^3$ full factorial design, and regression \texttt{m4} corresponds to the $2^{3-1}$ fractional factorial design. Regressions \texttt{m2} and \texttt{m3} are correctly specified, whereas regressions \texttt{m1} and \texttt{m4} are not. }
\end{figure}

Figure \ref{fig:ff_2}(a) shows the {violin plot of the differences between the estimated and true values} for the three main effects over 1,000 independent complete randomizations under the full and fractional factorial designs, respectively. 
Specification \texttt{m4} under the fractional factorial design is now considerably biased for estimating the main effects of factors A and B.
This is coherent with the fact that the main effects of factors A and B are aliased with the interaction between B and C and that between  A and C, respectively, and illustrates the potential problems with the fractional factorial design when its assumption is misspecified.

Specifications \texttt{m1}--\texttt{m3} under the full factorial design, on the other hand, show no visible empirical biases. 
This illustrates the advantage of the full factorial design when only a subset of the interaction effects is absent.
Importantly, specification \texttt{m1} shows no visible empirical biases  despite being misspecified. This is a direct consequence of  Proposition \ref{prop:bd}, and illustrates the robustness of factor-unsaturated specifications under full factorial designs when the treatment groups are of equal sizes.  
Figure \ref{fig:ff_2}(b) shows the corresponding results when we change the treatment group sizes under the full factorial design to $(5, 5, 15, 15, 5, 15, 5, 15)$ in lexicographical order of $(abc)\in\mt$. 
Specification \texttt{m1} now shows visible empirical biases for estimating the main effects of factors A and B. The biases are nevertheless still smaller than that of specification \texttt{m4} under the fractional design.
The full factorial design as such allows for more robust inference when the prior knowledge on negligible effects is less certain.

\section{A class of linear estimators}\label{sec:yb}
We introduce in this section a class of linear estimators of $\by$ that includes $\hys \ (\nfl)$ as special cases.
The results ensure the asymptotic efficiency of $\htl$ over a broader range,  and provide the basis for all theoretical results in the main paper.

Recall $\hx(q) = N_q^{-1}\sumiq x_i$ as the sample mean of covariates under treatment level $q \in\mt$.
For some prespecified $b_q \in \mathbb R^J$ that may depend on the data, define 
\begina 
\hat Y(q; b_q) = \hat Y(q) - \hx(q) ^\T b_q 
\enda 
as a covariate-adjusted variant of $\hy(q)$ for estimating $\by(q)$. Intuitively, it is the sample mean of the covariate-adjusted potential outcomes $Y_i(q; b_q) = Y_i(q) - \cxi ^\T b_q$.
A common choice is to take $b_q$ as the coefficient vector of $x_i$ from the \olss fit of $Y_i \sim 1 +x_i$ over $\{i: Z_i =q\}$, denoted by $\hg_q$ \citep{Lin13}.

Define
\begina
 \hyba= \big(\hat Y(1; b_1), \ldots, \hy(Q; b_Q) \big)^\T 
\enda as the corresponding covariate-adjusted estimator of $\by$, with $b = (b_1^\T, \ldots, b_Q^\T)^\T\in\bbr^{JQ}$. We focus on the set of  $b$'s that have finite probability limits under complete randomization:
\begina
\mb = \big\{b\in\bbr^{JQ}:  \text{$\plim b$ exists and is finite under complete randomization and Condition \ref{asym}}\big\}.
\enda
By definition, $\mb$ contains all constant  vectors in $\mathbb R^{JQ}$. 
The random vector $\hat\gamma = (\hat\gamma_1^\T, \ldots, \hat\gamma_Q^\T)^\T$ is also in $\mb$ with $ \hat\gamma_q = \gamma_q + \op$ for all $q \in \mt$.
Let
\beginy\label{eq:my}
\my = \big\{ \hyba: b\in\mb \big\}
\endy
summarize the corresponding covariate-adjusted linear estimators of $\by$.


Recall that $ \gamma = (\gamma_1^\T, \ldots, \gamma_Q^\T)^\T$ and $ \gp = \sumq e_q \gamma_q$.
Recall $\hgl$ as the coefficient vector of $t_i \otimes x_i$ from the \olss fit of \eqref{eq:lm_l}, with $\hgl = \hg$ by standard properties of \ols.
Further let $\hbf$  denote the coefficient vector of $x_i$ from the \olss fit of \eqref{eq:lm_f}. 
 Proposition \ref{prop:numeric_corr} below ensures that $ \hys \ (\nfl)$ are all elements in $\my$.

\begin{proposition}\label{prop:numeric_corr}
$\hy_* = \hybs $ for $\ms$ with $b_\neyman = 0_{JQ}$, $b_\fisher = 1_Q\otimes \hbf$, and $b_\lin = \hgl$.  
Further assume \creasym. Then 
$\hbf = \gp + \op$ and $\hgl = \bg + \op$ with $ \plim b_\fisher =  1_Q\otimes \gp$ and $\plim b_\lin =\gamma$. 
\end{proposition}

For $b   \in \mb$ with $\plim b = \bi = (b_{1,\infty}^\T, \ldots, b^\T_{Q, \infty})^\T$, define
\beginy\label{eq:vbi}
\vbi =  \diag( S _{b, \infty,qq} /\pq )_{q\in\mt} -  S_{b, \infty},
\endy 
where $S_{b,\infty} =  (S _{b, \infty,qq'})_{q,q'\in\mt}$ denotes the finite population covariance matrix of $\{Y_i(q; b_{q,\infty}): q\in\mt\}_{i=1}^N$.
By  Proposition \ref{prop:numeric_corr}, the $V_* \ (\nfl)$ in \eqref{eq:v_star} are all special cases of $\vbi$ with $b = b_*$. 
Lemma \ref{lem:yb} below ensures that the estimators in $\my$ are all consistent and asymptotic normal for estimating $\by$, and establishes the asymptotic efficiency of $\hyl$ over $\my$.

\begin{lemma}\label{lem:yb}
For all $b \in \mb$ with $\plim b = \bi$, we have 
\begine[(i)]
\item $\vbi \geq V_\lin$, where the equality holds if $\bi = \gamma$; 
\item $\sqrt N (\hyba-\bY)\rs\mn(0_Q, \vbi)$ under complete randomization and Condition \ref{asym}. 
\ende 
\end{lemma}

Lemma \ref{lem:yb} includes Lemmas \ref{lem:reg} and \ref{lem:v_r} as special cases, and establishes the asymptotic efficiency of $\hyl$ over all linear estimators in $\my$ for estimating $\by$. 
This ensures the asymptotic efficiency of $\htl$ over $\{C\hyba:  \hyba \in \my\}$ for estimating $\tau = C\by$, extending \citet[][Example 9]{LD20} to \mes.
In addition, $\hyba \asim \hyl$ if  $\bi = \gamma$. 

Moreover, 
let $\hyg = (\hy(1; \gamma_1), \ldots, \hy(Q; \gamma_Q))^\T$ be the linear estimator defined by $\gq =(\gamma_1^\T,\dots, \gamma_Q^\T)^\T$.
Lemma \ref{lem:yb} establishes $\hyg$ as the oracle estimator with fixed adjustment coefficients $b_q = \gamma_q \ (\qit)$. Intuitively, the properties of \olss ensure full reduction of the variability due to the covariates, and guarantee the efficiency of $\hyg $ over $ \my$.  
Recall that  $\gamma_q$ gives the target parameter of $\wiq  \, x_i$ in \eqref{eq:lm_l} from the derived linear model perspective.  The additive and fully interacted regressions can thus also be viewed as two ways to estimate the optimal adjustment coefficients $\gamma_q \ (q\in\mt)$. 
In particular, we can estimate $\gamma_q$ by the \olss coefficient of $\wiq  \, x_i$, namely $\hg_{\lin,q}$, from the fully interacted regression \eqref{eq:lm_l},  or by the \olss coefficient of $x_i$, namely $\hbf$, from the additive regression \eqref{eq:lm_f}. 
The resulting $\hyba$ equals $\hyl$ and $\hyf$, respectively,  by Proposition \ref{prop:numeric_corr}. 
Proposition \ref{prop:numeric_corr} further ensures that $\hg_{\lin,q}$ is consistent for $\gamma_q$, whereas $\hbf$ in general is not.
This gives the intuition behind the asymptotic efficiency of $\hyl$.

Theorem \ref{thm:optimal_gamma} below states a stronger result than Lemma \ref{lem:yb}, characterizing the asymptotic distance of $\hyba\in \my$ from the oracle $\hyg$.
Let 
$\db   = \diag\{(b_q - \gamma_q)^\T\}_{q\in\mt}\in\mathbb R^{Q\times JQ}$ with  
\begina
\hyba= \hyn - \{\diag(b_q^\T)_{q\in\mt}\} \hx =  \hyg - \db   \hx,  
\enda 
where $\hat x = (\hx(1)^\T, \ldots, \hx(Q)^\T)^\T$. 
Let  
$\dbi  = \plim D_b  = \diag\{(b_{q,\infty} - \gamma_q)^\T\}_{q\in\mt}$.

\begin{theorem}\label{thm:optimal_gamma} 
{\precre}
For $\hyba\in \my$ with $\plim b = \bi$, we have
\begina
\hyba \ \ \approxi  \ \ \hyg - \dbi   \hx  \ \ \preci\ \  \hyg
\enda with  $\vbi = N\covi\{\hyba\} = V_\lin + \dbi V_x \dbi^\T$. In particular, $\hyba \approxi \hyg$  if 
\begin{equation}\label{eq::condition-efficiency}
\dbi V_x \dbi^\T = 0 .
\end{equation}
A sufficient condition for \eqref{eq::condition-efficiency} is $\bi = \gamma$.  
\end{theorem}

Theorem \ref{thm:optimal_gamma} states the asymptotic equivalence of $\hyba$ and $\hyg -\dbi \hx = \hybi$, suggesting $\dbi\hx$ as the distance of $\hyba$ from the optimally adjusted $\hyg$.

\section{Additional results on restricted least squares}\label{sec:rls_app_res} 
\subsection{Additional results under the correlation-only and separable restrictions}
 
Recall the definition of $\mb$ and $\my$ from Section \ref{sec:yb}. 
Proposition \ref{prop:noy} below states the numeric equivalence between $\hyr$ and $\hybr$ under \rlss subject to the correlation-only restriction, and ensures that $\hyr$ is an element of $\my$. 
The design-based properties of $\hyr$ in Theorem \ref{thm:noy} then follow immediately from Lemma \ref{lem:yb}.

\begin{proposition}\label{prop:noy}
Assume \rlss subject to the correlation-only restriction \eqref{eq:rest_noy}. Then 
$
\hyr   =    \hybr ,
$
where $\hgr$ belongs to $\mb$ and satisfies $\plim \hgr = \gamma$  if \eqref{eq:rest_noy} is correctly specified.
\end{proposition}

Proposition \ref{prop:sep} below ensures that $\hyr$ is a linear function of $\hybr  \in \my$ under \rlss subject to the separable restriction.  This, together with Lemma \ref{lem:yb}, leads to the design-based properties of $\hyr$ in Theorem \ref{thm:sep}. 
Recall $U$ and $\mur $ in \eqref{eq:umu}.

\begin{proposition}\label{prop:sep}
Assume \rlss subject to the separable  restriction \eqref{eq:rest_sep}  with $\rhy \neq 0$. Then 
\begina
\hyr  - \by = U(\hybr  - \by) + \mur,
\enda
where $\hgr$ belongs to $\mb$ and satisfies  $\plim \hgr = \gamma$ if $\rhg\gamma = r_\gamma$ is correctly specified.   
\end{proposition}

Theorem \ref{thm:gm_a} below extends Theorem \ref{thm:gm}\eqref{item:gm} and gives the design-based counterpart of the classical Gauss--Markov theorem for \rlss  subject to the separable restriction.
The result extends \citet[][Theorem A5]{ZDa} on unadjusted estimators to the covariate-adjusted settings.

Let
\begina
\my' = \left\{ L\hyba+ a: \ \text{$\hyba \in \my$, $L \in \mathbb R^{Q\times Q}$, and $a \in \mathbb R^Q$ with $L\hyba+ a= \by + \op$} \right\}
\enda
denote a class of linear consistent estimators of $\by$, which contains all elements in  $ \my$ and hence $\hys \ (\nfl)$ as special cases.
By Lemma \ref{lem:yb}, all elements in $\my'$ are consistent and asymptotically normal  such that we can compare their asymptotic efficiency by comparing their asymptotic covariances.

\begin{theorem}\label{thm:gm_a}
Assume complete randomization, Conditions \ref{cond:sa}--\ref{asym}, and \rolss subject to the separable  restriction \eqref{eq:rest_sep} with $\rho_Y\neq 0$ being a contrast matrix with rows orthogonal to $1_Q$. 
If both $\rhy \by = r_Y$ and $\rhg\gamma = r_\gamma$ are correctly specified, then $\hyr$ is the best linear consistent estimator in $\my'$, in the sense that $\hyr$ 
is an element of $\my'$ and has the smallest asymptotic covariance over all estimators in $\my'$. 
\end{theorem}

\subsection{Properties under general restriction}
Write $R = (\ry, \rg)$, where $\ry \in \mathbb R^{m\times Q}$ and $\rg\in \mathbb R^{m\times JQ}$ denote the columns of $R$ corresponding to $\by$ and $\gamma$, respectively. 
Theorem \ref{thm:rls_g} below quantifies the asymptotic behaviors of $\hyr $ for general $R$, and suggests that $\hyr $ is in general not consistent for estimating $\by$ unless $\ry = 0$ or \eqref{eq:restriction} is correctly specified. 

Recall $\mr=  \ccinvl R^\T\{R \ccinvl R^\T\}^{-1}$ from Lemma \ref{lem:hthr}. 
Let $\xir$ be the first $Q$ rows of $- \mr\reslp $. 
Let $\mri = \plim \mr$ be the finite probability limit of $\mr$ under complete randomization and Condition \ref{asym}. 
Let $\sri$ be the upper-left $Q\times Q$ submatrix of $(I - \mri R) {\{N \covi(\hth_\lin)\}} (I - \mri R)^\T$.  We give the explicit forms of  $\mri$ and $N \covi(\hth_\lin)$ in  \eqref{eq:mri} and Lemma \ref{lem:hth}.
Recall $\emat = \diag(e_q)_{q\in\mt}$ with $e_q = N_q/N$. 
Let 
$
\dz  =[R \, \diag\{\einv, \esinv\} R^\T]^{-1}$.

\begin{theorem}\label{thm:rls_g}
Assume \rlss subject to \eqref{eq:restriction}. Then 
\begina
\hyr  = \hybr  -  \einv \ryt  \left[ R \left \{N\ccinvl \right\} R^\T \right]^{-1}(R\hthl-r).
\enda 
Further assume \creasym. Then 
\begine[(i)]
\item\label{item:rg} $\hgr = \gamma -   (\es )^{-1} \rgt \dz  (R\tl  - r)  + \op$; 
\item\label{item:yr} 
$\sqrtn(\hyr   -\by - \xir ) \rs  \mn(0_{Q}, \sri)$
with $ \xir=  - \einv  \ryt  \dz(R\tl  -r)+\op$. In particular,  $\xir = 0$, and hence   $\sqrtn(\hyr  -\by) \rs  \mn(0_{Q}, \sri)$, if \eqref{eq:restriction} is correctly specified.  
\ende
\end{theorem}

Theorem \ref{thm:rls_g}\eqref{item:rg} ensures that $\hgr$ has a finite probability limit under complete randomization and Condition \ref{asym} regardless of whether the restriction  $R\thl = r$  is correctly specified or not.
Theorem \ref{thm:rls_g}\eqref{item:yr} implies $\plim \xir = - \einv  \ryt  \dz(R\tl  -r)$ as the asymptotic bias of $\hyr$. Juxtapose Theorem \ref{thm:rls_g} with Theorems \ref{thm:noy} and \ref{thm:sep}. 
The consistency of $\hyr$ in general requires the whole restriction \eqref{eq:restriction} to be correctly specified,  whereas 
special structures like $R_Y = 0$ or $R = \diag(\rhy, \rhg)$ promise weaker sufficient conditions.
In particular, the {\go} restriction ensures consistency regardless of whether the restriction on $\gamma$ is correctly specified or not;  the {\sr} ensures consistency as long as the restriction on $\by$ is correctly specified.

%
%
%
%

\subsection{Numeric properties of restricted least squares}\label{sec:rls_app_n}

We present in this subsection some useful numeric properties about \rls. 
The results are stated in terms of general regression formulation to highlight their generality.
 
For $Y\in \mathbb R^N$, $X \in \mathbb R^{N\times p}$, $R \in \mathbb R^{m \times p}$, and $r \in \mathbb R^m$,
 consider the \rlss fit of 
\begina
Y = X \hat\beta + \hep \qquad \text{subject to} \ \ R\hat\beta = r,
\enda
where $\hb$ and $\hep = (\hep_1, \ldots, \hep_N)^\T$ denote the \rlss coefficient vector and residuals, respectively. 
To simplify the presentation, we suppress the subscript ``$\rr$'' for \rlss in this subsection when no confusion would arise.
We propose to estimate the sampling covariance of $\hb$ by
\beginy\label{eq:ehw_rls}
\hv = (I_p - \mr R) (X^\T X)^{-1} X^\T \big\{\diag(\hep^2_i)_{i=1}^N \big\} X (X^\T X)^{-1} (I_p - \mr R)^\T,
\endy
where $\mr =  (X^\T X)^{-1} R^\T\{R (X^\T X)^{-1} R^\T\}^{-1}$.
Refer to $\hv$ as the {\it double-decker-taco robust} covariance estimator from \rls.

Lemma \ref{lem:inv_rls}  below states the invariance of \rlss to {\ndt}.

\begin{lemma}\label{lem:inv_rls} 
For $Y\in \mathbb R^N$, $X  \in \mathbb R^{N\times p}$, and $X' = X\Gamma   \in \mathbb R^{N\times p}$ for some nonsingular  $\Gamma \in \mathbb R^{p\times p}$, consider the \rlss fits of 
\begina
\begin{array}{ll}
Y =  X\hbb+ \hep &\quad\text{subject to} \ \ R \hbb = r, 
 \\
Y =  X' \hbb' +\hep' &\quad \text{subject to} \ \ (R\Gamma) \hbb' = r,
\end{array}
\enda
where $(\hb, \hep)$ and $(\hb', \hep')$ denote the \rlss coefficient vectors and residuals. 
Further let $\hv$ and $\hv'$ denote the robust covariance estimators of $\hb$ and $\hb'$ by \eqref{eq:ehw_rls}, respectively.
Then
$$
\hbb = \Gamma\hbb',\qquad
\hep = \hep',\qquad
\hv = \Gamma \hv' \Gamma^\T.
$$ 
\end{lemma}

Lemma \ref{lem:inv_rls} holds for arbitrary choices of $R$ and $r$. 
Lemma \ref{lem:ehw_g} below presents a novel result on the numeric equivalence between the robust covariance \eqref{eq:ehw_rls} from the \rlss fit and that from the \olss fit of a corresponding restricted specification when $r=0$. 
The proof follows from direct linear algebra and is omitted.

\begin{lemma}\label{lem:ehw_g}
For $Y\in \mathbb R^N$, $X \in \mathbb R^{N\times p}$, and $R \in \mathbb R^{m \times p}$, consider the \rlss fit of 
\beginy\label{eq:fs}
Y = X \hb + \hep\qquad \text{subject to} \ \ R\hb = 0,
\endy
where $\hb \in \mathbb R^{p}$ and $\hep\in \mathbb R^N$ denote the \rlss coefficient vector and residuals, respectively.
Then 
\begine[(i)]
\item The corresponding restricted specification can be formed as 
\beginy\label{eq:rs}
Y = \big\{X\rpt(\rpp\rpt)^{-1}\big\}\hb_\ols + \hep_\ols,
\endy
where $ \rpp \in \mathbb R^{(p-m) \times p}$ is an orthogonal complement of $R$ in the sense that $(\rpt, R^\T)$ is nonsingular with $\rpp R^\T   = 0$.
Let $\hb_\ols\in  \mathbb{R}^{p-m}$ and $\hep_\ols \in  \mathbb{R}^N $ denote the coefficient vector and residuals from the \olss fit of \eqref{eq:rs}. 
Then 
\begina
\hb_\ols = \rpp \hb , \qquad \hep_\ols = \hep.
\enda 
\item Let $\hv_\rls = \rpp \hv \rpt$ denote the robust covariance estimator of $\rpp \hb$ from the \rlss fit of \eqref{eq:fs}, with $\hv$ given by \eqref{eq:ehw_rls}. 
Let $\hv_\ols$ denote the \ehws covariance estimator of  $\hb_\ols$ from the \olss fit of \eqref{eq:rs}. Then 
\begina
\hv_\rls = \hv_\ols.
\enda 
\ende
\end{lemma}

Lemma \ref{lem:ehw_g} includes Examples \ref{ex:ehw_equiv} and \ref{ex:ehw_equiv_2} below as special cases, which correspond to the unadjusted and additive regressions, respectively. 

\begin{example}\label{ex:ehw_equiv}
For  $ Y \in \mathbb R^{N}$,  $ X_1 \in \mathbb R^{N\times k}$, and  $ X_2\in \mathbb R^{N\times l}$, consider the \rlss fit of 
\begina
Y =  X_1 \hb_1 +  X_2 \hb_2 + \hep \qquad\text{subject to} \ \ \hb_2 = 0,
\enda
where $\hb_1$, $\hb_2$, and $\hep$ denote the \rlss coefficient vectors and residuals, respectively. 
Let 
$\hv_\rls$ denote the robust covariance estimator of $\hb_1$ by \eqref{eq:ehw_rls}.
Alternatively, consider the \olss fit of 
\begina
Y =  X_1 \hb_{1,\ols}  + \hep_\ols 
\enda
as the corresponding restricted specification, 
where $\hb_{1,\ols}$ and $\hep_\ols$ denote the \olss coefficient vector and residuals, respectively. 
Let $\hv_\ols$ denote the \ehws covariance estimator of $\hb_{1,\ols}$. 
Then 
\begina
\hb_1 = \hb_{1,\ols},\qquad \hep = \hep_\ols, \qquad
\hv_\rls = \hv_\ols.
\enda
\end{example}

\begin{example}\label{ex:ehw_equiv_2}
For  $ Y \in \mathbb R^{N}$,  $ X_0 \in \mathbb R^{N\times k}$, and  $ X_1, \ldots, X_Q \in \mathbb R^{N\times l}$,
consider the \rlss fit of 
\begina
Y =  X_0 \hb_0 +  X_1 \hb_1 +\cdots + X_{Q} \hb_Q + \hep \qquad\text{subject to} \ \ \hb_1 = \cdots = \hb_Q,
\enda
where $\hb_q \ (q =\zt{Q})$ and $\hep$ denote the \rlss coefficient vectors and residuals, respectively. 
Let 
$\hv_\rls$ denote the robust covariance estimator of $\hb_0$ by \eqref{eq:ehw_rls}. 
Alternatively, consider the \olss fit of 
\begina
Y =  X_0 \hb_{0,\ols}  + \xpp \hb_\ols + \hep_\ols 
\enda
as the corresponding restricted specification, where $\hb_{0,\ols}$, $\hb_\ols$, and $\hep_\ols$ denote the \olss coefficient vectors and residuals, respectively. Let $\hv_\ols$ denote the \ehws covariance estimator of $\hb_{0,\ols}$.  
Then 
\begina
\hb_0 = \hb_{0,\ols},\qquad \hb_q = \hb_{\ols} \quad (q=\ot{Q}), \qquad \hep = \hep_\ols, \qquad
\hv_\rls = \hv_\ols.
\enda
\end{example}

Recall that $\hsis$ and $\hssr$ denote the robust covariances of $\hys \ (\nf)$ from the \olss fits of \eqref{eq:lm_n}--\eqref{eq:lm_f} and the \rlss fits of \eqref{eq:lm_l}, respectively. The numeric equivalence between $\hsis$ and $\hssr$ follows immediately from Examples \ref{ex:ehw_equiv} and \ref{ex:ehw_equiv_2}. 

\begin{proposition}\label{prop:ehw_equiv}
$\hsis = \hssr$ for $\nf$.
\end{proposition}

\section{Additional results on $2^K$ factorial experiments}\label{sec:2k_app}
Recall that $\tmk = \cmk^\T\by$ denotes the standard factorial effect corresponding to $\mk\in\pk$ under the $2^K$ factorial experiment.
We will write $(\tmk, \cmk)$ as $(\tsmk, \csk)$ throughout this section to differentiate the standard factorial effects from variants under the \{0,1\} coding system.
Let $c_{\sss,\emptyset} =  c_{\emptyset} = 2^{-(K-1)} 1_Q$.  Let $\pk' = \{\emptyset\} \cup \pk$, and let $Z_{i, \emptyset} = 1$ for all $i$.

\subsection{Design-based theory for general regression specifications} \label{sec:2k_app_s}

For $\mfp \subseteq \pk$ and $\mfpp \subseteq \pk'$, 
consider the regression specification
\beginy\label{eq:lm_2k_g}
Y_i \sim 1 + \sum_{\mk \in \mfp} \zimks + \sum_{\mk \in \mfp'} \zimks \cdot \cxi, 
\endy 
where $Z_{i, \emptyset}\cdot x_i = x_i$ if $\mk = \emptyset \in \mfp'$. 
It is a general specification and includes \eqref{eq:lm_2k_n}--\eqref{eq:lm_l_2k_u} as special cases with different choices of $(\mfp, \mfpp)$, summarized in Table \ref{tb:specs_2k_app}. 
The resulting regression is factor-saturated if $\mfp = \pk$, and factor-unsaturated if $\mfp \subsetneq \pk$ with $  \mfm  = \pk \backslash \mfp \neq \emptyset$.

\begin{table}[t]
\renewcommand{\arraystretch}{1.3}
\caption{\label{tb:specs_2k_app}  The choices of $(\mfp, \mfp')$ in \eqref{eq:lm_2k_g} for the six factor-based regressions in Table \ref{tb:specs_2k}.  
We use $\emptyset$ to denote the empty set, and use $\{\emptyset\}$ to denote the set that contains the empty set as the only element. }
\begin{center}
 
\begin{tabular}{|c|l|c|c|}\hline
  & & &\\
base model & \multicolumn{1}{c|}{regression equation} & $\mfp$ & $\mfp'$  \\\hline
&\eqref{eq:lm_2k_n}: $Y_i \sim 1 + \sum_{\mk \in \pk} \zimks$ &   & $\emptyset$      \\\cline{2-2}\cline{4-4}
$Y_i \sim 1 + \sum_{\mk \in \pk} \zimks$ &\eqref{eq:lm_2k_f}: $Y_i \sim 1 + \sum_{\mk \in \pk} \zimks+ \cxi $  & $\pk$ & $\{\emptyset\}$      \\\cline{2-2}\cline{4-4}
(factor-saturated) &\eqref{eq:lm_2k_l}: $Y_i \sim 1 + \sum_{\mk \in \pk} \zimks + \cxi  + \sum_{\mk \in \pk} \zimks \cxi  $&   & $\pk'$  \\\hline
& \eqref{eq:lm_n_2k_u}: $Y_i \sim 1 + \sum_{\mk \in \mfp} \zimks$ &     & $\emptyset$  \\\cline{2-2}\cline{4-4}
{$Y_i \sim 1 + \sum_{\mk \in \mfp} \zimks$}&\eqref{eq:lm_f_2k_u}: $Y_i \sim 1 + \sum_{\mk \in \mfp} \zimks+\cxi $ & $\mfp$   & $\{\emptyset\}$  \\\cline{2-2}\cline{4-4}
(factor-unsaturated) & \eqref{eq:lm_l_2k_u}: $Y_i \sim 1 + \sum_{\mk \in \mfp} \zimks + \cxi  + \sum_{\mk \in \mfp} \zimks \cxi   $ &     & $\{\emptyset\} \cup \mfp$  \\\hline
\end{tabular}		

\end{center}
\end{table}

Assume that $\mfp$ is non-empty throughout this section. We state below the design-based properties of the \olss outputs from \eqref{eq:lm_2k_g}. The result includes those of   \eqref{eq:lm_2k_n}--\eqref{eq:lm_l_2k_u} as special cases. 

Recall $\tsmk = \csk^\T \by$ as the standard factorial effect corresponding to $\mk\in\pk$. 
Let
\begina
\tsp   = \{\tsmk   : \mk \in \mfp \} = \csp\by, \qquad
\tsm = \{\tsmk : \mk\in\mfm \} = \csm \by
\enda
concatenate the effects of interest and nuisance effects corresponding to $\mfp$ and $\mfm = \pk \backslash \mfp$, respectively.
To simplify the presentation, define $\csm = 0_J^\T$ if $\mfm$ is empty, with $\tsm = 0$.

Let $\mfm' = \pk' \backslash\mfp'$ analogous to $\mfm$, and let $\csm'$ concatenate rows of $\{\csmk:\mk\in \mfm'\}$ with $\csm' = 0_J^\T$ if $\mfm'$ is empty.

Recall that 
$2^{-1}\tsmk$, $2^{-1}(c^\T_{\sss,\emptyset} \otimes I_J) \gamma$, and $2^{-1}(\csk^\T\otimes I_J) \gamma$ give the target parameters of $\zimk$, $x_i$, and $\zimk \cdot x_i$ for $\mk \in \pk$ in \eqref{eq:lm_2k_l}, respectively; see the proof of Proposition \ref{prop:2k_g} in Section \ref{sec:factor_proof}.
Regression \eqref{eq:lm_2k_g} is thus a restricted variant of \eqref{eq:lm_2k_l}
subject to the separable restriction
\beginy\label{eq:rest_2k_g}
\csm \by = 0,\qquad (\csm' \otimes I_J) \gamma = 0,
\endy
and allows us to estimate $\tsmk$ by 2 times the \olss coefficient of $\zimk$ for $\mk\in\pk$. In particular, we say that there is no restriction on $\by$ if $\mfp = \pk$, with $\csm \by = 0$ in \eqref{eq:rest_2k_g} correctly specified by definition.  Likewise for there to be no restriction on $\gamma$ if $\mfp' = \pk'$, with $(\csm' \otimes I_J) \gamma = 0$ in \eqref{eq:rest_2k_g} correctly specified by definition.

Let $\ttrmk$ be 2 times the coefficient of $\zimk$ from the \olss fit of  \eqref{eq:lm_2k_g}, vectorized as $$\ttrp  = \{\ttrmk: \mk\in\mfp\}.$$ 
Let $\torp $ be the \ehws covariance estimator of $\ttrp$ from the same \olss fit. 
Let $\hyrs $ be the coefficient vector of $t_i$ from the \rolss fit of  \eqref{eq:lm_l} subject to \eqref{eq:rest_2k_g}, and let $\hat\Psi_{\rr,\sss}$ be the double-decker-taco robust covariance estimator of $\hyrs$ from Section \ref{sec:ehw_rols}.
Proposition \ref{prop:2k_g} below states the numeric correspondence between $(\ttrp, \torp )$ and $(\hyrs, \hat\Psi_{\rr,\sss})$. 

\begin{proposition}\label{prop:2k_g}
$\ttrp   =  \csp\hyrs$ and $\torp  = \csp\hat\Psi_{\rr,\sss}\csp^\T$. 
\end{proposition}

The design-based properties of $(\ttrp, \torp )$  then follow from those of $(\hyrs, \hat\Psi_{\rr,\sss})$ in  Lemma \ref{lem:yb} and Theorems \ref{thm:noy}--\ref{thm:ehw_rr}. 
Let $\hgrs$ and $\vrrs$ be the values of $\hgr$ and $\vrr$ associated with $\hyrs$. 
By Propositions \ref{prop:noy} and \ref{prop:sep},  $\hgrs$ belongs to $\mb$ regardless of whether the restriction is correctly specified or not, and satisfies $\plim \hgrs = \gamma$ if $(\csm' \otimes I_J) \gamma = 0$. 
Corollary \ref{cor:2k_g_1} below justifies the large-sample Wald-type inference of $\ts$ based on factor-saturated specifications for arbitrary choice of $\mfp'$.

 \begin{corollary}\label{cor:2k_g_1} For \eqref{eq:lm_2k_g} that is factor-saturated with $\mfp = \pk$, we have $$\ttrp   =  \cs\hybrs.$$
 Further assume \creasym.  Then 
\begine[(i)]
\item $\sqrt N(\ttrp - \ts ) \rs \mn(0, \cs \vrrs \cs^\T)$ with $\ttrp \preci \ttl$;
\item $\torp$  is asymptotically conservative for estimating the true sampling covariance of $\ttrp$;
\item $\ttrp \asim \ttl$  if 
\begin{equation}
\label{eq::condition-factorial}
(\csm' \otimes I_J) \gamma = 0.
\end{equation}
The condition \eqref{eq::condition-factorial} holds if Condition \ref{cond:equal} holds and \eqref{eq:lm_2k_g} includes $x_i$.
\ende 
\end{corollary}

Next, 
Corollary \ref{cor:2k_g_2} below justifies the large-sample Wald-type inference of $\tsp$ based on factor-unsaturated specifications when the nuisance effects excluded are indeed zero. 
From \eqref{eq:umu}, let 
\begina
\us =I_Q - \einv \csm^\T   (\csm \einv \csmt)^{-1} \csm, \qquad \murs =- \einv \csm^\T   (\csm \einv \csmt)^{-1} \tsm
\enda
%
%
be the values of $U$ and $\mur$ at $\rhy = \csm$, respectively, when $\mfm  \neq \emptyset$. 
Recall $\ttsp = \{\ttsmk: \mk\in\mfp\}$ as the estimators of $\tsp$ from the factor-saturated regressions \eqref{eq:lm_2k_n}--\eqref{eq:lm_2k_l} with $\ttlp \succi \ttnp, \ttfp$.

\begin{corollary}\label{cor:2k_g_2}
For \eqref{eq:lm_2k_g} that is factor-unsaturated with  $\mfp  \subsetneq \pk$ and $\mfm  \neq \emptyset$, we have 
$$
\ttrp  - \tsp -\csp\murs = \csp\us(\hybrs  - \by),$$
where $ \csp \murs = 0$ if $\tsm = 0$. 
Further assume \creasym. Then 
\begine[(i)]
\item $\sqrt N(\ttrp  - \tsp - \csp \murs) \rs \mathcal N(0, \csp \us  \vrrs  \us^\T \csp^\T)$; 
\item $\torp$ is asymptotically conservative for estimating the true sampling covariance of $\ttrp$; 
\item $
\ttrp  \succi \ttfp  \asim \ttlp \succi \ttnp$
if Condition \ref{cond:sa} holds and \eqref{eq:lm_2k_g} includes $x_i$. 
\ende
\end{corollary}

Recall that Condition \ref{cond:sa} of constant treatment effects implies Condition \ref{cond:equal} of equal correlations. Corollaries \ref{cor:2k_g_1}(iii) and \ref{cor:2k_g_2}(iii) together establish the asymptotic efficiency of additive regressions like $Y_i \sim 1+\sum_{\mk \in \mfp} \zimk + x_i$ under constant treatment effects.
Specifically, the resulting estimator is asymptotically as efficient as $\ttl$ if the specification is factor-saturated with $\mfp = \pk$ by Corollary \ref{cor:2k_g_1}(iii), and ensures additional efficiency over $\ttlp$ if the specification is factor-unsaturated with $\mfp \subsetneq \pk$ by Corollary \ref{cor:2k_g_2}(iii). 
 This illustrates the value of factor-unsaturated regressions in combination with covariate adjustment for improving efficiency.

Without constant treatment effects, a key limitation of the factor-unsaturated regressions with $\mfp \subsetneq \pk$ is that the consistency of  $\ttrp$ depends critically on the actual absence of the nuisance effects. 
Proposition \ref{prop:bd_g} below generalizes 
Proposition \ref{prop:bd} and ensures the consistency of  $\ttrp$ under equal-sized designs even when $\tsm \neq 0$. 
The asymptotic efficiency of $\ttrp$, on the other hand, still requires the restriction on $\gamma$ in \eqref{eq:rest_2k_g} to be correctly specified, and can no longer exceed that of $\ttlp$.

\begin{proposition}\label{prop:bd_g}
Assume Condition \ref{condition::equal-group-size}. 
Then 
\beginy\label{eq:bd}
\ttrp     = \csp \hybrs.
\endy
Further assume \creasym. Then 
\begine[(i)]
\item $\sqrt N(\ttrp - \tsp ) \rs \mn(0, \csp \vrrs \csp^\T)$ with  $\ttrp \preci \ttlp$; 
\item $\ttrp \asim \ttlp$ if $(\csm' \otimes I_J) \gamma = 0$. 
\ende
\end{proposition}

Recall the definition of $\hysr \ (\nfl)$ from Proposition \ref{prop:2k}.  
Let $\hbsr$ be the value of $\hgr$ associated with $\hysr$, with  $\hbnr = 0_{JQ}$. 
Corollary \ref{cor:bd_a} below is an immediate consequence of \eqref{eq:bd} and underlies the asymptotic results in Proposition \ref{prop:bd}.

\begin{corollary}\label{cor:bd_a}
Assume Condition \ref{condition::equal-group-size}. 
Then $
\ttsrp     = \csp \hy\la\hbsr\ra$ for $\nfl$ with $\ttnrp =  \csp \hyn = \ttnp$.
\end{corollary}

Condition \ref{condition::equal-group-size} depends on the design. 
\cite{ZDa} showed that weighted least squares can secure the same benefit as equal-sized designs, and ensures the consistency of $\ttnrp$ regardless of whether $\tsm$ equals zero or not  in covariate-free settings. 
The same result extends to the covariate-adjusted variant $\ttrp$ with minimal modification. We omit the details.

\subsection{Factorial effects under $\{0,1\}$-coded regressions}\label{sec:def}

Define 
$\zikz = 2^{-1}(\zik+1)$ as the counterpart of $\zik$ under the $\{0,1\}$ coding system. 

Replacing $\zimk = \prod_{\kik}\zik$ with $\zimkz = \prod_{\kik} \zikz$ in \eqref{eq:lm_2k_l} yields
\beginy\label{eq:lm_l_2k_0}
Y_i  \ \sim \ 1 + \sum_{\mk \in \pk} \zimkz + x_i  + \sum_{\mk \in \pk} \zimkz \cdot \cxi  
\endy 
 as the fully interacted specification under the $\{0,1\}$ coding system. Let
\begina
\Gamma_0 =  \otimes_{k=1}^K \beginp 1&0\\ -1 & 1\endp   = \beginp c_0^\T \\ C_0\endp 
\enda with $c_0 = (1, 0_{Q-1}^\T)^\T$ and $C_0 1_Q = 0_{Q-1}$. 
Let $[-]$ denote the treatment level $q = (-1, \ldots, -1) \in \mt$ that has $-1$ in all dimensions. 
The comments after Proposition \ref{prop:2k_g} extend here, and ensure that
 $\tau_0 = C_0\by$, $\gamma_{[-]}$, and $(C_0
\otimes I_J)\gamma$ give the target parameters of $(\zimkz)_{\mk\in\pk}$, $x_i$, and $(\zimkz \cdot x_i)_{\mk\in\pk}$ in \eqref{eq:lm_l_2k_0}, respectively; see the proof of Proposition \ref{prop:2k_0} in Section \ref{sec:factor_proof}.  
The elements of $\tau_0 = C_0\by$ define the analogs of the $2^K-1$ standard factorial effects under a different weighting scheme; see Remark \ref{rmk:baseline weighting} at the end of this section. 
This is also known as {\it baseline parameterization} in the experimental design literature; see, e.g., \cite{rahul08} and \cite{rahul12}.

Let $\czmk^\T$ be the row in $\cz$ that corresponds to $\zimkz$. 
Then $\tzmk  = \czmk^\T  \by$ and $   (\czmk^\T  \otimes I_J) \gamma $ give the target parameters of $\zimkz$ and $\zimkz \cdot \cxi$, respectively, for $\mk\in\pk$. 
Consider 
\beginy\label{eq:lm_2k_g_0}
Y_i \ \sim \ 1 + \sum_{\mk \in \mfp} \zimkz + \sum_{\mk \in \mfp'} \zimkz \cdot \cxi 
\endy 
as the $\{0,1\}$-coded analog of the general specification \eqref{eq:lm_2k_g}. 
Let 
\begina
\tzp   = \{\tzmk   : \mk \in \mfp \} = \czp\by, \qquad 
\tzm = \{\tzmk : \mk\in\mfm \} = \czm \by
\enda
 vectorize the effects of interest and nuisance effects corresponding to $\mfp$ and $\tz$, respectively, analogous to $\tsp$ and $\tsm$. 
Let $\czm'$ concatenate rows of $\{\czmk:\mk\in \mfm'\}$, with $c_{0,\emptyset} = c_0$ if $\mk = \emptyset \in \mfm'$.
Following the convention in Section \ref{sec:2k_app_s}, define $\czm = 0_J^\T$ when $\mfm$ is empty, and $\czm' = 0_J^\T$ when $\mfm'$ is empty, respectively.

Let $\ttrp^0$ be the coefficient vector of $\{\zimkz:\mk\in\mfp\}$ from the \olss fit of  \eqref{eq:lm_2k_g_0}, with $\torp^0$ as the associated \ehws covariance estimator.  
Let $\ttsp^0$ be the corresponding estimators of $\tzp$ from the $\{0,1\}$-coded analogs of the factor-saturated regressions \eqref{eq:lm_2k_n}--\eqref{eq:lm_2k_l} for $\nfl$, analogous to $\ttsp$. 
Let $(\hyrz, \hgrz)$ be the coefficient vectors of $(t_i, t_i \otimes x_i)$, respectively, from the \rolss fit of  \eqref{eq:lm_l} subject to the separable restriction  
$$
\czm \by = 0, \qquad (\czm' \otimes I_J) \otimes \gamma = 0.
$$ 
Let $V_{\rr,0}$, $U_0$, and $\mu_{\rr,0}$ be the corresponding values of $\vrr$, $U$, and $\mur$, respectively, paralleling $\vrrs$, $\us$, and $\murs$ from Section \ref{sec:2k_app_s}.

\begin{proposition}\label{prop:2k_0}
Proposition \ref{prop:2k_g} and Corollaries \ref{cor:2k_g_1}--\ref{cor:2k_g_2} hold for inference of $\tzp$ based on \eqref{eq:lm_2k_g_0} if we change (i) all $(\ttrp, \torp, \ttsp)$ to $(\ttrp^0, \torp^0, \ttsp^0)$ for $\nfl$, and (ii) all subscripts ``$\sss$'' to ``$0$''. 
\end{proposition}

A key distinction between the two coding systems is that the result in Proposition \ref{prop:bd_g} under equal-sized designs no longer holds here, owing to the loss of orthogonality between the contrast vectors that define $\tau_0$.
The consistency of $\ttrp^0$ thus in general requires the actual absence of the nuisance effects even under equal-sized designs. 

\begin{remark}\label{rmk:baseline weighting}
We can show that $\tzmk$ equals the effect of the factors in $\mk\in \pk$ when the rest of the factors are fixed at $-1$ \citep{ZDa}. 
Denote by $-_k$ and $+_k$ the $-1$ and $+1$ levels of factor $k$, respectively, when multiple factors are concerned. 
Let $(z_k, [-])$ denote the treatment combination with factor $k$ at level $z_k \in \{-_k, +_k\}$ and the rest of the factors all at $-1$.
Then 
$$\tau_{0, \{k\}} =  \by(+_k, [-])- \by(-_k, [-])$$
for $ \mk = \{k\}$, measuring the main effect of factor $k$ when the rest of the factors are fixed at $-1$. 
Let $(z_k, z_{k'}, [-])$  denote the treatment combination with factors $k$ and $k' \ (\neq k)$ at levels $z_k\in \{-_k, +_k\}$ and $z_{k'}\in \{-_{k'}, +_{k'}\}$, respectively,  and the rest of the factors all at $-1$. 
Then 
$$
\tau_{0, \{k,k'\}} = \by(-_k, -_{k'}, [-]) + \by(+_k, +_{k'}, [-]) -\by(-_k, +_{k'}, [-]) - \by(+_k, -_{k'}, [-])$$
for $\mk = \{k, k'\}$, measuring the interaction effect of factors $k$ and $k'$ when the rest of the factors are fixed at $-1$.
The intuition extends to general $\mk \in \pk$ and elucidates the causal interpretation of $\tzmk$.
\end{remark}

\section{Rerandomization using the Mahalanobis distance}\label{sec:rem}
\subsection{Overview}

Rerandomization discards treatment allocations that do not satisfy a prespecified covariate balance criterion in the design stage of experiments \citep{cox:1982, morgan2012rerandomization},  enforcing covariate balance for additional efficiency.
A special type of rerandomization, ReM, uses the \mhld between  covariate means by treatment group
 as the balance criterion under the treatment-control experiment, and accepts a randomization if and only if the distance does not exceed some prespecified threshold  \citep{morgan2012rerandomization, LD2018, LD20, ZDfrt}. 
\citet{branson} and \citet{AOS} extended the discussion to $2^K$ factorial experiments, yet did not consider regression adjustment in the analysis stage.

We provide in this section a unified theory for ReM and regression adjustment in \mes. 
Specifically, we quantify the impact of ReM on the asymptotic efficiency of $\htl$ and $\htr$, respectively, with $\hts \ (\nf)$ being special cases of $\htr$. The results are coherent with the existing theory under the treatment-control experiment \citep{LD20}:
ReM has no effect on $\htl$ asymptotically, yet improves the asymptotic efficiency of $\htlr$ when the restriction on the correlations between potential outcomes and covariates is misspecified and separate from that on the average potential outcomes. 
The resulting estimator, though still not as efficient as $\htl$ asymptotically, can have better finite sample performance when the sample size is moderate relative to the number of covariates or treatments.
This illustrates the duality between ReM and regression adjustment for improving efficiency under \mes, and 
further expands the theoretical guarantees of $\htr$. 
The combination of ReM and $\htr$, in addition to delivering all guarantees as under complete randomization, further reduces the loss in asymptotic efficiency when the restriction is misspecified. 

Let $$\|\delta\|_\mm = \delta^\T\{\cov(\delta)\}^{-1} \delta$$ denote the \mhl distance of a random vector $\delta$ from the origin.

\subsection{ReM under \mes}
Recall $\hx(q) = N_q^{-1}\sumiq x_i$ as the sample mean of covariates under treatment level $q \in\mt$.
ReM under the treatment-control experiment measures covariate balance by the Mahalanobis distance between $\hx(0)$ and $\hx(1)$, which is equivalent to the  \mhld of $\htx = \hx(1) - \hx(0)$ from the origin.
The presence of multiple treatment arms permits more flexible measures of  covariate balance.

Specifically, denote by $\hx = (\hx^\T(1) , \dots, \hx^\T(Q) )^\T \in \mathbb R^{JQ}$ the vectorization of $\hx(q)$'s  over $q \in\mt$. 
For a prespecified $Q\times 1$ contrast vector $g_1 = (g_{11}, \dots, g_{1Q})^\T$ with $\sumq g_{1q} = 0$, the contrast of $\hx(q)$'s, denoted by 
\begina
\hd_{g_1} = \sumq  g_{1q} \hx  (q) = ( g_1^\T \otimes \mI_J) \hx \in \mathbb R^J, 
\enda defines a measure of covariate balance across the $Q$ treatment arms. 
Intuitively, a well-balanced allocation would have $\hd_{g_1}$ that is close to $0_J$. The $\htx$ under the  treatment-control experiment
is a special case with $Q = 2$ and $g = (-1, 1)^\T$. 

Assume that $H \geq 1$ of such contrasts are of interest, denoted by $\hd_{g_h} = (g_h^\T \otimes I_J)\hx\in \mathbb R^J$ for $h = \ot{H}$,  with $(g_h)_{h=1}^H$ being $H$ prespecified, linearly independent contrast vectors. 
We can vectorize them as \begina
\hd=(  \hd_{g_1}^\T, \dots, \hd_{g_H}^\T )^\T  
= ( G\otimes  I_J)\hx \in \mathbb R^{JH} 
\enda 
with $G =  ( g_1,  \dots,   g_H)^\T$, and 
conduct ReM based on the \mhl distance between $\hd$ and $0_{JH}$.
We formalize the intuition in Definition \ref{def::rem-factorial} below.

\begin{definition}
[ReM]\label{def::rem-factorial} 
Draw an initial treatment allocation by the complete randomization in Definition \ref{def::complete-randomization}, and accept it  if and only if the resulting $\hd$ satisfies $\|\hd\|_\mm     \leq a$ for some prespecified threshold $a$. 
\end{definition}

Complete randomization is a special case with $a = \infty$. 
The linear independence of $g_h$'s limits the maximum number of contrast vectors  to $H \leq Q-1$.
In the treatment-control experiment, this implies that there can only be one contrast vector, proportional to $g=(-1,1)^\T$. The resulting $\hd$ is proportional to $\htx$, illustrating the uniqueness of balance criterion when $Q=2$. 
In the $2^K$ factorial experiment, \citet{branson} and \citet{AOS} considered $G = \cs$, with $H = Q-1$ contrasts as those that define the $Q-1$ standard factorial effects. 
Experimenters in practice may choose $G$ that differs from $\cs$, with $H \ll Q$ when $Q$ is large.
An intuitive choice is  $G = (c_{\{1\}}, \ldots, c_{\{K\}})^\T$, corresponding to the contrast vectors that define the $K$ main effects.  
This suggests the need for theory of ReM with general $G$.

\subsection{Asymptotic properties}

Recall the definition of $\hyba$ and $\my$ from \eqref{eq:my}, with $\hyba\asim \hyl$ if $\plim b = \gamma$ by Lemma \ref{lem:yb}. 
Proposition \ref{prop:rem_g_supp} and Theorem \ref{thm:rem_g} below clarify the utility of ReM for improving the asymptotic efficiency of suboptimally adjusted $\hyba$ with $\plim b \neq \gamma$.
Recall that 
\begine[(i)]
\item $\hys = \hybs \in\my$ for $\nfl$ from Proposition \ref{prop:numeric_corr};
\item $\hyr  = \hybr  \in\my$ under \rlss subject to the correlation-only restriction \eqref{eq:rest_noy} by Proposition  \ref{prop:noy}; 
\item  $\hyr  - \by = U (\hybr  - \by)$ with $\hybr  \in\my$ under \rlss subject to the separable restriction \eqref{eq:rest_sep} when $\rhy \neq 0$ and $\rhy\by =r_Y$ is  correctly specified by Proposition \ref{prop:sep}.
\ende
The results on $\hyba$ hence imply the impact of ReM on $\hts  \ ( \nfl)$ and $\htr$ as direct consequences. 

To begin with, 
Proposition \ref{prop:rem_g_supp} below states the asymptotic distribution of $\hyba$ under ReM.

Denote by $(\hyba\mid \ma)$, where $\ma = \{\remd\}$, the sampling distribution of $\hyba$ under {\rem}. 
Let  
\beginy\label{eq:vb}
\vpbi = \dbi  (\Phi\otimes \sxx)\dbi  ^\T, \qquad \vrbi= \vbi  - \vpbi
\endy with  $\Phi  =  \einv   G^\T ( G \einv    G^\T)^{-1}  G \einv$.
Following \cite{AOS}, let $(\vpbi)^{1/2}_{JH}$ denote a $Q\times JH$ matrix that satisfies $(\vpbi)^{1/2}_{JH} \{(\vpbi)^{1/2}_{JH}\}^\T = \vpbi$. 
Let $\epsilon \sim \mn(0_Q, I_Q)$ and $\ml \sim \ep' \mid  (\| \ep'\|_2^2 \leq a)$ be two independent standard and truncated normal random vectors with  $\ep' \sim \mn(0_{JH}, I_{JH})$. 
Then $\ml \succeq \ep'$ with mean  $0_{JH}$ and covariance $\nu_{JH, a}  I_{JH}< I_{JH}$.  

\begin{proposition}\label{prop:rem_g_supp}
{\prerem} For $\hyba\in \my$ with $\plim b = \bi$, we have
\begine[(i)]
\item 
$
\big\{ \sqrtn (\hyba-\bar Y) \mid \ma\big\} \  \rs \   (\vrbi)^{1/2} \cdot  \epsilon +  (\vpbi)^{1/2}_{JH} \cdot   \mL$, where 
the limiting distribution   satisfies
\beginy\label{eq:pk}
 \mn (0_Q,  V_\lin) \ \ \succeq \ \  (\vrbi)^{1/2} \cdot  \epsilon +  (\vpbi)^{1/2}_{JH} \cdot   \mL \ \  \succeq \ \ \mn (0_Q, \vbi). 
\endy
\item  
$
\{ \sqrtn (\hyba-\bar Y) \mid \ma\} \ \rs  \ \mn(0_Q, \vl)
$ if and only if 
\begin{equation}
\label{eq::condition-rem}
\dbi  (\Phi\otimes \sxx)\dbi  ^\T = 0 .
\end{equation}
A sufficient condition for \eqref{eq::condition-rem} is $\bi = \gamma$.
\ende 
\end{proposition}

Recall that $\sqrtn(\hyl - \by) \rs \mn(0_Q, V_\lin)$ and $\sqrtn(\hyba- \by) \rs \mn(0_Q, V_{b,\infty}) $ under complete randomization by Lemma \ref{lem:yb}. The relationship in  \eqref{eq:pk} provides the basis for comparing the asymptotic relative efficiency between $(\hyba\mid \ma)$, $\hyba$, and $\hyl$, which we formalized in Theorem \ref{thm:rem_g} below. 
The choice of $(\vpbi)^{1/2}_{JH}$ is not unique, but the asymptotic distribution is \citep{AOS}.

In addition, recall that $\hyba\approxi  \hyg - \dbi  \hx $ from Theorem \ref{thm:optimal_gamma}.
Lemma \ref{lem:yg} in Section \ref{sec:yb_app_lemma} further ensures that $\hyg$ is asymptotically independent of $\hx$. 
ReM based on $\hd = (G\otimes I_J)\hx$ thus affects only the $\dbi\hx$ part asymptotically, and increases its peakedness if $\bi \neq \gamma$. 
This gives the intuition for the asymptotic distribution of $\hyba$ being the convolution of a normal and a truncated normal when $\dbi \neq 0$, and ensures that the asymptotic sampling distribution of $\hyl = \hy\la b_\lin\ra$ remains unchanged under  ReM given $\plim b_\lin = \gamma$.

A key distinction from the treatment-control experiment is that choosing a small $a$ alone no longer suffices to ensure that $\hyba$ is asymptotically almost as efficient as $\hyl$ when $\bi \neq \gamma$ \citep{LD2018}.
In particular, a small $a$ ensures $N\covi\{\hyba\} \approx \vrbi$, with 
 $$
 \vrbi  - \Vl = \dbi \{(\einv -\po - \Phi)\otimes \sxx\} \dbi^\T \geq 0.
 $$
To have $\vrbi = \Vl$ thus entails additional conditions. 
Without further assumptions on  $\bi$,  this requires $\einv -\po - \Phi =0$, with a sufficient condition given by $H=Q-1$; see Lemma \ref{lem:gimmick} in Section \ref{sec:yb_app_lemma}. 
ReM under the treatment-control experiment is a special case with $H = 1 = Q-1$.

Theorem \ref{thm:rem_g} builds on Proposition \ref{prop:rem_g_supp} and quantifies the asymptotic relative efficiency of $\hyba$ under complete randomization and ReM.

\begin{theorem}\label{thm:rem_g}
{\prerem}
For $\hyba\in \my$ with $\plim b = \bi$, we have 
\begina
 \hyl \ \asim \ (\hyl \mid \ma) \ \succi \ (\hyba\mid \ma)  \ \succi\  \hyba
\enda
with $(\hyba\mid \ma) \asim \hyba\asim \hyl$ if $\bi = \gamma$.
\end{theorem}

Recall that $\plim \hgr = \gamma$ when the restriction on $\gamma$ is correctly specified and separate from that on $\by$. 
ReM thus has no effect on the asymptotic distribution of $\hyl$, or $\hyr$ when the restriction on $\gamma$ is correctly specified, but improves the asymptotic efficiency of $\hyr$ if $\rhg\gamma = r_\gamma$ is misspecified under the separable restriction.
Inference based on $\hyl$ under ReM can therefore use the same normal approximation as under complete randomization; likewise for that based on $\hyr$ when the restriction on $\gamma$ is correctly specified.
The same normal approximation, however, will be overconservative when $\rhg\gamma = r_\gamma$ is misspecified, suggesting the need of more  accurate ReM-specific inference. 
We relegate the details to Section \ref{sec:rem_plugin}.

In addition to the efficiency boost for the suboptimally adjusted $\hyba$'s, ReM also improves the coherence between estimators based on different regression adjustments. 
Let $\ei(\cdot\mid \ma)$ denote the asymptotic expectation under ReM.
\begin{corollary}\label{cor:close}
{\prerem} Then 
\begina
 \ei \big( \|\hyba - \hy\la b' \ra \|_2^2  \mid  \ma \big ) 
\ = \  \nu_{JH,a}   \ei\big( \|\hyba - \hy\la b' \ra \|_2^2  \big) 
\    \leq \     \ei \big( \|\hyba - \hy\la b' \ra \|_2^2\big)
\enda
  with $ \nu_{JH, a} =  P(\chi^2_{JH+2}< a)/P(\chi^2_{JH}< a) < 1$ for all $b \neq b' \in \mb$.  
\end{corollary}

Recall that $\hys \ (\nfl)$ are all special cases of $\hyba$ by Proposition \ref{prop:numeric_corr}.
Corollary \ref{cor:close} ensures that the discrepancy between the unadjusted and adjusted estimators is smaller under ReM than under complete randomization. This is a desirable property in empirical research.

\subsection{ReM-specific inference}\label{sec:rem_plugin}
Recall that ReM increases the peakedness of $\hyr$ when the restriction on $\gamma$ is misspecified. 
The usual normal approximation will therefore be overconservative. 
We consider below more  accurate large-sample inference based on ReM-specific sampling distributions. 

Recall from Proposition \ref{prop:rem_g_supp} the asymptotic distribution of $\hyba$ under ReM.  
With $\ep$ and $\ml$ both following known distributions, the only parts that are unknown are $\vrbi$ and $\vpbi$. 
We can estimate them using their respective sample analogs, denoted by $\hvrbi$ and $\hvpbi$, respectively, and conduct ReM-specific inference based on the distribution of 
\beginy\label{eq:plugin}
(\hvrbi)^{1/2} \cdot  \epsilon +  (\hvpbi)^{1/2}_{JH} \cdot   \mL.
\endy
Proposition \ref{prop:inference} below follows from \citet[][Lemma A10]{LD2018} and ensures the asymptotic validity of the inference based on \eqref{eq:plugin}. 

Recall that $\hg_q$ denotes the coefficient vector of $x_i$ from the \olss fit of  $Y_i \sim 1+ x_i$ over $\{i:Z_i = q\}$ and gives the sample analog of $\gamma_q$.
Let 
\begina
\hat S_{b, \infty, qq}  = (N_q-1)^{-1}\sumiq \left[Y_i - b_{q}^\T x_i - \{\hy(q) - b_{q}^\T \hx(q)\} \right]^2, \quad \hat V_{b,\infty} = \diag(\hat S_{b, \infty,qq}/e_q)_\qit 
\enda
be the sample analogs of $S_{b,\infty,qq}$ and $\vbi$, respectively. 
Then 
\begina
\hvpbi = \hat D_{b,\infty}(\Phi\otimes \sxx)  \hat D^\T_{b,\infty}, \qquad \hvrbi = \hat V_{b,\infty} - \hvpbi,
\enda 
where $  \hat D_{b,\infty} = \diag\{(b_q - {\hg_q} 
)^\T\}_{\qit}$.

\begin{proposition}\label{prop:inference}
{\prerem}
Then 
\begina
\hvpbi = \vpbi + \op, \qquad \hvrbi  = \vrbi + S_{b,\infty} + \op. 
\enda 
\end{proposition}

For $\hys = \hybs  \ (\nf)$ as direct outputs from \ols, 
we can also estimate $\vbi = V_*$ by $N\hat\Psi_*$, which is asymptotically equivalent to the $\hat V_{b,\infty}$ defined above.


\section{Proof of the results on  $\hyba\in \my$}\label{sec:yb_app}
Assume throughout this section that $b = (b_1^\T, \dots, b_Q^\T)^\T \in\mb$,  with $\plim b = \bi = (b^\T_{1,\infty}, \ldots, b^\T_{Q, \infty})^\T$ denoting its probability limit under complete randomization and Condition \ref{asym}.
We verify below the sampling properties of $\hyba$ under {\cre} and {\rem}, respectively. 
Recall that $\db   = \diag\{(b_q - \gamma_q)^\T\}_{q\in\mt}$ and $\dbi = \plim \db= \diag\{(b_{q,\infty} - \gamma_q)^\T\}_{q\in\mt}$. 
Recall $\Pi = \diag(e_q)_{q\in\mt}$.  
Let $\circ$ denote the Hadamard product of matrices.
Then $
\vbi =  \core \circ S_{b,\infty}$ for $b\in\mb$, with $V_* =  \diag( S _{*,qq} /\pq )_{q\in\mt} -  S_* = \core \circ S_*$ as special cases for $\nfl$.

\subsection{Lemmas}\label{sec:yb_app_lemma}
\begin{lemma}\citep[][Theorems 3 and 5]{DingCLT}\label{lem:Ding17}
Assume the {\cre} in Definition \ref{def::complete-randomization}. 
Let $Y_i(q)$ be the $L\times 1$ potential outcome vector of unit $i$ under treatment $q$.
Let $\bar Y(q) = \meani Y_i(q)$  and $S_{qq'} = (N-1)^{-1} \sum_{i=1}^N \{Y_i(q) - \bar Y(q)\}\{Y_i(q') - \bar Y(q')\}^\T$ be the finite population mean and covariance for $q, q'\in \mt$, respectively.
Let $\bt =  \sumq   \Gamma_q \bar Y(q)$, where $\Gamma_q$ is an  arbitrary $K\times L$ coefficient matrix for $q\in \mt$. Then 
\begine[(i)]  
\item $\hbt =  \sumq   \Gamma_q \hY(q)$ has mean $\bt$ and covariance 
$\cov(\hbt) = \sumq N_q^{-1}  \Gamma_q  S_{qq}  \Gamma_q^\T - N^{-1}  S_{\bt}^2$, 
where $S_{\bt}^2$ is the finite population covariance of $\{\bt_i =  \sumq   \Gamma_q Y_i(q): i=\ot{N}\}$; 
\item 
if for all $q, q' \in \mt$, $S_{qq'}$ has a finite limit, $N_q/N$ has a limit in $(0,1)$, and $\max_{i = \ot{N}} \|Y_i(q)-\bar Y(q)\|^2_2 /N =o(1)$, then 
$N\cov(\hbt)$ has a limiting value, denoted by $ V$, and 
$$\sqrtn(\hbt  - \bt) \rightsquigarrow \mathcal{N}( 0,  V).$$
\ende
\end{lemma}

\begin{lemma}\label{lem:cov_xy}
Assume \cre.  
Then  
\begin{eqnarray*}
N\cov(\hyn) &=&    \core \circ  S, \\
V_x = N\cov(\hx) &=&  \core  \otimes  \sxx,\\
N\cov(\hyn, \hx) 
&=& 
 \left(
 \begin{array}{cccc}
 (e_1 ^{-1}-1)\bg_1^\T & -\bg_1^\T & \dots & - \bg_1^\T \\
 - \bg_2^\T & (e_2 ^{-1}-1)\bg_2^\T & \dots & - \bg_2^\T\\
 \vdots & \vdots &&\vdots\\
 - \bg_Q^\T & -\bg_Q^\T & \dots &  (e_Q ^{-1}-1)\bg_Q^\T
 \end{array}
 \right) \otimes \sxx  .
\end{eqnarray*}
Further assume Condition \ref{asym}. Then 
$\sqrtn( (\hyn - \by)^\T, \hx^\T)^\T \rs \mn(0, V)$, where $V$ is the finite limit of $ N\cov( (\hyn^\T, \hx^\T)^\T)$. 
\end{lemma}

\begin{proof}[Proof of Lemma \ref{lem:cov_xy}]
See covariates as potential outcomes unaffected by the treatment. Define 
$$
\begin{pmatrix}
Y_i(q) \\
x_i
\end{pmatrix}  \qquad (i=1, \ldots, N)
$$
as the pseudo potential outcome vectors under treatment $q$. The result follows from Lemma \ref{lem:Ding17} with 
\begina
 \Gamma_q = \left( \begin{array}{c c}
 a_{\cdot q}  &    \\ 
  &  a_{\cdot q}\otimes  I_J 
 \end{array} \right),\qquad   \text{where $ a_{\cdot q}$  denotes the $q$th column of $\mI_Q$}.
\enda
We omit the algebraic details. 
\end{proof}

\begin{lemma}\label{lem:yg}
Assume {\cre}. Then 
\begina
E(\hyg) = \by, \qquad \cov(\hyg) =  N^{-1}V_\lin, \qquad \cov(\hyg ,\hx)= 0.
\enda
Further assume Condition \ref{asym}. Then 
\begina
\sqrtn \beginp
\hyg - \by\\
\hx
\endp \rs \mn\left\{ 0 , \beginp V_\lin & 0   \\ 0   & V_x \endp \right\},
\enda
where $V_x = N\cov(\hx) =  \core  \otimes  \sxx$. 
\end{lemma}

\begin{proof}[Proof of Lemma \ref{lem:yg}]
Recall that $\hyg = \hyn - \{\diag(\gamma_q^\T)_{q\in\mt}\} \hx$. 
The result follows from Lemma \ref{lem:cov_xy} with  $\cov(\hyg , \hx) = \cov(\hyn, \hx) - \{\diag(\gamma_q^\T)_{q\in\mt}\} \cov(\hx) = 0$. 
\end{proof}

\begin{lemma}\label{lem:gimmick}
Recall  from \eqref{eq:vb} that  $ \coremat   = \emat ^{-1}  G^\T ( G\emat ^{-1}   G^\T)^{-1}   G \emat ^{-1}$, where $G$ is an $H \times Q$ contrast matrix that has full row rank. 
We have 
$\coremat = \emat ^{-1}  - \po$ if $H = Q-1$. 
\end{lemma}

The proof of Lemma \ref{lem:gimmick} follows from direct linear algebra and is omitted.

%
%

\subsection{Results under complete randomization}

\begin{proof}[Proof of Lemma \ref{lem:yb} and Theorem \ref{thm:optimal_gamma}] 
The results  follow from $\hyba= \hyg - \db \hx \approxi \hyg - \dbi \hx$ by Lemma \ref{lem:cov_xy} and Slutsky's theorem. 
\end{proof}

\subsection{Results under ReM} 
We verify below Proposition \ref{prop:rem_g_supp}, which implies Theorem \ref{thm:rem_g} as a direct consequence. 
For random variables $A$ and $B$, let $$\proj(A \mid  1, B) = E(A) + \cov(A, B)\{\cov(B)\}^{-1}\{B-E(B)\}$$  denote the linear projection of $A$ onto $(1, B)$, 
with residual  $\res(A \mid  1,B) = A - \proj(A \mid  1,B)$.

Let 
\begina
\mu_x = \proj(\hx \mid  1, \hd) = \vxd\vdd^{-1}\hd, \qquad r_x = \res(\hx\mid 1, \hd) = \hx - \mu_x
\enda with 
$ \vxd =N \cov(\hx,\hd)$ and $\vdd =N \cov(\hbd)$. 
We have
\begina
&& \vxd = N\cov(\hx) (G^\T \otimes I_J) 
= (\emat ^{-1}G^\T)\otimes \sxx,\\
&&\vdd   = (G \otimes \mI_J)\{ N\cov(\hx)\} (G ^\T \otimes \mI_J) = (G  \emat ^{-1} G ^\T) \otimes \sxx
\enda
by $\hd = (G\otimes I_J) \hx$, $N\cov(\hx) =  (\emat ^{-1} - \po) \otimes \sxx$ from Lemma \ref{lem:cov_xy}, and $  \po G ^\T =  0$. This ensures
\beginy\label{eq:algebra_delta}
\cov(\mu_x) = N^{-1} \vxd\vdd^{-1}\vxd^\T = N^{-1}( \Phi \otimes \sxx), 
\endy
where $ \coremat   = \emat ^{-1}  G^\T ( G\emat ^{-1}   G^\T)^{-1}   G \emat ^{-1}$ as defined in \eqref{eq:vb}. 

\begin{proof}[Proof of Proposition \ref{prop:rem_g_supp}]
Let $\rb  = \hyg - \by - \dbi r_x$ and $\mub  = - \dbi \mu_x$ with 
\beginy\label{eq:AB}
\rb  + \mub =  \hyg  - \dbi \hat x - \by  \approxi   \hyba- \by
\endy
by Theorem \ref{thm:optimal_gamma}. This ensures
\beginy
 \{ \sqrtn (\hyba- \by) \mid \ma\} & \asim& \{\sqrtn (\rb +\mub ) \mid \ma\}.  \label{eq:rem_dist_ss}
\endy
We derive below the asymptotic distribution of $\{ \sqrtn(\rb +\mub ) \mid \ma\}$. 

First,  it follows from Lemma \ref{lem:yg} that $(\hyg, r_x, \mu_x)$ are pairwise uncorrelated in finite samples, and asymptotically independent and jointly normally distributed. This ensures that $\rb $ and $ \mub $ are uncorrelated in finite samples, and asymptotically independent and jointly normally distributed. 

Second, $A+B = \hybi  - \by$ with $\cov(\hybi) = N^{-1}\vbi$. We have  
\beginy\label{eq:va}
&&\cov(\mub )  = \dbi  \cov(\mu_x) \dbi ^\T= N^{-1} \vpbi, \nonumber\\  
&&\cov(\rb ) = \cov(\hybi) - \cov(\mub ) = N^{-1} \vrbi
\endy
by \eqref{eq:algebra_delta} and the uncorrelatedness of $A$ and $B$.
This, together with  $E(\rb ) =E(\mub ) = 0_Q$, ensures
\begina
(\sqrtn \rb  \mid \ma) \ \  \asim \ \  (\vrbi)^{1/2}\cdot   \epsilon,  \qquad 
(\sqrtn \mub  \mid \ma) \ \ \asim  \ \   (\vpbi)^{1/2}_{JH} \cdot \mL,
\enda
and hence 
$$
 \{ \sqrtn (\rb  + \mub ) \mid \ma \}  \asim   (\vrbi)^{1/2}\cdot   \epsilon + (\vpbi)^{1/2}_{JH} \cdot \mL$$ by the asymptotic independence between $\rb $ and $(\hd, \mub )$.  
The asymptotic distribution of $(\hyba\mid \ma)$ then follows from \eqref{eq:rem_dist_ss}. 

In addition, it follows from the definition of $A$ and the uncorrelatedness of $\hyg$ and $r_x$ that
\begina
\cov(\rb ) = \cov(\hyg) + \dbi \cov(r_x) \dbi = N^{-1}V_\lin + \dbi \cov(r_x) \dbi.
\enda
Juxtapose this with \eqref{eq:va} to see that $\vrbi = \vl$ if and only if  
\beginy\label{eq:ss}
\dbi \cov(r_x) \dbi^\T  = 0.\endy
Without restrictions on $b$, condition \eqref{eq:ss} requires $\cov(r_x) = 0$, which is equivalent to $$\coremat = \emat ^{-1}-\po$$ by 
$
\cov(r_x) = \cov(\hx) - \cov(\mu_x) =  (\emat ^{-1}-\po - \coremat )\otimes \sxx$ from \eqref{eq:algebra_delta}. 
The sufficiency of $H = Q-1$ follows from Lemma \ref{lem:gimmick}.
The necessity of $H =Q-1$ follows from $ \rank(\emat ^{-1}-\po) = Q-1$ given $\rank(\emat ^{-1}-\po) + \rank(\po) \geq \rank(\emat ^{-1}) = Q$. 
\end{proof}

\begin{proof}[Proof of Corollary \ref{cor:close}] 
Direct algebra shows that
\begina
  \ei \big( \|\hyba - \hy\la b' \ra \|_2^2   \big)
&=&
 \ei \left( \tr\left[ \big(\hyba - \hy\la b' \ra\big)\big(\hyba - \hy\la b' \ra\big)^\T \right] \right)\\
&=&
   \tr  \left[  \covi  \big(\hyba - \hy\la b' \ra \big)  \right]
\enda
and likewise 
\begina
   \ei \big( \|\hyba - \hy\la b' \ra \|_2^2  \mid \ma \big) &=&\tr \left[  \covi  \big(\hyba - \hy\la b' \ra  \mid \ma\big)  \right].
 \enda
 The result then follows from    
$ \covi (\hyba - \hy\la b' \ra  \mid \ma) = \nu_{JH,a}\cdot \covi  (\hyba - \hy\la b' \ra )$.
\end{proof}

%
\section{Proof of the results on treatment-based regressions}\label{sec:ols_app}
%
\subsection{Notation and useful facts}\label{sec:ols_app_notation}
%
%
%
%
Let 
$Y = (Y_1, \dots, Y_N)^\T$, $ T  = ( t_1,\dots, t_N)^\T$, $ X =  \left(  x _{1},\dots, x_N \right)^\T$, and $T_x =(t_1 \otimes x_1, \dots, t_N \otimes x_N )^\T$.  We can unify the treatment-based regressions \eqref{eq:lm_n}--\eqref{eq:lm_l} in matrix form as 
\begina
Y = \mss \hths +\heps \qquad (\ms)
\enda
with design matrices
\begina
\cnn  = T, \qquad \cff  = (T, X), \qquad \cll  = (T, T_x), 
\enda
\olss coefficients 
\begina
 \hthn = \hyn,\qquad\hthf = (\hyf^\T, \hbf ^\T)^\T,\qquad  \hthl = (\hyl^\T, \hgl^\T )^\T,
\enda
and residuals $\heps = (\hep_{*,1}, \ldots, \hep_{*, N})^\T$. 
Denote by
\begin{equation}
\label{eq::Sigma-M}
\hsigs =  \ccinvs \Ms \ccinvs \text{ with } M_* =  \chi_* ^\T  \diag(\hat\epsilon_{*,1}^2, \dots, \hat\epsilon_{*,N}^2) \chi_* \quad \quad   (\nfl)
\end{equation}
the \ehws covariance estimator of $\hths$ from the \olss fit of $ Y =  \chi_* \hths  +  \hep_*$.
The \ehws covariance estimator of $\hys$, namely $\hsi_*$, is then the upper-left $Q\times Q$ submatrix of $\hsigs$.

For $\nfl$, let $\hys(q)$ denote the $q$th element of $\hys$, with $\hyn(q) = \hy(q)$, $\hyf(q) = \hy(q) - \hx(q)^\T\hbf$, and $\hyl(q) = \hy(q) - \hx(q)^\T\hglq = \hy(q) - \hx(q)^\T\hg_q$ by Proposition \ref{prop:numeric_corr}.

Let 
\begina
\hsxyq =  N_q ^{-1}\sumiq  x_i Y_i, \qquad \tsx(q) =  N_q ^{-1}\sumiq  x_i x_i^\T\qquad(q\in\mt)
\enda be the sample means of $\{x_i Y_i(q)\}_{i=1}^N$ and $(x_ix_i^\T)_{i=1}^N$, respectively, with 
$\hsxyq = \sxyq+\op$ and $\tsx(q) =\sxx +\op$ under complete randomization and Condition \ref{asym}. 
Let $\hat X = (\hx(1), \dots, \hx(Q))^\T$ be the $Q\times J$ matrix with $\hx^\T(q)$ as the $q$th row vector. We have  
\renewcommand{\arraystretch}{1.5}
\beginy \label{eq:algebra_n}
&& 
\left\{\begin{array}{l}N^{-1} \cnn ^\T \cnn  = N^{-1}T^\T   T = \emat ,\\ 
N^{-1} \cnn ^\T  Y =N^{-1} T^\T  Y =  \emat  \hyn = \emat \by + \op;\end{array}\right. \\
\label{eq:algebra_f} &&
\left\{
\begin{array}{l}
N^{-1} \cff ^\T \cff   
=
\begin{pmatrix}
N^{-1} T^\T  T & N^{-1} T^\T X  \\
N^{-1} X^\T T & N^{-1} X^\T X
\end{pmatrix} = \begin{pmatrix}
\emat  & \emat  \hat X\\
\hat X^\T \emat  & \kappa \sxx
\end{pmatrix} 
= \begin{pmatrix}
\emat  &   \\
   &  \sxx
\end{pmatrix} + \op,\\
N^{-1}\cff ^\T Y = 
\begin{pmatrix}
N^{-1} T^\T  Y  \\
N^{-1} X^\T  Y
\end{pmatrix} 
=
\begin{pmatrix}
\emat  \hyn \\
 \sumq \pq  \hsxyq
\end{pmatrix} = \begin{pmatrix}
\emat  \by \\
\sxx \gp 
\end{pmatrix} + \op,
\end{array}\right.
\endy
where $\kappa = 1-N^{-1}$; and 
\beginy\label{eq:algebra_l}
\qquad \left\{\begin{array}{ll}
N^{-1} \cll ^\T \cll   
&=  
\begin{pmatrix}
N^{-1} T^\T  T & N^{-1} T^\T T_x  \\
N^{-1} T_x^\T T & N^{-1} T_x^\T T_x
\end{pmatrix} \\
&=  \begin{pmatrix}
\emat  & \emat  \diag\{ \hxt(q)\}_{q\in\mt}\\
\diag\{\hx(q)\}_{q\in\mt}\emat  & \diag\{\pq   \tsx(q)\}_{q\in\mt}
\end{pmatrix} 
= 
\begin{pmatrix}
\emat  & \\
  & \es 
\end{pmatrix} + \op,\\
N^{-1}\cll ^\T Y &= 
\begin{pmatrix}
N^{-1} T^\T  Y  \\
N^{-1} T_x^\T  Y
\end{pmatrix} =  
\begin{pmatrix}
\emat  \hyn \\
(\emat \otimes I_J) \hsxy 
\end{pmatrix} = \begin{pmatrix}
\emat  \by \\
(\emat \otimes I_J) \sxy
\end{pmatrix} + \op, 
\end{array}\right.
\endy
where $\hsxy  = (\hat S_{xY(1)}^\T, \dots, \hat S_{xY(Q)}^\T)^\T$ and 
$
\sxy = (S_{xY(1)}^\T, \dots, S_{xY(Q)}^\T)^\T$.

\subsection{Asymptotics of $\hth_\textup{L}$}\label{sec:ols_app_lemma}
We show below the asymptotic normality of $\hth_\lin $ under complete randomization. 
The result underlies the asymptotic normality of $\hyr$ in Theorem \ref{thm:rls_g}.

Recall that $\hyg$ vectorizes $\hy(q; \gamma_q) = \hy(q) -  \hx(q)^\T \gamma_q = \meaniq Y_i(q; \gamma_q)$ for $q \in \mt$:
$$
\hy(q; \gamma_q) 
= \begin{pmatrix}
N_1^{-1}\sum_{i:Z_i=1} Y_i(1; \gamma_1) \\
\vdots\\
N_Q^{-1}\sum_{i:Z_i=Q}  Y_i(Q; \gamma_Q)
\end{pmatrix} .
$$
Define 
\begina
\hat\psi
= 
\begin{pmatrix}
N_1^{-1}\sum_{i:Z_i=1} x_i \{Y_i(1;\gamma_1) - \by(1)\} \\
\vdots\\
N_Q^{-1}\sum_{i:Z_i=Q}  x_i\{Y_i(Q;\gamma_Q) - \by(Q)\}
\end{pmatrix}. 
\enda
With a slight abuse of notation, define 
\beginy\label{eq:hthb}
\hth 
= 
\begin{pmatrix}
\hyg   \\ 
\hat\psi 
\end{pmatrix}
\endy
as the sample analog of 
\begina
\theta = \begin{pmatrix}
N^{-1}\sumi Y_i(1; \gamma_1)\\
\vdots\\
N^{-1}\sumi  Y_i(Q; \gamma_Q)\\\hline
N^{-1}\sumi x_i \{Y_i(1;\gamma_1) - \by(1)\}\\
\vdots\\
N^{-1}\sumi  x_i \{Y_i(1;\gamma_Q) - \by(Q)\}
\end{pmatrix}
 = 
\begin{pmatrix}
\by\\
0_{JQ}
\end{pmatrix} .
\enda

For $q, q'\in\mt$, let $S_{Y, xY} (q, q'; \gamma) $ be the  finite population covariance of $Y_i(q; \gamma_q)$ and $x_i \{Y_i(q'; \gamma_{q'}) - \by(q')\}$, summarized in $S_{Y, xY}\la \gamma \ra  = (S_{Y, xY} (q, q';\gamma) )_{q,q'\in\mt}$.
Let  $S_{xY, xY} (q, q'; \gamma) $ be the finite population covariance of $x_i  \{Y_i(q; \gamma_{q}) - \by(q)\}$ and $x_i  \{Y_i(q'; \gamma_{q'}) - \by(q')\}$, summarized in $S_{xY, xY}\la \gamma \ra  = (S_{xY, xY} (q, q'; \gamma) )_{q,q'\in\mt}$.

\begin{lemma}\label{lem:hth_cov}
Assume \cre. Then 
\begina
\Sigma_\theta   = N\cov(\hth) = \beginp
V_\gamma & V_{Y, xY}\\
V_{Y, xY}^{\T}& V_{xY, xY}
\endp,
\enda
where $V_{Y, xY} = \diag\{e_q^{-1} S_{Y, xY}(q, q; \gamma)\}_{q\in\mt} - S_{Y, xY}\la \gamma \ra $ and  $V_{xY, xY}= \diag\{e_q^{-1} S_{xY, xY}(q, q; \gamma)\}_{q\in\mt} - S_{xY, xY}\la \gamma \ra $.
Further assume Condition \ref{asym}. Then 
\begina
\sqrtn (\hth - \theta) \rs \mn(0,\Sigma_\theta  ).
\enda
\end{lemma}

\begin{proof}[Proof of Lemma \ref{lem:hth_cov}]
Let  
\begina
 \begin{pmatrix}
Y_i(q; \gamma_q) \\
x_i \{Y_i(q; \gamma_q) - \by(q)\} 
\end{pmatrix}
\enda
be a potential outcome vector analogous to the $ (Y_i(q), x_i^\T)^\T$ in the proof of Lemma \ref{lem:cov_xy}. 
The result then follows from Lemma \ref{lem:Ding17} with $\Gamma_q = \diag(a_{\cdot q}, a_{\cdot q}\otimes I_J)$ for $q \in \mt$ identical to those defined in the proof of Lemma \ref{lem:cov_xy}.

\end{proof}

\begin{lemma}\label{lem:hth}
Under \creasym, 
\begina
\sqrtn (\hth_\lin - \thl ) \rs \mn(0, \lmd  \Sigma_\theta  \lmd)
\enda
with   
$
\lmd  = \diag\{I_Q, I_Q\otimes \sxxinv\}$.
\end{lemma}

\begin{proof}[Proof of Lemma \ref{lem:hth}]
Recall the definitions of $\hth$ and $\theta$ from \eqref{eq:hthb}.
It follows from $\chil = (T, T_x)$ and $
Y-\chi_\lin\thl  =  \{Y_i - \by(Z_i) - x_i^\T\gamma_{Z_i} \}_{i=1}^N$
that
\begina
N^{-1}\chi_\lin^\T (Y - \chi_\lin\thl ) 
&=& 
N^{-1} \begin{pmatrix}
T^\T(Y - \chil\thl)\\
T_x^\T(Y - \chil\thl)
\end{pmatrix} 
= 
N^{-1} \begin{pmatrix}
 \sum_{i:Z_i=1} \{Y_i(1) - \by(1) - x_i^\T \gamma_1\}\\
\vdots\\
 \sum_{i:Z_i=Q}  \{Y_i(Q) - \by(Q) - x_i^\T \gamma_Q\}\\\hline
 \sum_{i:Z_i=1} x_i \{Y_i(1) - \by(1) - x_i^\T \gamma_1\}\\
\vdots\\
 \sum_{i:Z_i=Q}  x_i \{Y_i(Q) - \by(Q) - x_i^\T \gamma_Q\}\\
\end{pmatrix}
\\
&=& \begin{pmatrix}
\emat  &\\
& \emat \otimes I_J
\end{pmatrix} 
\beginp
\hyg - \by \\
\hat\psi 
\endp 
= \begin{pmatrix}
\emat  &\\
& \emat \otimes I_J
\end{pmatrix} (\hth - \theta).
\enda
This, together with $ (N^{-1}\chi_\lin^\T\chi_\lin)^{-1} = \diag\{\emat ^{-1}, \emat ^{-1}\otimes \sxxinv\} + \op$ from \eqref{eq:algebra_l}, ensures 
\begina
\hth_\lin - \thl  
&=&  (N^{-1}\chi_\lin^\T\chi_\lin)^{-1} 
\{N^{-1}\chi_\lin^\T (Y - \chi_\lin\thl )\} \\
&=& (N^{-1}\chi_\lin^\T\chi_\lin)^{-1} 
\begin{pmatrix}
\emat  &\\
& \emat \otimes I_J
\end{pmatrix} (\hth - \theta)\\
&\dsim&
\begin{pmatrix}
\emat ^{-1} &\\
& \emat ^{-1}\otimes \sxxinv
\end{pmatrix}
\begin{pmatrix}
\emat  &\\
& \emat \otimes I_J
\end{pmatrix} (\hth - \theta)
= 
\lmd  (\hth - \theta) 
\enda
by Slutsky's Theorem. 
The result  then follows from Lemma \ref{lem:hth_cov}.

\end{proof}

\subsection{Results on outputs from ordinary least squares}\label{sec:ols_ehw_proof}
We next verify the results on $(\hys, \hsi_*) \ (\nfl)$.

\begin{proof}[Proof of Proposition \ref{prop:numeric_corr}]
We verify below the results on $\hys$ for $\nfl$, respectively. 

\prg{Unadjusted regression ($* = \textup{\footnotesize N}$).} 
The result follows from $\hth_\neyman = (\chin^\T\chin)^{-1}(\chin^\T Y) = \hyn$ by \eqref{eq:algebra_n}. 

\prg{Additive regression  ($* = \textup{\footnotesize F}$).} 
By \eqref{eq:algebra_f}, the first-order condition of \olss ensures   
\begina
 ( \cff ^\T  \cff )\hthf = 
\cff ^\T Y 
\quad \Longleftrightarrow\quad
 \begin{pmatrix}
\emat  & \emat  \hat X\\
\hat X^\T \emat  & \kappa \sxx
\end{pmatrix} 
\beginp
\hyf\\
\hbf
\endp =
\begin{pmatrix}
\emat  \hyn \\
 \sumq \pq  \hsxyq
\end{pmatrix}.
\enda
Compare the first row to see that $\Pi \hyf + \Pi \hX \hbf = \Pi \hyn$ and hence 
$\hyf = \hy\la 1_Q\otimes \hbf\ra$.

The probability limit of $\hbf $ then follows from 
$
(  \hyf^\T ,\hbf^\T  )^\T = 
 ( \cff ^\T  \cff )^{-1} \cff ^\T Y$,  
  where   
  $ (N^{-1} \cff ^\T  \cff )^{-1} = \diag\{\emat ^{-1}, \sxxinv\}+ \op$ 
  and 
  $N^{-1}\cff^\T Y = ( ( \emat \bY)^\T, (\sxx \gp )^\T)^\T + \op$ by \eqref{eq:algebra_f}. 

\prg{Fully interacted regression  ($* = \textup{\footnotesize L}$).}
By \eqref{eq:algebra_l}, the first-order condition of \olss ensures
\begina
 ( \cll ^\T  \cll )\hthl = 
\cll^\T Y 
\quad \Longleftrightarrow\quad
 \begin{pmatrix}
\emat  & \emat  \diag\{ \hxt(q)\}_{q\in\mt}\\
\diag\{\hx(q)\}_{q\in\mt}\emat  & \diag\{\pq   \tsx(q)\}_{q\in\mt}
\end{pmatrix} 
\begin{pmatrix}
\hyl\\
\hgl 
\end{pmatrix}
=  
\begin{pmatrix}
\emat  \hyn \\
(\emat \otimes I_J) \hsxy 
\end{pmatrix}.
\enda
Compare the first row to see that $\Pi \hyl + \Pi \diag\{ \hxt(q)\}_{q\in\mt}\hgl = \Pi \hyn$ and hence  $\hyl = \hy\la \hgl\ra$. 

The probability limit of $\hg_{\lin}$ then follows  from $(  \hyl^\T ,\hgl^\T  )^\T= (\cll^\T\cll)^{-1}\cll^\T Y$ and \eqref{eq:algebra_l}. 
\end{proof}

\begin{proof}[Proof of Lemma \ref{lem:reg}]
The asymptotic distributions of $\hys$  follow from Proposition \ref{prop:numeric_corr} and Lemma \ref{lem:yb}. 
We verify below the asymptotic conservativeness of $\hsi_*$.
Given $V_*  = \diag (S_{*,qq}/e_q )_{\qit}  - S_* $,  the results are equivalent to
\begina
N \hsis - \diag (S_{*,qq}/e_q )_{\qit} = \op\qquad (\nfl).
\enda
We verify below this for $* = \nm, \lin$. The proof for $* = \fisher$ is almost identical to that for $* = \lin$ and thus omitted. 

A useful fact is that
\beginy\label{eq:meat_asym}
N^{-1} T^\T \{\diag(\hesi^2)_{i=1}^N\} T = N^{-1}\diag \left(\sumiq \hesi^2 \right)_{\qit} = \diag\left(e_q S_{*, qq}  \right)_{\qit} + \op 
\endy
 for $\nfl$. The second equality in \eqref{eq:meat_asym} follows from 
\begina
\heni &=& Y_i - \hyn(Z_i),\\
\hefi &=&  Y_i - \hyf(Z_i) -  x_i^\T  \hbf,\\
 \heli &=& Y_i - \hyl(Z_i) -  x_i^\T  \hg_{\lin,Z_i}  
\enda
by Proposition \ref{prop:numeric_corr}, 
such that $\meaniq \hesi^2 = S_{*, qq} + \op$ by standard results.

\prg{Unadjusted regression ($* = \textup{\footnotesize N}$).} It follows from $\chi_\nm = T$ and $N^{-1}T^\T T = \Pi$ that
\begina
N\hsi_\nm =N (T^\T T)^{-1}   T^\T \{\diag(\heni^2)_{i=1}^N\} T  (T^\T T)^{-1} =   \Pi^{-1} \left[ N^{-1} T^\T \{\diag(\heni^2)_{i=1}^N\} T \right]  \Pi^{-1}.
\enda
The result then follows from \eqref{eq:meat_asym}. 

\prg{Fully interacted regression ($* = \textup{\footnotesize L}$).}
First, 
\begina 
M_\lin \, = \,
\beginp
T^\T\\
T_x^\T 
\endp\big \{\diag(\heli^2)_{i=1}^N \big\}
(T, T_x)
=
\left(\begin{array}{cc} M_1 &  M_2 \\  M_2^\T & M_3 \end{array}\right),
\enda 
where 
\begina
M_1 =  T^\T \big \{\diag(\heli^2)_{i=1}^N \big \} T,
\qquad 
M_2 = T^\T \big\{\diag(\heli^2)_{i=1}^N \big\} T_x, 
\qquad 
M_3 = T_x^\T \big\{\diag(\heli^2)_{i=1}^N \big\} T_x.
\enda

Next, it follows from \eqref{eq:meat_asym} and analogous algebra that 
\beginy\label{eq:M_1_l}
\qquad \ninv{M_1}  = \diag(e_q S_{\lin,qq})_{q\in\mt} + \op, \quad N^{-1} M_2 =\Op(1), \quad N^{-1} M_3 = \Op(1). 
\endy
This, together with $ (I_Q, 0_{Q\times JQ}) (\ninv \ccl)^{-1}   =(\emat^{-1}, 0_{Q\times JQ}) + \op$ by \eqref{eq:algebra_l}, ensures 
\beginy\label{eq:ehw_l}
N\hsi_\lin &=&  (I_Q, 0_{Q\times JQ}) (N\hsigl) (I_Q, 0_{Q\times JQ})^\T\nonumber\\
& =&   (I_Q, 0_{Q\times JQ})  ( N^{-1} \chil ^\T    \chil )^{-1} (N^{-1} M_\lin ) ( N^{-1}  \chil^\T    \chil  )^{-1}(I_Q, 0_{Q \times JQ})^\T\nonumber\\
&=& \beginp \emat ^{-1}, \ 0_{Q\times JQ} \endp  
\beginp   N^{-1}M_1  & N^{-1}M_2  \\ N^{-1}M_2^\T  & N^{-1}M_3 \endp 
\beginp\emat ^{-1} \\ 0_{JQ\times Q}  \endp  + \op \nonumber\\
&=& \emat ^{-1}(N^{-1}M_1) \Pi^{-1} + \op \nonumber\\
&=& \diag( S_{\lin ,qq}/e_q)_{q\in\mt} + \op
\endy 
from \eqref{eq:M_1_l}.
\end{proof}

Recall that we propose to use
$
W = (R \hthl - r)^\T (R   \hat\Sigma_\lin R^\T)^{-1}(R \hthl - r)$
as the test statistic for testing $H_0: R\thl = r$, and compute a one-sided $p$-value by comparing $W$ to $\chi^2_m$ with $m=\rank(R)$. 
Theorem \ref{thm:test} below follows from Lemma \ref{lem:hth} and ensures that the resulting $p$-value preserves the nominal type one error rates asymptotically \citep{DD18}.

\begin{theorem}\label{thm:test}
Assume complete randomization and Condition \ref{asym}. Under $H_0: R\thl = r$, the asymptotic distribution of $W$ is stochastically dominated by $\chi^2_m$. 
\end{theorem}

\begin{proof}[Proof of Theorem \ref{thm:test}] 
Recall $\hsig_\lin$ and  $M_\lin$ in \eqref{eq::Sigma-M}. 
Similar algebra as the proof of Lemma \ref{lem:reg} ensures that $\hsig_\lin$ is asymptotically conservative for estimating the true sampling covariance of $\hthl$: 
\begina
\plim N \left\{ \hsig_\lin -  \cov(\hthl) \right\} = \lmd \beginp
S_\lin & S_{Y, xY}\la\gamma\ra\\
S_{Y, xY}\la\gamma\ra^\T & S_{xY, xY}\la\gamma\ra  
\endp \lmd \geq 0. 
\enda
The result then follows from $\sqrtn (R\hthl - R\thl) \rs \mn(0, R\lmd\Sigma_\theta\lmd^\T R^\T)$ by Lemma \ref{lem:hth}. 
\end{proof}

\begin{remark}\label{rmk:nay_F}
Let $\rss_\lin$ and $\rss_\rr$ denote the residual sum of squares from the \olss fit of \eqref{eq:lm_l} and the \rlss fit of \eqref{eq:lm_l} subject to $R\thl = r$, respectively. 
The classical model-based framework tests $H_0: R\thl = r$ by an $F$-test with 
\beginy\label{eq:F}
F = \frac{(\rss_\rr - \rss_\lin)/m}{\rss_\lin/(N- Q - JQ)}
\endy
as the test statistic, and compares it against the $F$-distribution with degrees of freedom $m$ and $N-Q-JQ$, denoted by $F_{m, N-Q-JQ}$, to obtain the one-sided $p$-value. 

We do not pursue this route because the validity of the above $F$-test depends critically on the classical linear model assumptions of  uncorrelated and homoskedastic sampling errors.
The design-based framework violates both assumptions and renders the sampling distribution of the $F$-statistic in \eqref{eq:F} no longer $F_{m, N-Q-JQ}$ under $H_0$ even asymptotically. The  $F$-test  as a result does not preserve the correct type one error rates even asymptotically.
See \cite{DD18} and \cite{wuanding2020jasa} for related discussions.
\end{remark}

\subsection{Results on outputs from restricted least squares}\label{sec:rls_proof}
\subsubsection{RLS subject to general restriction}

\black

\begin{proof}[Proof of Theorem \ref{thm:rls_g}]
We verify below the numeric expression and asymptotic results, respectively. 

\prg{Numeric expression.}
The numeric result follows from the method of Lagrange multipliers.
In particular, the Lagrangian for the restricted optimization problem equals
\beginy\label{eq:lag}
(Y - \chil \theta)^\T (Y - \chil \theta) - 2 \lambda^\T(R\theta  - r)
\endy and yields the first-order condition as
\begina
\ccl \hthlr = \chil^\T Y - R^\T\lambda,  
\enda
where $\lambda = \{R \ccinvl R^\T\}^{-1}(R\hthl-r)$.
This, together with \eqref{eq:algebra_l}, ensures  
\begina
\begin{pmatrix}
\emat  & \emat  \diag\{ \hxt(q)\}_{q\in\mt}\\
\diag\{\hx(q)\}_{q\in\mt}\emat  & \diag\{\pq   \tsx(q)\}_{q\in\mt}
\end{pmatrix} 
\beginp
\hyr \\
\hgr
\endp = \begin{pmatrix}
\emat  \hyn \\
(\emat \otimes I_J) \hsxy 
\end{pmatrix} - N^{-1}
\beginp
\ryt \lambda\\
\rgt \lambda
\endp.  
\enda 
Extract the first row of both sides to see $$ \hyr  +  \diag\{ \hxt(q)\}_{q\in\mt} \hgr =  \hyn -  \einv\ryt (N^{-1}\lambda).$$ 
The result then follows from $\hyn -  \diag\{ \hxt(q)\}_{q\in\mt} \hgr = \hybr $ such that 
\beginy\label{eq:hyr_numeric}
\hyr = \hybr  -  \einv\ryt (N^{-1}\lambda).
\endy

\prg{Asymptotic results.} By \eqref{eq:algebra_l}, we have 
 $N \ccinvl  = \diag \{ \emat ^{-1}, (\es )^{-1} \} + \op$ 
and hence
\begina
N \{R \ccinvl R^\T\}^{-1} = \dz + \op.
\enda
This ensures
\beginy\label{eq:mri} 
  \mri  = \plim \mr = \begin{pmatrix}
\einv  & \\ & \esinv 
\end{pmatrix} R^\T  \dz  = \begin{pmatrix}
 \einv  \ryt  \dz  \\   (\es )^{-1} \rgt \dz 
\end{pmatrix}
\endy
for arbitrary $R$. 
The probability limits of $\hyr$ and $\hgr$ then follow from 
$$\hthlr - \theta_\lin = - \mri\reslp +  \op$$
by Lemma \ref{lem:hthr} and the fact that $\hthl = \thl + \op$ by Proposition \ref{prop:numeric_corr}. 

The asymptotic normality of $\hyr  - \xir$  follows from extracting the first $Q$ rows of 
\beginy\label{eq:hthr_lim}
\hthlr - \thl + \mr(R\thl-r) = (I - \mr R)(\hthl-\thl) \approxi  (I- \mri R)(\hthl-\thl).  
\endy
The first equality in \eqref{eq:hthr_lim} follows from Lemma \ref{lem:hthr}. 
The asymptotic equivalence $\approxi$ follows from Slutsky's theorem and the asymptotic normality of $\sqrtn (\hth_\lin  - \tl )$ by Lemma \ref{lem:hth}.
\end{proof}

\subsubsection{RLS subject to  {\go} restriction}
\begin{proof}[Proof of Proposition \ref{prop:noy} and Theorem \ref{thm:noy}]
Observe that $R\theta_\lin -r = \rho_\gamma \gamma - r_\gamma$ under the \gor.
Proposition \ref{prop:noy} follows from Theorem \ref{thm:rls_g}, which ensures $\hyr = \hybr $  when $\ry = 0$, with $\plim \hgr = \gamma$ if \eqref{eq:rest_noy} is correctly specified.

Theorem \ref{thm:noy} then follows from Lemma \ref{lem:yb}, which ensures $\sqrtn(\hybr  - \by) \rs \mn(0, \vrr)$, with $\vrr = V_\lin$ if $\plim \hgr = \gamma$. 
\end{proof}

\subsubsection{RLS subject to separable restriction with $\rhy \neq 0$}
Recall that 
\begina
U =  I_Q-   \einv \rhyt (\rhy \einv \rhyt)^{-1} \rhy
\enda 
with $\Pi = \diag(e_q)_{q\in\mt}$
for $\rhy \neq 0$.

\begin{proof}[Proof of Proposition \ref{prop:sep} and Theorem \ref{thm:sep}]
Recall that $\chi_\lin = (T, T_x)$, with $T = (t_1, \ldots, t_N)^\T$ and $T_x = (t_1\otimes x_1, \ldots, t_N\otimes x_N)^\T$. 
Denote by $\theta = (\theta_Y^\T, \theta_\gamma^\T)^\T$ the dummy for the model coefficients, with $\theta_Y$ and $\theta_\gamma$ corresponding to $T$ and $T_x$, respectively: $Y = T\theta_Y + T_x\theta_\gamma + \epsilon$. 
Under the separable restriction with $\rho_Y \neq 0$ and $\rho_\gamma \neq 0$, $R\theta -r$ simplifies to the vectorization of $\rho_Y\theta_Y - r_Y$ and $\rho_\gamma\theta_\gamma - r_\gamma$, and we can write $\lambda = (\lambda_Y^\T, \lambda_\gamma^\T)^\T$, with $\lambda_Y$ and $\lambda_\gamma$ corresponding to $\rho_Y\theta_Y - r_Y$ and $\rho_\gamma\theta_\gamma - r_\gamma$, respectively.
Under the separable restriction with no restriction on $\theta_\gamma$, $R\theta -r$ simplifies to $\rho_Y\theta_Y - r_Y$, and we write can $\lambda = \lambda_Y$. 
The Lagrangian in \eqref{eq:lag} simplifies to
\begina
&&(Y - T\theta_Y - T_x\theta_\gamma)^\T(Y - T\theta_Y - T_x\theta_\gamma) - 2\lambda_Y^\T(\rhy \theta_Y - r_Y) - 2\lambda_\gamma^\T(\rhg \theta_\gamma - r_\gamma)\\
\text{or}\quad&&(Y - T\theta_Y - T_x\theta_\gamma)^\T(Y - T\theta_Y - T_x\theta_\gamma) - 2\lambda_Y^\T(\rhy \theta_Y - r_Y)
\enda
depending on if there is restriction on $\theta_\gamma$ or not. 
The first-order condition with regard to $\theta_Y$ equals
\begina
&&T^\T(Y - T \hyr - T_x \hgr) - \rhyt \lambda_Y = 0\\
&\Longleftrightarrow & \hyr = (T^\T T)^{-1}T^\T Y -  (T^\T T)^{-1} T^\T T_x \hgr -  (T^\T T)^{-1} \rhy^\T \lambda_Y.
\enda
This, together with $(T^\T T)^{-1}T^\T Y = \hyn$, $(T^\T T)^{-1} T^\T T_x  =  \diag\{\hx^\T(q)\}_{q\in\mt}$, and $(T^\T T)^{-1} = N^{-1}\einv$  by \eqref{eq:algebra_l}, ensures 
\beginy\label{eq:hylr_sep}
\hyr  = \hyn - \left[\diag\{\hx^\T(q)\}_{q\in\mt}\right]\hgr - N^{-1}\einv \rhy^\T \lambda_Y = \hybr  - N^{-1}\einv \rhy^\T \lambda_Y.
\endy
The restriction $\rhy \hyr  = r_Y$ further suggests 
$\lambda_Y = N (\rhy \einv \rhyt)^{-1}\{ \rhy \hybr  - r_Y\}$. 
Plugging this in \eqref{eq:hylr_sep} verifies 
\beginy\label{eq:sss}
\hyr   = \hybr  - \einv \rhy^\T   (\rhy \einv \rhyt)^{-1}\{ \rhy \hybr  - r_Y\} =U \hybr   + \einv \rhy^\T   (\rhy \einv \rhyt)^{-1} r_Y
\endy
with $\hyr - \by = U(\hybr  - \by) + \mur$. 

The probability limit of $\hgr$ follows from 
$
\plim \hgr = \gamma -  (\es )^{-1} \rgt \dz  (R\theta_\lin - r)$ by Theorem \ref{thm:rls_g}.
Specifically, we have (i) $    \rg   = 0$ if $R = (\rhy, 0)$ with no restriction on $\gamma$, and (ii) $\rg = (0, \rhg^\T)^\T$ and thus 
$$
\rgt \dz (R\theta_\lin - r) =\rhgt \{\rhg \esinv \rhgt\}^{-1} (\rhg \gamma - r_\gamma)$$ if $R = \diag(\rhy, \rhg)$ with non-empty restrictions on both $\by$ and $\gamma$. 

The asymptotic normality follows from 
$\sqrt N (\hybr  - \by) \rs \mn(0_Q, \vrr )$ by Lemma \ref{lem:yb}.  
\end{proof}

We next verify the asymptotic bias-variance trade-off and the design-based Gauss--Markov theorem under constant treatment effects.

\begin{lemma}\label{lem:gm}
Assume that $\rhy$ is a contrast matrix that has full row rank. For $U =  I_Q-   \einv \rhyt (\rhy \einv \rhyt)^{-1} \rhy$ and $L = I_Q + \ary $, where $A$ is an arbitrary matrix, we have 
\begina
U\core U^\T \leq L \core L^\T.
\enda
\end{lemma}

\begin{proof}[Proof of Lemma \ref{lem:gm}]
Let $\uy   = I_Q-U=  \einv \rhyt (\rhy \einv \rhyt)^{-1} \rhy$. 
Then 
$
L - U  = U_Y + A \rhy =  \{\einv \rhyt (\rhy \einv \rhyt)^{-1} +A\}\rhy
$
with
\begina
(L-U) \core U^\T &=& \big \{\einv \rhyt (\rhy \einv \rhyt)^{-1} +A\big\} \rhy \core U^\T\\
&=& 0;
\enda
the last equality follows from $\rhy\einv U^\T  = 0$ and hence $\rhy \core U^\T = 0 $. 
This ensures  
\begina
L \core L^\T 
&= &  (L-U + U)\core(L-U + U)^\T  \\
&= &  (L-U)\core(L-U)^\T + U\core (L-U)^\T \\
 && + \ (L-U) \core U^\T  + U\core U^\T\\
 &= &  (L-U)\core(L-U)^\T  + U\core U^\T\\
&\geq & U\core U^\T. 
\enda

\end{proof}

\begin{proof}[Proof of Theorems \ref{thm:gm} and \ref{thm:gm_a}]
Assume that the restriction on $\gamma$ is correctly specified. 
Then  $\plim \hgr = \gamma$ by Proposition \ref{prop:sep} with 
$\sqrt N(\hyr  - \by-\mur) \rs \mathcal N(0_Q, U \Vl U^\T)$ 
by Theorem \ref{thm:sep}.  

Condition \ref{cond:sa} further ensures that 
\beginy\label{eq:vl}
\Vl = N\covi(\hyl) = \sz \core,
\endy where $\sz$ denotes the common value of $S_{\lin,qq'}$ for all   $q,q'\in\mt$ under Condition \ref{cond:sa}. 
We verify below the asymptotic bias-variance trade-off result in Theorem \ref{thm:gm}\eqref{item:bv} and the design-based Gauss--Markov result in Theorem \ref{thm:gm_a}, respectively.

\prg{Asymptotic bias-variance trade-off.} The result follows from 
\begina
 N\covi(\hyr   -  \mur  ) =
   U \vl U^\T  \leq \vl  = N\covi(\hyl)
\enda
by \eqref{eq:vl} and Lemma \ref{lem:gm}. 

\prg{Gauss--Markov theorem when \eqref{eq:rest_sep} is correctly specified.} 
For an arbitrary $L\hyba+a \in \my'$ that is consistent for $\bar Y$, the definition of consistency implies that 
$L\by + a = \by$
for all $\by$ that satisfies $\rhy \by = r_Y$. 
Let $\by_0$ be a solution to $\rhy \by = r_Y$ with $\rhy \by_0 = r_Y$ and hence $L\by_0 + a = \by_0$.
Then $
(L - I_Q) (\by - \by_0) = 0$ for all $\by$ that satisfy $\rhy\by=r_Y$. This ensures
$L - I_Q = \ary$ for some matrix $A$. 

The result then follows from 
\begina
N\covi(L\hyba+a) = L \vbi L^\T \geq  L V_\lin L^\T, \qquad N\covi(\hyr )  =U \vl U^\T,
\enda
with $L V_\lin L^\T \geq  U \vl U^\T$ by \eqref{eq:vl} and Lemma \ref{lem:gm}. 

\end{proof}

\subsubsection{Robust covariance estimator}
Recall that 
$
\hsigr  =  \ccinvl  \{ \chi_\lin ^\T  \diag(\hat\epsilon_{\lr,1}^2, \dots, \hat\epsilon_{\lr,N}^2) \chi_\lin \}\ccinvl$, where $(\hat\epsilon_{\lr,i})_{i=1}^N$ are the residuals from the \rlss fit of \eqref{eq:lm_l}. 
 Let $
 \hsigrq$ denote the upper-left $Q\times Q$ submatrix of $\hsigr$. 
Let $\hyr(q)$ denote the $q$th element of $\hyr$.

 \begin{lemma}\label{lem:M_rls}
 Assume complete randomization, Condition \ref{asym},  and general $R$. Then
 \beginy \label{eq:holr_lim}
 N\hsigr  =   \left(\begin{array}{cc}  N\hsigrq & \opo  \\ \opo  & \opo \end{array}\right) ,
\endy 
where $
N  \hsigrq =  \diag (S_{\rr,qq}/e_q )_{\qit}+ \diag[\{\hyr(q) - \by(q)\}^2 /e_q ]_{\qit}+ \op$.
 \end{lemma}

\begin{proof}[Proof of Lemma \ref{lem:M_rls}]
Recall \eqref{eq::Sigma-M}, where $M_\lin  $ is the matrix in the middle for computing the \ehws covariance estimator of $\hyl$ from the \olss fit. 
Define
$
M_{\lin,\rr} = \chi_\lin ^\T  \diag(\hat\epsilon_{\lr,1}^2, \dots, \hat\epsilon_{\lr,N}^2) \chi_\lin 
$
as a variant of $M_\lin$ based on the \rolss residuals such that 
 \begina
 \hsigr  = \ccinvl M_{\lin,\rr} \ccinvl. 
 \enda
The same algebra as in the proof of Lemma \ref{lem:reg} ensures 
\beginy\label{eq:M1_rr}
\mlr 
=
\left(\begin{array}{cc} M_1 &  M_2 \\  M_2^\T & M_3 \end{array}\right),
\endy 
where $
M_1  = \diag   (\sumiq   \heri^2  )_{q\in\mt}$,  
$M_2 = \diag  (\sumiq    \heri^2  x_i^\T)_{q\in\mt}$, and $ M_3   = \diag (\sumiq   \heri^2 x_i  x_i^\T )_{q\in\mt}$.

Recall that $\brq = \plim \hgrq$, with $S_{\rr,qq} = (N-1)^{-1}\sumi\{Y_i(q) - x_i^\T \brq- \by(q) \}^2$. 
Then  
\begina
\heri &=& Y_i(q) -\hyr(q) - x_i^\T \hgrq  \\
&=& \left\{Y_i(q) - x_i^\T \brq- \by(q) \right\} - \big\{\hyr(q) - \by(q)\big\} - x_i^\T \big(\hgrq - \brq\big) 
\enda
for $i$ with $Z_i = q$. This ensures
\beginy\label{eq:ssss}
\meaniq \hepri^2 
&=&  \meaniq \left\{Y_i(q) - x_i^\T \brq- \by(q) \right\} ^2  + \  \big\{\hyr(q) - \by(q)\big\}^2 \nonumber\\
&&+ \ \big(\hgrq - \brq\big) ^\T \hsxq  \big(\hgrq - \brq\big) \nonumber \\
&& - \ 2 \big\{\hyr(q) - \by(q)\big\} \big\{\hy(q)  - \hx^\T(q) \brq - \by(q)\big\}\nonumber\\
&&  - \ 2 \big\{\hyr(q) - \by(q)\big\}  \big(\hgrq - \brq\big)^\T \hx(q) \nonumber\\
&& - \ 2 \big(\hgrq - \brq\big) ^\T  \big\{\hsxyq - \hsxq \brq  - \hx(q) \by(q) \big\} \nonumber \\
&=&  S_{\rr,qq} +   \big\{\hyr(q) - \by(q)\big\}^2  + \op, 
\endy
where the last equality follows from
\begina
 \meaniq \left\{Y_i(q) - x_i^\T \brq- \by(q) \right\} ^2 = S_{\rr,qq} + \op,\quad 
 \hgrq = \brq + \op, \quad \hyr(q) - \by(q)=\opo
 \enda in addition to
$
 \hy(q) = \by(q) + \op$, $ \hx(q) = \op$, $ \hsxq = \sxx + \op$, and $ \hsxyq = \sxyq + \op$. 
Plugging \eqref{eq:ssss} in the definition of $M_1$ ensures
\beginy\label{eq:M1_rr_lim}
\ninv{M_1}
&=&   \diag  \left(e_q \meaniq   \heri^2 \right)_{q\in\mt} \\
&=& \diag(e_q S_{\rr,qq})_\qit +\diag\left[ e_q \big\{\hyr(q) - \by(q)\big\}^2 \right]_{q\in\mt} + \op.  \nonumber
\endy

Similar algebra ensures that  $\meaniq \heri^2 x_i = \opo$ and $\meaniq \heri^2 x_i x_i^\T = \opo$, such that  $N^{-1} M_2 =\Op(1)$ and $N^{-1} M_3 = \Op(1)$.
It then follows from the same reasoning as in \eqref{eq:ehw_l} that  
\begina
N\hsigrq &=& (I_Q, 0_{Q\times JQ}) (N\hsigr) (I_Q, 0_{Q\times JQ})^\T \\
& =& \einv(\ninv M_1) \einv+ \op \\
&=&  \diag ( S_{\rr,qq} / e_q)_\qit +\diag \left[  \big\{\hyr(q) - \by(q)\big\}^2/ e_q \right]_{q\in\mt} + \op
\enda 
by \eqref{eq:M1_rr_lim}.
\end{proof}

\begin{proof}[Proof of Theorem \ref{thm:ehw_rr}]
We verify below the result for the {\go} and separable restrictions, respectively. 

\prg{Correlation-only restriction.} 
When the restriction satisfies \eqref{eq:rest_noy}, \eqref{eq:mri} simplifies to 
\begina
\mri =\beginp
0\\
(\es )^{-1} \rgt \dz
\endp
\enda
such that 
\begina
I - \mri   R   
= I - \beginp 0\\    (\es )^{-1} \rgt \dz  \endp (0, \rg) = \beginp I_Q&   \\   &I_{JQ} -   (\es )^{-1} \rgt \dz  \rg \endp. 
\enda 
This, together with \eqref{eq:holr_lim}, ensures 
\begina
&& N  (I-\mr R )\hsigr (I-\mr R )^\T 
\\
&=&  
\begin{pmatrix}
I_Q&   \\ 
  & \oo 
\end{pmatrix} 
\beginp N\hsigrq & \opo  \\ \opo  & \opo  \endp
 \begin{pmatrix}
 I_Q &   \\ 
   & \oo 
\end{pmatrix}+ \op\\
 &=& 
 \beginp N\hsigrq  & \opo \\ \opo & \opo \endp + \op,
\enda
and hence $N \hsr = N\hsigrq +  \op$. 

Theorem \ref{thm:noy} further ensures that $\hyr(q) = \by(q) + \op$ under the {\co} restriction. This ensures $N\hsigrq = \diag( S_{\lr, qq}/ \pq )_{q\in\mt}  + \op = \vrr + S_{\lr} + \op$ by Lemma \ref{lem:M_rls} and the definition of $\vrr$. 

\prg{Separable restriction with $\rhy \neq 0$.} Consider first the case with $\rhy \neq 0$ and $\rhg \neq 0$.
Then $R = \diag(\rhy, \rhg)$, and  it follows from 
\begina
  \mri  =
  \beginp
\einv  & \\ & \esinv 
\endp
R^\T \dz =
  \beginp
\einv \rhyt (\rhy \einv \rhyt)^{-1} &  \\
 & \esinv \rhgt \{\rhg \esinv \rhgt\}^{-1} \endp
\enda
by \eqref{eq:mri} that 
\begina
I - \mri   R   
=\diag\{U, O(1)\}.
\enda 
This, together with \eqref{eq:holr_lim}, ensures that 
\begina
&& N  (I-\mr R )\hsigr (I-\mr R )^\T\\ 
&=&  
\begin{pmatrix}
U &    \\ 
  & \oo 
\end{pmatrix} 
\beginp N\hsigrq & \opo  \\ \opo  & \opo  \endp
 \begin{pmatrix}
 U^\T &    \\ 
   & \oo 
\end{pmatrix}+ \op\\
 &=& 
 \beginp U\{ N\hsigrq\} U^\T & \opo \\ \opo & \opo \endp + \op,
\enda
and hence $N \hsr = U\{ N\hsigrq\} U^\T+ \op$  with 
\begina
 N\hsigrq - \vrr = S_{\lr }  +  \diag[\{\hyr(q) - \by(q)\}^2 /e_q ]_{\qit}  + \op 
\enda 
by Lemma \ref{lem:M_rls}. 
Theorem \ref{thm:sep} further ensures that $\hyr(q) = \by(q) + \murq + \op$ under the {\so} restriction when $\rhy \neq 0$. This verifies the result for $R = \diag(\rhy, \rhg)$. 

The proof for the case with $\rhg = 0$ is analogous with $R = (\rhy, 0)$, $\Delta_0 = \rhyt (\rhy \einv \rhyt)^{-1}$, 
\begina
&&  \mri  =
  \beginp
\einv  & \\ & \esinv 
\endp
R^\T \dz =
  \beginp
\einv \rhyt (\rhy \einv \rhyt)^{-1}\\
0  \endp,
\enda
and $I - \mri R = \diag(U, 0_{JQ\times JQ})$. We omit the details.  
\end{proof}

\section{Proof of the results on factor-based regressions}\label{sec:factor_proof}

\subsection{Standard factorial effects}
We verify below Proposition \ref{prop:2k_g}, Corollaries \ref{cor:2k_g_1}--\ref{cor:2k_g_2}, and Proposition \ref{prop:bd_g} in Section \ref{sec:2k_app}. 
The results of Propositions \ref{prop:2k_saturated}--\ref{prop:2k}, Corollary \ref{cor:2k}, and Proposition \ref{prop:bd} then follow as special cases.

\begin{proof}[Proof of Proposition \ref{prop:2k_g}]
Recall that $\zik^0 = 2^{-1}(\zik +1)$ gives the $\{0,1\}$-counterpart of $\zik$. 
Then 
\beginy\label{eq:t_k}
t_i = \otimes_{k=1}^K (1-\zikz, \zikz)^\T
\endy gives the vector of $\{1(Z_i = q): q\in \mt\}$ from the treatment-based regressions \eqref{eq:lm_n}--\eqref{eq:lm_l}. 
Let 
\begina
f_i = \otimes_{k=1}^K (1, \zik)^\T
\enda vectorize $\{1, \zimks: \mk \in \pk\}$ from the factor-based regressions \eqref{eq:lm_2k_n}--\eqref{eq:lm_2k_l}.
It
follows from 
\begina
 \beginp
 1 - \zikz\\
 \zikz
 \endp 
 =  
 2^{-1} \beginp
 1-\zik\\
 1+\zik
 \endp 
 =
2^{-1} \beginp
 1 & -1\\
 1 & 1
 \endp 
 \beginp
 1 \\
 \zik
 \endp \qquad (k = \ot{K}) 
\enda
that  $t_i = \Gamma^\T f_i$ for nonsingular
\begina
\Gamma =  2^{-K} \left\{\otimes_{k=1}^K \beginp 1&1\\ -1 & 1\endp \right\}  = 2^{-K} \left\{2^{K-1} \beginp c_{\sss,\emptyset}^\T \\ \cs \endp \right\} = 2^{-1} \beginp c_{\sss,\emptyset}^\T \\ \cs \endp,
\enda where $c_{\sss,\emptyset} = 2^{-(K-1)} 1_Q$ and $\cs$ is the contrast matrix corresponding to $\ts$. 
The regressor vectors of the treatment-based and factor-based fully interacted regressions \eqref{eq:lm_l} and \eqref{eq:lm_2k_l} thus satisfy
\beginy\label{eq:linear transformation_standard}
\beginp
t_i \\
t_i\otimes x_i
\endp =
\beginp
 \Gamma^\T & \\
& \Gamma^\T\otimes I_J
 \endp  \beginp
f_i \\
f_i\otimes x_i
\endp.
\endy

Consider the \rlss fit of \eqref{eq:lm_2k_l} subject to \eqref{eq:rest_2k_g}. 
Let $\ttrp'$ denote 2 times the coefficient vector of $\{\zimk: \mk \in \mfp\}$, with $\torp'$ as the associated double-decker-taco robust covariance estimator by \eqref{eq:ehw_rls}.
Lemma \ref{lem:inv_rls} ensures that 
\begina
\ttrp' = \csp \hyrs, \qquad \torp' = \csp \hsr \cspt
\enda by \eqref{eq:linear transformation_standard}. 
The result then follows from $\ttrp = \ttrp'$ and $\torp = \torp'$ by Example \ref{ex:ehw_equiv}.

In addition, \eqref{eq:linear transformation_standard} ensures that
\begina
t_i^\T \by + (t_i \otimes \cxi)^\T\gamma &=& f_i^\T(\Gamma\bY) + (f_i \otimes \cxi) ^\T \{(\Gamma\otimes I_J) \gamma\}
\enda
in \eqref{eq:dlm_l}. 
This justifies the forms of target parameters in the comments  after Proposition \ref{prop:2k_g}.
\end{proof}

\begin{proof}[Proof of Corollary \ref{cor:2k_g_1}]
The numeric result follows from 
$
\ttrp  = \csp \hyrs = \csp \hybrs
$
by  Proposition \ref{prop:2k_g} and Proposition \ref{prop:noy}.
This ensures the asymptotic distribution by Lemma \ref{lem:yb}, and the asymptotic conservativeness of $\torp$ by Proposition \ref{prop:2k_g} and Theorem \ref{thm:ehw_rr}. 

Further assume 
Condition \ref{cond:equal}. 
Then $(\csm' \otimes I_J)\gamma = 0$ is correctly specified as long as $\csm' $ is a contrast matrix, which is equivalent to $\mk = \emptyset \not\in \mfm'$ with \eqref{eq:lm_2k_g} including $x_i$.
When this is true, we have $\plim  \hgrs  = \gamma$ with $\hybrs \asim \hyl$ and $\ttrp \asim \ttlp$ by Proposition \ref{prop:noy} and Lemma \ref{lem:yb}.
 \end{proof}
 
 \begin{proof}[Proof of Corollary \ref{cor:2k_g_2}] 
The numeric result follows from 
\begina
\ttrp - \tsp - \csp \murs = \csp (\hyrs - \by - \murs) = \csp  \us \{\hybrs  - \by\}
\enda
by Proposition \ref{prop:2k_g} and Proposition \ref{prop:sep}. This ensures the asymptotic distribution by Lemma \ref{lem:yb}, and the asymptotic conservativeness of $\torp$ by Proposition \ref{prop:2k_g} and Theorem \ref{thm:ehw_rr}. 

Further assume Condition \ref{cond:sa}. The same reasoning as above ensures that $\vrrs = \vl$ as long as \eqref{eq:lm_2k_g} includes $x_i$.
When this is true,  the result on $\ttrp \succi \ttfp \asim \ttlp \succi \ttnp$ then follows from 
\begina
N\covi(\ttrp) = 
\csp \us\vrrs \ust  \csp^\T,\qquad N\covi(\ttsp) = \csp V_* \csp^\T \qquad  (\nfl)
\enda
with 
$\us\vrrs \ust = \us \vl  \ust \leq \vl = V_\fisher \leq V_\neyman$ by Lemma \ref{lem:gm}. 

\end{proof} 

 \begin{proof}[Proof of Proposition \ref{prop:bd_g}]
Recall $\hyrs $ as the coefficient vector of $t_i$ from the \rolss fit of  \eqref{eq:lm_l} subject to \eqref{eq:rest_2k_g}. 
By \eqref{eq:sss}, we have
\begina
\hyrs   = \hybrs    -  \einv \rhy^\T   (\rhy \einv \rhyt)^{-1} \rhy\hybrs, 
\enda
where $\rhy = \csm$. 
The equal treatment sizes further ensure that $\Pi = Q^{-1}I_Q$ such that $\csp \einv \rho_Y^\T = Q \csp   \csm^\T =0$. 
This, together with $\ttrp   =  \csp\hyrs $ by Proposition \ref{prop:2k_g}, ensures
\begina
\ttrp     = \csp \hybrs - \csp\einv \rhy^\T   (\rhy \einv \rhyt)^{-1} \rhy\hybrs  =  \csp \hybrs.
\enda 

The asymptotic results then follow from Lemma \ref{lem:yb}. 
\end{proof}

\subsection{Factorial effects under $\{0,1\}$-coded regressions}
First, it follows from 
\begina
\Gamma_0 1_Q = \left\{\otimes_{k=1}^K \beginp 1&0\\ -1 & 1\endp \right\} \left\{ \otimes_{k=1}^K \beginp 1\\1\endp \right\} = \otimes_{k=1}^K \beginp 1&0\\ -1 & 1\endp\beginp 1\\1\endp   = \otimes_{k=1}^K \beginp 1\\0\endp = \beginp 1\\0_{Q-1}\endp
\enda
that $C_0$ is indeed a contrast matrix. 

\begin{proof}[Proof of Proposition \ref{prop:2k_0}]
Recall from \eqref{eq:t_k} that 
$
t_i 
= \otimes_{k=1}^K (1-\zikz, \zikz)^\T$. 
Let $f_i^0 = \otimes_{k=1}^K (1, \zikz)^\T  $ be the analog of $f_i = \otimes_{k=1}^K (1, \zik)^\T$, vectorizing $\{1, \zimkz: \mk\in\pk\}$ from \eqref{eq:lm_l_2k_0}. 
It follows from
\begina 
 \beginp
 1 - \zikz\\
 \zikz
 \endp =  
 \beginp
 1 & -1\\
 0 & 1
 \endp 
 \beginp
 1 \\
 \zikz
 \endp \qquad (k = \ot{K}) 
\enda
that 
$
t_i = \Gamma_0^\T f_i^0
$. 
The regressor vectors of \eqref{eq:lm_l} and \eqref{eq:lm_l_2k_0} thus satisfy
\begina
\beginp
t_i \\
t_i\otimes x_i
\endp =
\beginp
 \Gamma_0^\T & \\
& \Gamma_0^\T\otimes I_J
 \endp  \beginp
f^0_i \\
f^0_i\otimes x_i
\endp,
\enda
analogous to \eqref{eq:linear transformation_standard}. 
All results in Proposition \ref{prop:2k_g} and Corollary \ref{prop:2k_g} then follow from the same reasoning as that under the $\{-1, +1\}$ coding system. 
\end{proof}

\end{document}